\documentclass[10pt]{elsarticle}

\usepackage{amssymb}

\usepackage[T1]{fontenc}
\usepackage[utf8]{inputenc}
\usepackage[english]{babel}
\usepackage{times}
\usepackage{graphicx}
\usepackage[a4paper,twoside,width=145mm,top=35mm,bottom=35mm,bindingoffset=16mm]{geometry} 

\usepackage{float}
\usepackage{multirow}
\usepackage{amsmath}
\usepackage{amsmath,amssymb}
\usepackage{epsfig}
\usepackage{feynmp}
\usepackage{hyperref}
\usepackage{color}
\hypersetup{
	linktoc=all
}

\def\bk{\mathbf{k}}

\def\bx{\mathbf{x}}

\def\cD{{\mathcal D}}
\def\cJ{{\mathcal J}}
\def\cH{{\mathcal H}}
\def\cL{{\mathcal L}}

\def\cN{{\mathcal N}}

\def\cP{{\mathcal P}}

\def\cZ{{\mathcal Z}}

\def\mR{{\mathbb{R}}}

\def\mZ{{\mathbb{Z}}}

\def\ol{\overline}
\def\ul{\underline}
\def\wt{\widetilde}

\def\tJ{{\tilde{J}}}
\def\tw{{\tilde{\omega}}}
\def\hn{\widehat{n}}
\def\hP{\widehat{P}}

\def\btimes{\hbox{\boldmath $\times$}}

\def\be{\begin{equation}}
\def\ee{\end{equation}}
\def\lb{\label}

\journal{Physica D}

\begin{document}

\begin{frontmatter}



\title{\Huge 4-wave dynamics in\\ kinetic wave turbulence}


\author{Sergio Chibbaro\corref{cor1}}
\address{Sorbonne University, UPMC Univ Paris 06, CNRS, UMR 7190, Institut Jean Le Rond d'Alembert, F-75005 Paris, France}
\ead{chibbaro@ida.upmc.fr}
\cortext[cor1]{Corresponding author}
\author{Giovanni Dematteis}
\address{Dipartimento di Scienze Matematiche, Politecnico di Torino, Corso Duca degli Abruzzi 24, I-10129 Torino, Italy}
\author{Lamberto Rondoni}

\address{Dipartimento di Scienze Matematiche, Politecnico di Torino, Corso Duca degli Abruzzi 24, I-10129 Torino, Italy;\\
INFN, Sezione di Torino, Via P. Giuria 1, I-10125, Torino, Italy;\\
MICEMS, Universiti Putra Malaysia, 43400 Serdang Selangor, Malaysia}

\begin{abstract}
A general Hamiltonian wave system with quartic resonances is considered, in the standard kinetic limit of 
a continuum of weakly interacting dispersive waves with random phases. The evolution equation for the 
multimode characteristic function $Z$ is obtained within an ``interaction representation'' and a 
perturbation expansion in the small nonlinearity parameter. A frequency renormalization is performed 
to remove linear terms that do not appear in the 3-wave case. Feynman-Wyld diagrams are used to average 
over phases, leading to a first order differential evolution equation for $Z$. 
A hierarchy of equations, analogous to the Boltzmann hierarchy for low density gases is derived, which 
preserves in time the property of random phases and amplitudes. This amounts to a general
formalism for both the $N$-mode and the 1-mode PDF equations for 4-wave turbulent systems,
suitable for numerical simulations and for investigating intermittency. Some of the main results which are developed here in details have been tested numerically in a recent work.
\end{abstract}

\begin{keyword}
Weak wave turbulence \sep Kinetic theory \sep Intermittency


\end{keyword}

\end{frontmatter}


\section{Introduction}

Wave Turbulence (WT) theory concerns the dynamics of dispersive waves that interact nonlinearly over 
a wide range of scales~\cite{Nazarenko11}. 
In general the nonlinear interaction can be considered small, 
allowing a perturbative analysis
and then an asymptotic closure for statistical observables~\cite{newell2011wave}. 
For this reason, sometimes one then talks about Weak Wave Turbulence (WWT).
Until recently, most of the attention was given  to
the energy spectrum, which is governed by a kinetic
equation. Wave turbulence also 
provides exact solutions 
of the kinetic equation, which are related to equipartition, Rayleigh-Jeans solution, or stationary cascade, Kolmogorov-Zakharov solutions~\cite{Zakharovetal92}.
Many physical phenomena are studied within this general framework, for instance gravity~\cite{Hasselman,Falcon,Onorato,nazarenko2016wave}, capillary or Alfv\`en waves~\cite{zakcap67,PusZak96,CapFauve09,Galtier}, non-linear optics~\cite{Suret}  and elastic 
plates~\cite{Dur_06,Bou_08,Mor_08}.  Furthermore, applications of WT to non dispersive systems such as the acoustic waves~\cite{zakharov1970spectrum,l1997statistical} exist, even though the necessary statistical closure is subtler in such cases~\cite{newell1971semidispersive,shi2016resonance}.

In the last years, many experiments and numerical simulations were performed to verify the predictions of 
WT. The picture is relatively clear in the case of the capillary waves on a fluid surface 
(water, ethanol, liquid hydrogen or liquid helium): both experiments and numerical simulations confirm the 
Kolmogorov-Zakharov spectrum predicted by WT in this case. 
For other cases, {\em e.g.}
surface gravity waves or waves in vibrating elastic plates, the picture is more complicated: both numerics and 
experiments showed deviations from theoretical predictions, and the presence of intermittency~\cite{Frisch,during2009breakdown,Fal_10,meyrand2015weak}. 
This was unexpected, since WT appears as a mean-field theory, based on an
initial ``quasi-gaussianity'', previously believed to prevent sensible deviations from gaussianity.

An important step forward in this context has been the development
of a more efficient formalism for non-gaussian wavefields~\cite{choi2004probability,Choietal05b,JakobsenNewell04,Nazarenko11}. 
In particular, these works pointed out that probability density functions (PDF) are the 
relevant statistical objects to be analysed, reviving the interest in the
study of PDFs in WT, that dates back to the works of Peierls, Brout, Prigogine, Zawlaski and Sagdeev~\cite{peierls1929kinetischen,brout1956statistical,zaslavskii1967limits}. These authors had considered 
waves in anharmonic crystals, which constitute a special case of 3-wave systems. In the recent developments a diagrammatic approach was proposed~\cite{Nazarenko11}, based on Zakharov's pioneering work~\cite{monin2013statistical,zakharov1975statistical},
to analytically investigate PDF equations. Importantly, this has also clarified the role of the
different assumptions needed for the statistical closure.
In particular, the 3-wave resonant 
systems has been studied in details and a Peierls 
equation for the N-particles PDF has been proposed~\cite{choi2004probability,Choietal05b,Nazarenko11}. 

Nevertheless, the Peierls equation does not 
guarantee the strict preservation of the independence of phases and amplitudes, even though it can be argued that the property of \textit{random phases and amplitudes} (RPA) is preserved in a weaker form~\cite{Choietal05a,Nazarenko11}.
Starting from these premises, it has been shown that a proper normalization of the wave 
amplitudes is necessary for 3-wave resonant systems, in order to obtain a finite spectrum in 
the infinite-box limit, that leads to an amplitude density, dependent on the continuous variable 
$\bk$~\cite{Eyink}. In particular, the original amplitudes must be normalized by a factor 
scaling as ${1}/{V}$, where $V$ is the volume of the box. Adopting such a point of view, the 
Peierls equation for the multimode PDFs is not the leading-order asymptotic equation of the continuum 
limit of weakly interacting, incoherent waves. 
In Ref.\cite{Eyink}, then, new multimode equations were derived, that importantly have the factorized exponential solutions excluded by the Peierls equation. This is equivalent to the preservation of the RPA property. In turn, the preservation of exponential solutions 
implies a law of large numbers (LLN) for the empirical spectrum at times $\tau>0$, which is analogous 
to the propagation of chaos of the BBGKY hierarchy in the kinetic theory of gases. This LLN 
implies that the empirical spectrum satisfies  the wave-kinetic closure equations for 
nearly every initial realization of random phases and amplitudes, without necessity of averaging. Just as 
the Boltzmann hierarchy  has factorized solutions for factorized initial conditions, so does the kinetic
wave hierarchy for all multi-point spectral correlation functions. An $H$-theorem corresponding to 
positive entropy variation holds as well.
On the other hand, using these multimode equations, Ref.\cite{Eyink} shows that the 1-mode PDF 
equations is not altered by the different normalization,
if the modes initially enjoy the RPA property.

The 4-wave case has not yet been dealt with, although a formal analogy has been used to propose a possible 
extension of the 3-wave result to the 4-wave case~\cite{Choietal05a}.
Therefore, the present paper is devoted to the case of 4-wave interactions, which is of particular interest. As a matter of fact, most of the known violations of gaussianity arise in gravity waves and in vibrating elastic plates, which 
are 4-wave resonant systems. 
Following the same diagrammatic approach of Ref.\cite{Nazarenko11}, and using the normalization proposed in
Ref.\cite{Eyink}, we first explicitly derive the
continuos multimode equations, and then we obtain the equation for the $M$-mode PDF equation. 
These equations are different from the Peierls equations obtained by the formal analogy of Ref.\cite{Choietal05a};
they constitute instead a direct extension of the 3-wave case treated in Ref.\cite{Eyink}. The 
relation between the Peierls and our equations is thus discussed, showing the limit in which they 
coincide. Our framework also sheds some light on the issue of WT intermittency, as demonstrated by
a companion paper~\cite{chibbaroetal2017}, in which the equations obtained here are confirmed by 
numerical simulations of two 4-wave resonant Hamiltonian systems.

This work is organized as follows. First, we describe our model and notation, which are 
consistent with previous works~\cite{Nazarenko11,Eyink}. Section 2 discusses the probabilistic properties of RPA
fields. The main results of this paper are reported in sections 3 and 4, where the multimode 
equations are derived and discussed.  
In Section 3 the spectral generating functional and correlation functions are considered,
while Section 4 concerns the PDF generating function and the multipoint PDFs.
Section 5 summarizes our results.
Technical details are provided in Appendix  A, Appendix  B, and Appendix C, in which we also
briefly explain the diagrams used to calculate the  averages.

\subsection{Model and notation} \lb{model}
Similarly to~\cite{Eyink}, we consider a complex wavefield $u(\textbf{x},t)$ 
in a $d$-dimensional periodic cube with side $L$. This field is a linear combination of the canonical 
coordinates and momenta.
It is assumed that there is a maximum wavenumber $k_{\max},$
to avoid ultraviolet divergences. This can be achieved by a lattice regularization with spacing $a=L/M,$ 
for some large integer $M,$ so that $k_{\max}=\pi/a.$ The location variable $x$ then ranges over the physical space
\begin{equation}
\Lambda_{L}=a \mathbb{Z}_M^d,
\end{equation}
with the usual notation $\mathbb{Z}_M$ for the field of integers, modulo $M.$ This space has volume $V=L^d.$ 
The dual space of wavenumbers is 
\begin{equation}
\Lambda^*_{L}= \frac{2\pi}{L}\mathbb{Z}_M^d
\end{equation}
with $k_{\min}=2\pi/L.$ The total number of modes is $N=M^d$, so that $V=N a^d.$
The following index notation will be used:
\begin{equation}
 u^\sigma(\textbf{x}) = \left\{\begin{array}{ll}
                                            u(\textbf{x})\, & \sigma=+1\cr
                                            u^*(\textbf{x}) & \sigma=-1
                                            \end{array} \right.
\end{equation}
for $u$ and its complex-conjugate $u^*$. Likewise, we adopt the
convention for (discrete) Fourier transform:                                      
\begin{equation}
A^\sigma(\textbf{k})= \frac{1}{N}\sum_{\textbf{x}\in\Lambda_L}u^\sigma(\textbf{x},t)\exp(-i\sigma \textbf{k}\cdot\textbf{x})
\end{equation}
so that $A^+(\textbf{k})$ and $A^-(\textbf{k})$ are complex conjugates. 
This quantity converges to the continuous Fourier transform $ \frac{1}{L^d} \int_{[0,L]^d}
d^dx\,\,u^\sigma(\textbf{x},t)\exp(-i\sigma \textbf{k}\cdot\textbf{x})$ in the limit $a\rightarrow 0.$ The discrete inverse transform is 
\begin{equation}
u^\sigma(\textbf{x}) = \sum_{\textbf{k}\in\Lambda_L^*} A^\sigma(\textbf{k})\exp(i\sigma \textbf{k}\cdot\textbf{x}).
\end{equation}
The dynamics is assumed to be canonical Hamiltonian, with a $4^{th}$ power term in the Hamiltonian density 
(energy per volume) describing 4-wave interactions. 
As in~\cite{Zakharovetal92} and with lattice regularization, we write:
\begin{equation}
H=H_0+\delta H ~, \quad H_0 = \frac{1}{2}\sum_{\bk\in\Lambda^*} \omega_\bk  |A_\bk|^2 
\ee

Taking the most general Hamiltonian with any kind of 4-wave interactions,~\cite{Zakharovetal92}, one can 
write $\delta \mathcal{H}$ in the symmetrized compact form:
\begin{equation}\label{eq: deltaH}
\delta H=\epsilon \sum_{1234} \mathcal{H}^{\sigma_1 \sigma_2 \sigma_3 \sigma_4}_{1234}\:A_1^{\sigma_1}A_2^{\sigma_2}A_3^{\sigma_3} A_4^{\sigma_4} \:\delta_{1234}
\end{equation}
with the coefficients satisfying the general relations:
\begin{equation}
\left(\mathcal{H}^{\sigma_1 \sigma_2 \sigma_3 \sigma_4}_{1234}\right)^* = 
\mathcal{H}^{-\sigma_1 -\sigma_2 -\sigma_3 -\sigma_4}_{1234} ~, \quad
\mathcal{H}^{\sigma_1 \sigma_2 \sigma_3 \sigma_4}_{1234} = 
\mathcal{H}^{\Pi \left(\sigma_1 \sigma_2 \sigma_3 \sigma_4\right)}_{\Pi \left(1234\right)}. 
\label{eq: propH}
\end{equation}
$\Pi \in S^4$ represents any permutation of the four elements.
Introducing further notation:
\begin{eqnarray}\label{eq: defvec} 
&&\underline{\sigma} \doteq \left(\sigma_1, \sigma_2, \sigma_3, \sigma_4\right) ~ , \qquad 
\underline{\mathbf{k}}  \doteq \left(\mathbf{k}_1, \mathbf{k}_2, \mathbf{k}_3, \mathbf{k}_4\right) ~ , \qquad
\delta_{\underline{\sigma}\cdot \underline{\mathbf{k}},\mathbf{0}} = \delta_{\sigma_1 \mathbf{k}_1+\sigma_2 \mathbf{k}_2+\sigma_3 \mathbf{k}_3+\sigma_4 \mathbf{k}_4,\mathbf{0}} \nonumber \\
&&\omega_{1}    \doteq \omega \left(\mathbf{k}_1\right) ~ , \qquad
A_1                              \doteq A\left(\mathbf{k}_1\right) ~ , \qquad
\sum_{1}                      \doteq \sum_{\sigma_1=\pm1} \sum_{\mathbf{k}_1 \in \Lambda^*}
\end{eqnarray}
the Hamiltonian can be written as:
\begin{equation}\label{eq: H}
H=\frac{1}{2}\sum_{1}\omega_{1} A^{\sigma_1}_{1} A^{-\sigma_1}_{1} + \epsilon \sum_{1234} \mathcal{H}^{\underline{\sigma}}_{\underline{\mathbf{k}}}\:A_1^{\sigma_1}A_2^{\sigma_2}A_3^{\sigma_3} A_4^{\sigma_4} \:\delta_{\underline{\sigma}\cdot \underline{\mathbf{k}},\mathbf{0}}
\end{equation}
which leads to
\be 
\frac{\partial A^{\sigma}_{\mathbf{k}}}{\partial t}=i \sigma \omega_{\mathbf{k}} A^{\sigma}_{\mathbf{k}} +\epsilon \sum_{234}\mathcal{L}^{\sigma \sigma_2 \sigma_3 \sigma_4}_{\mathbf{k}234}A_2^{\sigma_2}A_3^{\sigma_3} A_4^{\sigma_4}\delta_{- \mathbf{k}+\sigma_2 \mathbf{k}_2+\sigma_3 
\mathbf{k}_3+\sigma_4 \mathbf{k}_4,\mathbf{0}} \label{dineq} 
\ee
where 
\be \lb{LH}
\mathcal{L}^{\sigma \sigma_2 \sigma_3 \sigma_4}_{\mathbf{k}234} 
\doteq 4i\sigma\mathcal{H}^{(-\sigma) \sigma_2 \sigma_3 \sigma_4}_{\mathbf{k}234}.
\ee
Changing $\mathbf{k}\rightarrow \mathbf{k}_1$ and introducing the interaction representation\footnote{Such 
a representation eliminates the fast linear oscillations, giving a variable
$a_\bk^\sigma$ that does not oscillate on fast scales.}
$A^{\sigma}_{\mathbf{k}}=a^{\sigma}_{\mathbf{k}} e^{i \sigma \omega_{\mathbf{k}} t}$,
one obtains\footnote{In our derivation, for simplicity and no loss of generality, we consider $\sigma=+1$. 
Trivially, the equations with $\sigma=-1$ are redundant, because obtained by complex conjugation of the 
ones with $\sigma=+1$. From now $a_1$ stands for $a_1^+$.}:
\begin{eqnarray} \lb{eq: after-int-repr}
\frac{\partial a_1}{\partial t}&=&\epsilon \sum_{234}\mathcal{L}^{+ \sigma_2 \sigma_3 \sigma_4}_{1234}a_2^{\sigma_2}a_3^{\sigma_3} a_4^{\sigma_4} \nonumber\\
&& \btimes \exp \left[i \left(-\omega_1+ \sigma_2 \omega_2 + \sigma_3 \omega_3 + \sigma_4 \omega_4 \right) t\right] \delta_{- \mathbf{k}_1+\sigma_2 \mathbf{k}_2+\sigma_3 \mathbf{k}_3+\sigma_4 \mathbf{k}_4,\mathbf{0}}
\end{eqnarray}
With notation~\cite{Eyink}:
\begin{eqnarray}
\mathcal{L}_{1234}  &\doteq& \mathcal{L}^{+ \sigma_2 \sigma_3 \sigma_4}_{1234} ~, \qquad
\omega^1_{234} \doteq -\sigma_1 \omega_1+ \sigma_2 \omega_2 + \sigma_3 \omega_3 + \sigma_4 \omega_4 \nonumber \\
\delta^1_{234}        & \doteq& \delta_{- \sigma_1 \mathbf{k}_1+\sigma_2 \mathbf{k}_2+\sigma_3 \mathbf{k}_3+\sigma_4 \mathbf{k}_4,\mathbf{0}}\nonumber \\
\end{eqnarray}
the dynamical equation of motion with 4-wave interactions now reads:
\begin{equation}\label{eq: dineq}
\dot{a}_1=\epsilon \sum_{234}\mathcal{L}_{1234}a_2^{\sigma_2}a_3^{\sigma_3} a_4^{\sigma_4} \exp \left(i \omega^1_{234} t\right) \delta^1_{234}
\end{equation}

\section{Fields with random phases and amplitudes}\lb{RPA}
In derivations of wave kinetic equations, it is often assumed that initial fields have Fourier coefficients 
with random statistically independent phases and amplitudes (RPA).  This property is 
expected to be preserved in time, in some suitable sense, in the wave-kinetic limit.

Let $N$ complex-valued random variables $a_\bk,$ $\bk\in \Lambda_L^*$ be the Fourier 
coefficients of a random field:
\be u_L(\bx) = \sum_{\bk\in\Lambda_L^*} a_\bk\exp(i\bk\cdot\bx).\ee
Here $a_\bk$ corresponds to $a^+_\bk$, i.e. $A^+_\bk$ in the previous section (no distinction need be made between 
the two at time $t=0$).
It will be crucial in the following to work with normalized variables
\be \wt{a}_{\bk}= \left(\frac{L}{2\pi}\right)^{d/2} a_{\bk} \ee 
which are assumed to remain finite in the large-box limit $L\rightarrow\infty.$ This normalization is 
sufficient for the spectrum of the random field to be well defined in that limit, as first pointed 
out in~\cite{Eyink}.
It is convenient to write the complex variables in polar coordinates (action-angle variables, or amplitudes and phases)
\be a_\bk=\sqrt{J_\bk}e^{i\varphi_\bk}=\sqrt{J_\bk}\psi_\bk \ee
with normalized action defined by 
\be \wt{J}_{\bk}= \left(\frac{L}{2\pi}\right)^{d} J_{\bk}.\ee
We denote by $s_\bk$ and $\xi_\bk$ for possible values of the random variables $\wt{J}_\bk\in \mR^+$
and $\psi_\bk=e^{i\varphi_\bk}\in S^1.$
\be 
d\mu(s,\xi)  = \prod_{\bk\in \Lambda^*_L} ds_\bk  \frac{|d\xi_\bk|}{2\pi} 
\lb{liouville2} 
\ee
suitably normalized.
The $N$-mode joint probability density function $\cP^{(N)}(s,\xi)$ is defined
with respect to the Liouville measure, such that the average of the random variable $f_{\tJ\psi}(s,\xi)$ is given by
\be\langle f_{\tJ\psi}\rangle=\int d\mu(s,\xi)\, \cP^{(N)}(s,\xi) f(s,\xi) \ee
where the integral is over $(s,\xi)$ in the product space $\left(\mR^+\right)^N\times \left(S^1\right)^N.$

The field $u_L(\bx)$ is called a {\it random-phase field} (RP) if for all $\bk\in \Lambda_L^*$ the
$\psi_\bk=e^{i\varphi_\bk}$ are independent and identically distributed (i.i.d.)  
random variables, uniformly distributed over the unit circle $S^1$ in the complex plane~\cite{Nazarenko11}. 
For the joint 
PDF, this is equivalent to:  
\be
\cP^{(N)}(s,\xi)=\cP^{(N)}(s)
\ee
Note that an RP $u_L(\bx)$ is a homogeneous random field on $\Lambda_L$, 
statistically invariant under space-translations by the finite group $a\mZ_M^d$.
In the limit $L\rightarrow\infty$ the field $u_L(\bx)$ defined with appropriately chosen $\tJ_{\bk,L}$
will converge to a homogeneous random field $u(\bx)$ invariant under translations by $a\mZ^d$. The
standard definition of the spectrum $n(\bk)=\lim_{L\rightarrow\infty}(L/2\pi)^d\langle |a_{\bk, L}|^2\rangle$
implies that one must choose \\
\be 
\lim_{L\rightarrow \infty}\langle \tJ_{\bk_L,L}\rangle=n(\bk), 
\lb{spectrum} 
\ee
for $\bk\in \Lambda^*=[-k_{\max},+k_{\max}]^d,$ where $\bk_L= \frac{\bk L}{2\pi} ({\rm mod}\,M)
\cdot \frac{2\pi}{L}\in \Lambda_L^*$ converges to $\bk$ as $L=aM\rightarrow\infty$ (for fixed $a$). 
So, $u_L(\bx)$ converges in distribution to a homogeneous field 
$u(\bx)$ with spectrum $n(\bk)$. 

Let $u_L(\bx)$ be a {\it random-phase and amplitude field}
(RPA) if $u_L(\bx)$ is RP and if also $\tJ_\bk$ are mutually independent random variables for all $\bk\in \Lambda_L^*.$ 
This is equivalent to the factorization of the $N$-mode PDF into a product of 1-mode PDFs:
\be
\cP^{(N)}(s)=\prod_{\bk\in \Lambda_L^*} P(s_\bk;\bk). 
\ee
All homogeneous Gaussian random fields are RPA. Conversely, for any sequence of RPA fields satisfying condition
(\ref{spectrum}) the spatial field $u_L(\bx)$ converges in distribution to the homogeneous Gaussian field with mean 
zero and spectrum $n(\bk)$ as $L\rightarrow\infty$ \cite{Kurbanmuradov95}. Here we note only that 
\be u_L(\bx) = \left(\frac{2\pi}{L}\right)^{d/2} \sum_{\bk\in\Lambda_L^*} \sqrt{\tJ_{\bk,L}} \exp(i\bk\cdot\bx+i\varphi_\bk) \ee
is a sum of $N$ independent variables scaled by $1/\sqrt{N}.$ It is important to emphasize that the Fourier coefficients 
$\wt{a}_{\bk,L}$ can remain far from Gaussian in this limit. In physical space also there are non-vanishing cumulants 
for large but finite $L.$

Let us define the characteristic  functional, containing information about the statistical distribution of amplitudes and phases:
\be \cZ_L(\lambda,\mu)=\left\langle \exp\left[\int d\bk (i\lambda_\bk J_\bk + i \mu_\bk\varphi_\bk)\right]\right\rangle \label{genfuncintro}\ee

A most important result for RPA fields is that the {\it empirical spectrum}  
\be\hn_L(\bk) = \left(\frac{2\pi}{L}\right)^d \sum_{\bk_1\in \Lambda_L^*} \tJ_{\bk_1,L}\delta^d(\bk-\bk_1), 
\,\,\,\, \bk\in \Lambda^* \ee
converges under the condition (\ref{spectrum}) to the deterministic spectrum $n(\bk)$ with probability going to 1 
in the limit $L\rightarrow\infty$ (weak LLN). One can show that $\int d^dk\,\,\lambda(\bk)\;\hn_L(\bk)$ 
converges in probability to $\int d^dk\,\,\lambda(\bk)n(\bk)$ for every bounded, continuous $\lambda.$
This is sufficient to infer that the amplitude characteristic function defined in (\ref{genfuncintro}) satisfies
\be  
\lim_{L\rightarrow\infty} \cZ_L(\lambda)=\exp\left(i\int d^dk\,\,\lambda(\bk)n(\bk)\right) 
\lb{Zexp0} 
\ee
with $n(\bk)$ the deterministic spectrum. The LLN means that 
for RPA fields the empirical spectrum $\hn_L(\bk)$ coincides with $n(\bk)$ at large $L$ for almost 
every realization of the random phases and amplitudes.  

Notice that for the above result one does not actually need the full independence assumption in RPA, but it suffices that 
\be  \lim_{L\rightarrow \infty} [\cN^{(2)}_L(\bk_1,\bk_2)-\cN^{(1)}_L(\bk_1) \cN^{(1)}_L(\bk_2)]=0, \lb{stosszahl} \ee
where the M-th order correlation functions are defined as
\be\cN^{(M)}_L(\bk_1,...,\bk_M)=\langle \hn_L(\bk_1)\cdots \hn_L(\bk_M)\rangle. \lb{correl} \ee
Property (\ref{stosszahl}) is analogous to Boltzmann's {\it Stosszahlansatz} for his kinetic equation.  
Under this assumption, the $M$-th order correlations that exist will factorize in the large-box limit
\cite{Lanford75,Lanford76}:
\be  
\lim_{L\rightarrow\infty} \cN^{(M)}_L(\bk_1,...,\bk_M)=\prod_{m=1}^M n(\bk_m). \lb{spect-fac} 
\ee
Our results indicate that properties RP and (\ref{spectrum},\ref{stosszahl}) for the initial
wave field, suffice for the wave kinetic 
equation and for the LLN for the empirical spectrum to hold at positive times.  

RPA fields whose Fourier amplitudes possess the full independence property satisfy the even
stronger LLN for the {\it empirical 1-mode PDF} 
\be 
\hP_L(s; \bk) = \left(\frac{2\pi}{L}\right)^d \sum_{\bk_1\in \Lambda_L^*} 
     \delta(s-\tJ_{\bk_1})\delta^d(\bk-\bk_1).  
\lb{emp-pdf-1} 
\ee
Assume that the limiting random variables $\tJ_\bk=\lim_{L\rightarrow\infty}\tJ_{\bk_L, L}$ 
of an RPA field exist and have PDFs $P(s;\bk)$ which are continuous in $\bk.$ Then, 
the random functions $\hP_L(s; \bk)$ converge to $P(s;\bk)$
with probability approaching 1 as $L\rightarrow\infty.$ This implies the previous LLN for the 
spectrum, since $\hn_L(\bk)=\int_0^\infty ds\, s \hP_L(s;\bk)$ and $n(\bk)=\int_0^\infty ds\, s P(s;\bk).$
Although the ``empirical PDF'' defined in (\ref{emp-pdf-1}) is mathematically very convenient, 
it is not a PDF for finite $L$. It is therefore more intuitive to use an alternative definition    
\be  \hP_L(s; \Delta) = \frac{1}{N_L(\Delta)}\sum_{\bk\in \Lambda_L^*\cap \Delta} \delta(s-\tJ_\bk), 
\lb{emp-PDF-2} \ee
for any open set $\Delta \subset \Lambda^*$ and with $N_L(\Delta)$ the number of elements 
in $\Lambda_L^*\cap \Delta$. This quantity is nearly the same as $\frac{1}{|\Delta|} \int_\Delta 
d^dk \,\, \hP_L(s; \bk)$ for large $L$ but it has the advantage that it defines a probability measure 
in $s$ for each fixed $\Delta$ and $L.$ Definition (\ref{emp-PDF-2}) also has a simple intuitive 
meaning, since it represents the instantaneous distribution of amplitudes of the large number of 
Fourier modes that reside in the set $\Delta$ for large box-size $L.$ Under the same assumptions 
as above, it follows with probability going to 1 that 
 \be 
\lim_{L\rightarrow\infty} \hP_L(s;\Delta) =\frac{1}{|\Delta|} \int_\Delta d^dk \,\,P(s;\bk)\equiv P(s; \Delta). 
\ee
Strict independence is not necessary for this to hold; 
factorization of {\it multimode PDFs} for 
$\bk_1,...,\bk_M\in \Lambda^*$ is required:
\be \cP^{(M)}_L(s_1,...,s_M;\bk_1,...,\bk_M)
=\langle \delta(s_1-\tJ_{\bk_{1,L},L})\cdots \delta(s_M-\tJ_{\bk_{M,L},L})\rangle. \ee
The factorization property for all pairs of distinct $\bk_1,\bk_2\in \Lambda^*$
\be  \lim_{L\rightarrow \infty} [\cP^{(2)}_L(s_1,s_2;\bk_1,\bk_2)-\cP_L^{(1)}(s_1; \bk_1)\cP_L^{(1)}(s_2;\bk_2)]=0
 \ee
suffices for the LLN for the empirical PDF and also the factorization of the multimode PDFs 
\be \lim_{L\rightarrow \infty}\cP^{(M)}_L(s_1,...,s_M;\bk_1,...,\bk_M)=
    \prod_{m=1}^M P(s_m;\bk_m) \lb{fact-pdf2}\ee
for all integers $M>2$ and distinct $\bk_1,...,\bk_M\in \Lambda^*.$  The asymptotic independence
is considerably weaker than RPA, permitting statistical dependence between 
Fourier modes at finite $L.$  
In the following, we show that properties 
(\ref{spect-fac}), (\ref{fact-pdf2}) are preserved by the limiting kinetic hierarchies of 
WT.

\section{Multimode hierarchy equations}\label{Sec: dinmulti}
In this section we formally derive the multimode kinetic equations for the 4-wave dynamics of our system.
Our analysis differs from those of previous works~\cite{Eyink,Choietal05a, Choietal05b}
mainly because of the nonlinear frequency shift, and because of the details of the
$L\rightarrow\infty$ and $\epsilon\rightarrow 0$ limits.

The action-angle variables (amplitudes and phases) for linear dynamics are defined as 
$J_{\bk}=|A_{\bk}^\sigma|^2$ and $\varphi_\bk=\sigma \arg(A_\bk^\sigma),$
so that $A_{\bk}^\sigma=\sqrt{J_{\bk}}\psi_{\bk}^\sigma$, where $\psi_{\bk}=\exp(i\varphi_{\bk})$. 
Then, the Liouville measure $\mu$ conserved by the Hamiltonian flow can be written as 
\be 
d\mu=\prod_\bk dQ_\bk dP_\bk = \prod_\bk \frac{1}{i}dA^+_\bk dA^-_\bk = \prod_\bk \frac{1}{i}da^+_\bk da^-_\bk 
=\prod_\bk dJ_\bk d\varphi_\bk 
\ee
The canonical momenta and coordinates are given by real and imaginary parts of
$A^\sigma_\bk=\frac{1}{\sqrt{2}}(P_\bk + i\sigma Q_\bk)$, and $A^\sigma_\bk$ and $a^\sigma_\bk$ 
are linked by the simple rotation in the complex plane used to obtain (\ref{eq: after-int-repr}).
Consistently with the general definition (\ref{genfuncintro}),
the generating function of amplitudes and phases for finite box-size $L$ is:
\be 
\cZ_L[\lambda,\mu,T]\doteq \left \langle \exp \left( \sum_{\bk \in \Lambda_L^*}\lambda_\bk J_\bk(T) \right) 
\prod_{\bk \in \Lambda_L^*} \psi_\bk^{\mu_\bk}(T)\right \rangle 
\label{eq: genfunct}
\ee
$$
\lambda_\bk \in \mR,\;\; \mu_\bk \in \mZ \quad \quad \forall \bk \in \Lambda^*_L
$$

\subsection{Power Series Expansion in the Dynamical Equation}\label{Sec: expansion}
\subsubsection{The frequency shift}\label{nonlinfreq}
Let us perturbatively expand the solution of Eq.(\ref{eq: dineq}) in 
$\epsilon$ at finite $L$. As explained in \cite{Eyink} and \cite{Nazarenko11}, we consider an 
intermediate time between the ``linear time'', that is the wave period, and the ``nonlinear time'' 
that represents the time scale of evolution of the wave amplitude statistics. To 
consider the long-time behavior of the wave field expanding in $\epsilon$ the solution of the 
dynamical equation, we need to renormalize the frequency \cite{Choietal05a, Nazarenko11}. 
The equation for the order zero in $\epsilon$ has a constant solution:
\be 
a_1^{(0)}(T)=a_1(0),
\label{eq: a0}
\ee
Thus, the terms like 
$\sum_{234} \cL_{1234} a_2^{(0)} a_3^{(0)} a_4 \exp \left(i \omega^1_{234} t\right) \delta^1_{234}$, 
for $\bk_2=\bk_3$, $\sigma_2=-\sigma_3$ and $\bk_4=\bk_1$, play the role of linear terms in 
$a_1$, responsible for fast oscillations. We want to remove all 
terms of this kind, using an interaction representation and a frequency renormalization \cite{Choietal05a}:
\begin{eqnarray}
\left.\sum_{234}\right.^{**}&\doteq&\sum_{\sigma_2\sigma_3\sigma_4}\sum_{\bk_2\bk_3\bk_4}\delta_{\sigma_2,\sigma_1}\delta_{\sigma_3,-\sigma_4}\delta_{\bk_2,\bk_1}\delta_{\bk_3,\bk_4} + (2\leftrightarrow3)+(2\leftrightarrow4)\label{star1a}\\
\left.\sum_{234}\right.^*&\doteq&\left(\sum_{\sigma_2\sigma_3\sigma_4}\sum_{\bk_2\bk_3\bk_4}-\left.\sum_{234}\right.^{**}\right)
\label{star1}\end{eqnarray}
Recalling Eq.(\ref{eq: dineq}), we can write:
\begin{eqnarray}
\dot{a}_1&=& \epsilon \bigg(\left.\sum_{234}\right.^*+\left.\sum_{234}\right.^{**}\bigg)\mathcal{L}_{1234}^{+\sigma_2\sigma_3\sigma_4}a_2^{\sigma_2}a_3^{\sigma_3} a_4^{\sigma_4} \exp \left(i \omega^1_{234} t\right) \delta^1_{234} \nonumber\\
&=& \epsilon \left.\sum_{234}\right.^*\mathcal{L}_{1234}^{+\sigma_2\sigma_3\sigma_4}a_2^{\sigma_2}a_3^{\sigma_3} a_4^{\sigma_4} \exp \left(i \omega^1_{234} t\right) \delta^1_{234}+i\,\Omega_1\, a_1+\epsilon^2 \cD_1 a_1 \label{nolinear0}
\end{eqnarray}
where
\be i\Omega_1 \doteq  \epsilon  \sum_{\sigma_2=\pm1}\sum_{\bk_2}\mathcal{L}_{1122}^{++\sigma_2-\sigma_2} \big|a_2^{(0)}\big|^2+(2\leftrightarrow3)+(2\leftrightarrow4)  \label{Omega}\ee
and $\cD_1=O(1)$.
Introducing a new interaction representation with
\be 
b_\bk= a_\bk e^{-i \Omega_\bk t} 
\label{ba}
\ee
Eq.(\ref{nolinear0}) becomes:
\be
\dot{b}_1= \epsilon \left.\sum_{234}\right.^*\mathcal{L}_{1234}^{+\sigma_2\sigma_3\sigma_4}\,b_2^{\sigma_2}b_3^{\sigma_3} b_4^{\sigma_4}\, e^{i \tw^1_{234} t}\, \delta^1_{234} + \epsilon^2\cD_1 b_1 \label{dineqb}
\ee
where the renormalized frequency with a shift is given by \cite{Choietal05a,Nazarenko11}:
\be 
\tw_\bk \doteq \omega_\bk+\Omega_\bk 
\label{tw}
\ee

\subsubsection{$2^{nd}$ order equations} \label{Sec: expansion2}
Considering an intermediate time between the linear and the nonlinear time 
($\frac{2\pi}{\tw_\bk}<<T<<\frac{2\pi}{\epsilon^2\tw_\bk}$), the solution of Eq.(\ref{dineqb}) 
to second order in $\epsilon$ is:
\be 
b_\bk(T)=b^{(0)}_\bk(T)+\epsilon b^{(1)}_\bk(T)+\epsilon^2 b^{(2)}_\bk(T)+O(\epsilon^3) 
\label{eq: powerseriesb}
\ee
which implies
\begin{align}
b_1^{(0)}(T)&=b_1(0) \label{eq: b0} \\
b_1^{(1)}(T)&=\left.\sum_{234}\right.^* \cL_{1234} b_2^{(0)}b_3^{(0)}b_4^{(0)} \Delta_T(\tw^1_{234}) \delta^1_{234} \label{eq: b1}\\
 b_1^{(2)}(T)&= \left.\sum_{234567}\right.^* \cL_{1234}\cL_{4567}\; b_2^{(0)}b_3^{(0)}b_5^{(0)}b_6^{(0)}b_7^{(0)}  E_T\left(\tw^1_{23567}, \tw^1_{234}\right) \delta^1_{234} \delta^4_{567}  \nonumber \\
& \qquad+ (4 \leftrightarrow 3) + (4 \leftrightarrow 2) + \int_0^T \cD_1 b_1^{(0)} dt \label{eq: b2}
\end{align}
where
\be 
\Delta_T(x) \doteq \int_0^T \exp (i x t) dt ~, \quad 
E_T(x,y) \doteq \int_0^T \Delta_t(x-y)\exp (i y t) dt 
\ee
and
\begin{eqnarray}
\left.\sum_{234567}\right.^{**}&\doteq&\sum_{\sigma_2\sigma_3...\sigma_7}\sum_{\bk_2\bk_3...\bk_7}\delta_{\sigma_2,\sigma_1}\delta_{\sigma_3,-\sigma_4}\delta_{\sigma_4,\sigma_5}\delta_{\sigma_6,-\sigma_7}\nonumber \\
&& \btimes\delta_{\bk_2,\bk_1}\delta_{\bk_3,\bk_4}\delta_{\bk_4,\bk_5}\delta_{\bk_6,\bk_7}+(2\leftrightarrow3)+(2\leftrightarrow4)\nonumber\\
&&+(5\leftrightarrow6)+(5\leftrightarrow7)+(2\leftrightarrow3,5\leftrightarrow6)+(2\leftrightarrow4,5\leftrightarrow6)\nonumber\\
&&+(2\leftrightarrow3,5\leftrightarrow7)+(2\leftrightarrow4,5\leftrightarrow7) \label{star2a}\\
\left.\sum_{234567}\right.^*&\doteq&\left(\sum_{\sigma_2\sigma_3...\sigma_7}\sum_{\bk_2\bk_3...\bk_7}-\left.\sum_{234567}\right.^{**}\right) \label{star2}\end{eqnarray}
\be \cD_1\doteq \sum_{\sigma_2=\pm1}\sum_{\bk_2}\mathcal{L}_{1122}^{++\sigma_{2}-\sigma_2} \Big(b_2^{(0)}b_2^{(1)*}+b_2^{(1)}b_2^{(0)*}\Big) +(2\leftrightarrow3)+(2\leftrightarrow4) \label{D1} \ee

\subsection{Phase averaging: Feynman-Wyld Diagrams}
In this section, we carry out the phase averaging using diagrammatic techniques, which are in essence 
those used in the 3-wave case in Ref.\cite{Choietal05b}. However, here we describe them in details, for completeness and also because 
we have introduced spin terms, $\sigma_i$, absent in~\cite{Choietal05b}.

An expansion like (\ref{eq: powerseriesb}) for the original normal variables $A_\bk$ may be written as
\be 
A_\bk(T)=A^{(0)}_\bk(T)+\epsilon A^{(1)}_\bk(T)+\epsilon^2 A^{(2)}_\bk(T)+O(\epsilon^3) 
\label{eq: powerseriesA}
\ee
where:
\be b_\bk^{(i)}=A_\bk^{(i)}e^{-i\tw t}, \quad i=0,1,2 \ee
and a similar expansion 
Equation (\ref{eq: powerseriesA}) leads to:
\be
J_\bk(T)={|A_\bk(T)|}^2={|b_\bk(T)|}^2 \doteq J^{(0)}_\bk + \epsilon J^{(1)}_\bk  + \epsilon^2 J^{(2)}_\bk + O(\epsilon^3)\label{eq: powerseriesJ}
\ee
Definition (\ref{eq: genfunct}) shows that $\cZ_L$ satisfies the simmetry:
\be 
\cZ_L\left[\lambda,\mu,T\right]=\cZ_L^*\left[\lambda,-\mu,T\right] 
\label{minusconj}
\ee
Therefore, writing
\be 
\cZ_L\left[\lambda,\mu,T\right]=\chi_L\left\{\lambda,\mu,T\right\}+\chi_L^*\left\{\lambda,-\mu,T\right\} 
\label{eq: Zchi}
\ee
one eventually gets:
\be 
\chi_L\left\{\lambda,\mu,T\right\}=\chi_L\left\{\lambda,\mu,0\right\} + \left\langle\prod_{\bk\in\Lambda_L^*}e^{\lambda_\bk J_\bk^{(0)}}\left[\epsilon \cJ_1 + \epsilon^2\left(\cJ_2+\cJ_3+\cJ_4+\cJ_5\right)\right]\right\rangle_J 
\label{eq: chi}
\ee
where \cite{Choietal05b}:
\begin{eqnarray}
&& \cJ_1\doteq\Big\langle\prod_{\bk}\psi_{\bk}^{(0) \mu_\bk}\sum_1 \Big(\lambda_1+\frac{\mu_1}{2J_1^{(0)}}\Big)b_1^{(1)}b_1^{(0)*}\Big\rangle_\psi \label{J1} \\
&& \cJ_2\doteq\frac{1}{2}\Big\langle\prod_{\bk}\psi_{\bk}^{(0)\mu_\bk}\sum_1\Big(\lambda_1+\lambda_1^2 J_1^{(0)}-\frac{\mu_1^2}{4 J_1^{(0)}}\Big)|b_1^{(1)}|^2\Big\rangle_\psi \label{J2} \\
&& \cJ_3\doteq\Big\langle\prod_{\bk}\psi_{\bk}^{(0)\mu_\bk}\sum_1 \Big(\lambda_1+\frac{\mu_1}{2J_1^{(0)}}\Big)b_1^{(2)}b_1^{(0)*}\Big\rangle_\psi \label{J3} \\
&& \cJ_4\doteq\Big\langle\prod_{\bk}\psi_{\bk}^{(0)\mu_\bk}\sum_1\Big(\frac{1}{2}\lambda_1^2+\frac{\mu_1}{4J_1^{(0)2}}(\frac{\mu_1}{2}-1)+
\frac{\lambda_1\mu_1}{2J_1^{(0)}}\Big)(b_1^{(1)}b_1^{(0)*})^2\Big\rangle_\psi \label{J4} \\
&& \cJ_5\doteq\frac{1}{2}\Big\langle\prod_{\bk}\psi_{\bk}^{(0)\mu_\bk}\sum_{1\neq 2}\Big(\lambda_1\lambda_2(b_1^{(1)}b_1^{(0)*}+b_1^{(1)*}b_1^{(0)})b_2^{(1)}b_2^{(0)*}\cr
&& \,\,\,\,\,\,\,\,\,\,\,\,\,\,\,\,\,\,\,\,\,\,\,\,\,\,\,\,\,\,\,\,\,\,
+(\lambda_1+\frac{\mu_1}{4J_1^{(0)}})\frac{\mu_2}{J_2^{(0)}}(b_2^{(1)}b_2^{(0)*}-b_2^{(1)*}b_2^{(0)})b_1^{(1)}b_1^{(0)*}\Big)\Big\rangle_\psi
\label{J5} 
\end{eqnarray}
The averages over phases and amplitudes have been separated. Furthermore,
\be \chi_L\left\{\lambda,\mu,0\right\} \doteq \left\langle \prod_\bk\exp\left[\lambda_\bk J^{(0)}_\bk\right] \right\rangle_{J}\ee

\subsubsection{Rules for \textit{phase-averaging}} \label{sec: rules}
The terms in the perturbative solution of the equation of motion can be represented by Wyld diagram expansions
\cite{Choietal05a,Eyink,Wyld61,ZakharovLvov75}. The main rules for such diagrams and 
for averages over phases follow.
\begin{itemize}
\item[\textit{Rule 1}] \textbf{How to build the basic diagrams}\\
The various contributions are represented by tree diagrams illustrated in Figs. 1-3, for the
zeroth-, first- and second-order terms, that we call ``basic diagrams''.
\begin{itemize}
\item Lines: a solid line labeled by an integer $j$ represents factor $b^{(0)}_j$;
dashed line indicates the absence of such a factor.
An arrow added to a solid line, pointing away from $j$, indicates $\sigma_j=+1$ (source);
if the arrow points toward $j$, it corresponds to $\sigma_j=-1$ (sink).
\item Vertices: the vertex labelled $1234$ represents  
$\cL^{\sigma_1,\sigma_2,\sigma_3,\sigma_4}_{1234}e^{\tw^1_{234}t}\delta^1_{234}$ 
with $\sigma_1=+1$ when the arrow points out of the vertex and $\sigma_1=-1$
when the arrow points into the vertex. The times at each vertex are ordered causally, with 
the latest times at the root of the tree, labelled by $1$. Integrating from time $0$ to $T$, 
one gets the various contributions to the perturbative solution.
\end{itemize}
\begin{figure}[htbp]
	\begin{center}
	    \unitlength = 1mm
	    \begin{fmffile}{figure0}
	        \begin{fmfgraph*}(35,10)
	            \fmfleft{i1}
	            \fmfright{o1}
	            \fmflabel{$1$}{i1}
	            \fmf{fermion}{i1,o1}
	        \end{fmfgraph*}
	    \end{fmffile} $\qquad\qquad$
	    \unitlength = 1mm
	    \begin{fmffile}{figure1}
	        \begin{fmfgraph*}(35,10)
	            \fmfleft{i1}
	            \fmfright{o1}
	            \fmflabel{$1$}{i1}
	            \fmf{fermion}{o1,i1}
	        \end{fmfgraph*}
	    \end{fmffile}
	    \caption{$b_1^{(0)+}$ and $b_1^{(0)-}$}
	    \label{Fig.: figure1}
	    \end{center}
\end{figure}
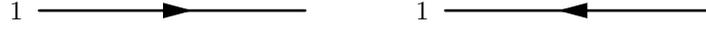
\begin{figure}[htbp]
	\begin{center}
	    \unitlength = 1mm
	    \begin{fmffile}{figure2a}
	        \begin{fmfgraph*}(35,35)
	            \fmfleft{i1}
	            \fmfright{o3}
	            \fmftop{o2}
	            \fmfbottom{o4}
	            \fmflabel{$1$}{i1}
	            \fmflabel{$2$}{o2}
	            \fmflabel{$3$}{o3}
	            \fmflabel{$4$}{o4}
	            \fmf{dashes_arrow}{v1,i1}
	            \fmf{plain}{v1,o2}
	            \fmf{plain}{v1,o3}
	            \fmf{plain}{v1,o4}
	        \end{fmfgraph*}
	    \end{fmffile} $\qquad\qquad$
	    \unitlength = 1mm
	    \begin{fmffile}{figure2b}
	        \begin{fmfgraph*}(35,35)
	            \fmfleft{i1}
	            \fmfright{o3}
	            \fmftop{o2}
	            \fmfbottom{o4}
	            \fmflabel{$1$}{i1}
	            \fmflabel{$2$}{o2}
	            \fmflabel{$3$}{o3}
	            \fmflabel{$4$}{o4}
	            \fmf{dashes_arrow}{i1,v1}
	            \fmf{plain}{v1,o2}
	            \fmf{plain}{v1,o3}
	            \fmf{plain}{v1,o4}
	        \end{fmfgraph*}
	    \end{fmffile}
	    \caption{$b_1^{(1)+}$ and $b_1^{(1)-}$}
	    \label{Fig.: figure2}
	    \end{center}
\end{figure}
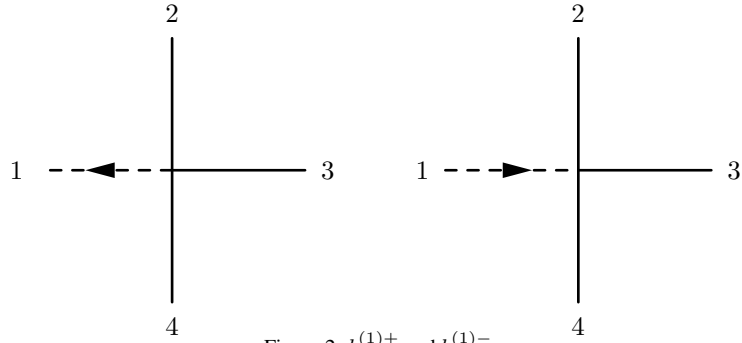
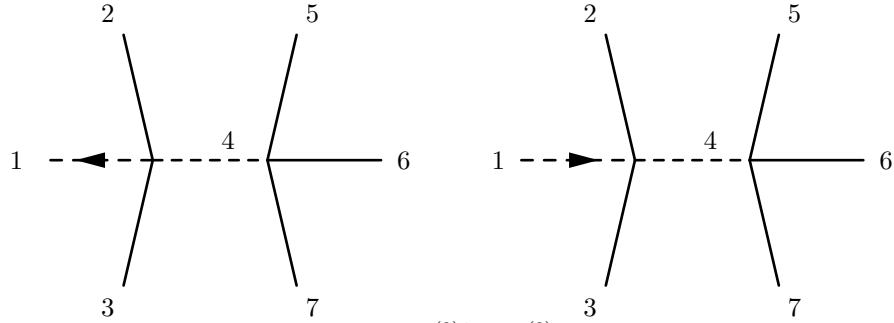
\begin{figure}[htbp]
	\begin{center}
	    \unitlength = 1mm
	    \begin{fmffile}{figure3a}
	        \begin{fmfgraph*}(45,35)
	            \fmfleft{i1}
	            \fmfright{o6}
	            \fmftop{i2,o2,i4,o5,i5}
	            \fmfbottom{i3,o3,i4b,o7,i7}
	            \fmflabel{$1$}{i1}
	            \fmflabel{$2$}{o2}
	            \fmflabel{$3$}{o3}
	            \fmflabel{$4$}{o4}
	            \fmflabel{$5$}{o5}
	            \fmflabel{$6$}{o6}
	            \fmflabel{$7$}{o7}
	            \fmf{dashes_arrow}{v1,i1}
	            \fmf{phantom}{i2,o2}	            
	            \fmf{plain}{o2,v1}
	            \fmf{phantom}{v1,i4,v2}
	            \fmf{phantom}{v1,i4b,v2}
	            \fmf{phantom}{i3,o3}
	            \fmf{plain}{o3,v1}
	            \fmf{dashes}{v1,o4,v2}
	            \fmf{plain}{v2,o5}
	            \fmf{phantom}{o5,i5}
	            \fmf{plain}{v2,o7}
	            \fmf{phantom}{o7,i7}
	            \fmf{plain}{v2,o6}
	        \end{fmfgraph*}
	    \end{fmffile} $\qquad\qquad$
	    	    \unitlength = 1mm
	    \begin{fmffile}{figure3b}
	        \begin{fmfgraph*}(45,35)
	            \fmfleft{i1}
	            \fmfright{o6}
	            \fmftop{i2,o2,i4,o5,i5}
	            \fmfbottom{i3,o3,i4b,o7,i7}
	            \fmflabel{$1$}{i1}
	            \fmflabel{$2$}{o2}
	            \fmflabel{$3$}{o3}
	            \fmflabel{$4$}{o4}
	            \fmflabel{$5$}{o5}
	            \fmflabel{$6$}{o6}
	            \fmflabel{$7$}{o7}
	            \fmf{dashes_arrow}{i1,v1}
	            \fmf{phantom}{i2,o2}	            
	            \fmf{plain}{o2,v1}
	            \fmf{phantom}{v1,i4,v2}
	            \fmf{phantom}{v1,i4b,v2}
	            \fmf{phantom}{i3,o3}
	            \fmf{plain}{o3,v1}
	            \fmf{dashes}{v1,o4,v2}
	            \fmf{plain}{v2,o5}
	            \fmf{phantom}{o5,i5}
	            \fmf{plain}{v2,o7}
	            \fmf{phantom}{o7,i7}
	            \fmf{plain}{v2,o6}
	        \end{fmfgraph*}
	    \end{fmffile}
	    \caption{$b_1^{(2)+}$ and $b_1^{(2)-}$}
	    \label{Fig.: figure3}
	    \end{center}
\end{figure}
For completeness, observe that:
\be
b_1^{(1)}(T)=\left.\sum_{234}\right.^* \cL_{1234} b_2^{(0)}b_3^{(0)}b_4^{(0)} \Delta_T(\tw^1_{234}) 
\delta^1_{234}
\ee
\begin{eqnarray}
\hskip -8pt
b_1^{(2)}(T)\hskip -7pt&=& \hskip -9pt\left.\sum_{234567}\right.^* \cL_{1234}\cL_{4567}\; b_2^{(0)}b_3^{(0)}b_5^{(0)}b_6^{(0)}b_7^{(0)} 
\int_0^T\Delta_t(\tw^4_{567}) \exp \left(i \tw^1_{234} t\right) dt
\delta^1_{234} \delta^4_{567}  \nonumber \\
&& + (4 \longleftrightarrow 3) + (4 \longleftrightarrow 2) + \int_0^T \cD_1 b_1^{(0)} dt 
\end{eqnarray}
\item[\textit{Rule 2}] \textbf{How to combine basic diagrams before phase-averaging}\\
Before averaging over phases, the various contributions (\ref{J1})-(\ref{J5}) can be represented by diagrams (see next section), 
combining the tree diagrams in Figs. \ref{Fig.: figure1}-\ref{Fig.: figure3}. The combination of two basic diagrams graphically represents the product of the two analytical terms to which the diagrams are associated, and this is performed by joining the trees with the same 
``root'' indices, over which there must be a sum. Each of the integer labels indicates an index to be summed over independently of the others, except for the constraints imposed by Kronecker deltas at the vertices.
\end{itemize}
From now, we omit superscripts, as they are $(0)$.
\begin{itemize}
\item[\textit{Rule 3}] \textbf{Phase-averaging: diagrams closed by internal or external couplings}\\
The only contributions that survive the average over phases have phases summing to zero 
before averaging. Then, each $b^{(0)}$ either pairs with another $b^{(0)}$ so that their 
phases sum to zero or belong to a set of $b^{(0)}$'s that pair with a $\psi_\bk^{(0)\mu_\bk}$ making 
the sum of their phases vanish. The first is an internal coupling, represented by 
a solid line connecting the paired indices $ij$ that contribute a factor 
$\delta_{\sigma_i+\sigma_j,\,0}\delta_{\bk_i,\bk_j}$
after phase averaging. The second is an external coupling, represented by joining all 
solid lines with indices $i_1,i_2,...,i_p$ to a blob $\bullet$ labeled $a$, that represents the 
phase $\psi_{\bk_a}^{(0)\mu_{\bk_a}}$ which contributes a factor 
$\delta_{\sigma_{i_1}+\cdots\sigma_{i_p}+\mu_a,\,0}\prod_{j=1}^p \delta_{\bk_j,\bk_a}$
after phase averaging. We will say that the blob (Kronecker delta) makes the wavenumber 
$\bk_j$ pinned to the value $\bk_a$. Conventionally, we omit the letters labeling the blobs:
factors such as $\delta_{\bk_j,\bk_a}\delta_{\mu_a+\sigma_j,0}$ are denoted by 
$\delta_{\mu_j+\sigma_j,0}$, meaning that $\bk_j$ is constrained to the value $\bk_a$ 
because of external coupling \cite{Choietal05b}.
\end{itemize}
Call bridge the line connecting two vertices, labeled with just one number: e.g. the line 
labeled with $1$ in presence of the factor 
$\cL_{1234}^{+\sigma_2\sigma_3\sigma_4}\cL_{1567}^{-\sigma_5\sigma_6\sigma_7}$.
We distinguish between {\em in-internal coupling}, with two lines starting from the same vertex 
closed together, and {\em cross-internal} coupling, when two lines starting from two different 
vertices are bound.
Let the number of degrees of freedom (or number of free wavenumbers) be the number of summations 
over all $N$ modes, cf.\ Appendix B.
\vskip 5pt \noindent
\textbf{Lemma}
{\em Let us assume the initial wavefield is an RP field.
Consider the phase average 
$$
\Big\langle\prod_l\psi_l^{\mu_l}\psi_1\cdot\cdot\cdot\psi_p\psi_{p+1}^*\cdot\cdot\cdot\psi_{q}^*\Big\rangle
$$
and all the possible associated diagrams giving non-null contributions.
Then, the degrees of freedom of each closed diagram are equal to the total number of internal couplings 
in the diagram, no matter if ``in-'' or ``cross-internal'' couplings. No degrees of freedom must be counted 
for a bridge.}
\vskip 5pt \noindent
This implies a new rule for the phase-averaging method.
\begin{itemize}
\item[\textit{Rule 4}] \textbf{Distinguishing leading order graphs}\\
The terms with a larger number of internal couplings are greater in order, so the leading contributions
come from the terms with the maximum number of internal couplings. 
Therefore, we can subdivide the diagrams in four different types:
type 0 diagrams with three free wavenumbers; type I diagrams two; type II diagrams
with one; type III with no free wavenumbers. The leading contributions are then
given by type 0 or type I diagrams and, in some cases, by type II diagrams.
\end{itemize}
The symbol ${\sum}^*$ expresses the fact that the combinations of 
$\bk_i$'s and $\sigma_i$'s giving linear terms inside the interaction term
are separated. Then, the interaction representation (\ref{ba}) allows us to remove such linear terms 
from the interaction, implying:
\begin{itemize}
\item[\textit{Rule 5}] \textbf{Diagram ``eliminated" by frequency renormalization}\\
Definition (\ref{star1}) implies that the Kronecker delta's inside (\ref{star1a}) vanish in $\sum_{234}^*$ for any allowed configuration.
Definition (\ref{star2}) implies that the delta's inside (\ref{star2a}) also vanish in the term $\sum_{2...7}^*$. 
Thus, a diagram for $b_1^{(1)}$ implying the arguements of the delta's inside (\ref{star1a}) 
to be simultaneously equal to zero is not contributing. The same holds for a graph for $b_1^{(2)}$ 
whose particular state requires null arguements for the delta's inside (\ref{star2a}).
\end{itemize}

\subsubsection{Contributions $\cJ_1-\cJ_5$}\label{Sec: Feynman}
The graph associated to $\cJ_1$ before phase-averaging is represented in Fig. \ref{Fig.: figure4}, and 
analitically expressed by:
\begin{eqnarray}
\cJ_1 &=&\Big\langle\prod_{\bk}\psi_{\bk}^{\mu_\bk}\left.\sum_{1234}\right.^* \Big(\lambda_1+\frac{\mu_1}{2J_1}\Big)\cL^{+\sigma_2\sigma_3\sigma_4}_{1234} a_1^-a_2^{\sigma_2}a_3^{\sigma_3}a_4^{\sigma_4} \nonumber \\
&& \;\;\;\btimes \Delta_T\left(-\omega_1+\sigma_2\omega_2+\sigma_3\omega_3+\sigma_4\omega_4\right)\delta_{\bk_1,\sigma_2\bk_2+\sigma_3\bk_3+\sigma_4\bk_4}\Big\rangle_\psi
\end{eqnarray}
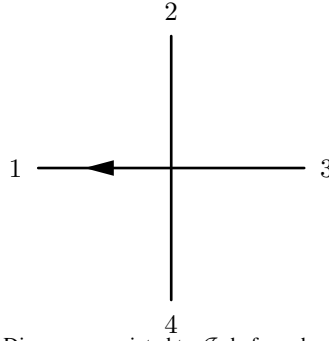
\begin{figure}[htbp]
	\begin{center}
	    \unitlength = 1mm
	    \begin{fmffile}{figure4}
	        \begin{fmfgraph*}(35,35)
	            \fmfleft{i1}
	            \fmfright{o3}
	            \fmftop{o2}
	            \fmfbottom{o4}
	            \fmflabel{$1$}{i1}
	            \fmflabel{$2$}{o2}
	            \fmflabel{$3$}{o3}
	            \fmflabel{$4$}{o4}
	            \fmf{fermion}{v1,i1}
	            \fmf{plain}{v1,o2}
	            \fmf{plain}{v1,o3}
	            \fmf{plain}{v1,o4}
	        \end{fmfgraph*}
	    \end{fmffile}
	    \caption{Diagram associated to $\cJ_1$ before phase-averaging}
	    \label{Fig.: figure4}
	    \end{center}
\end{figure}
\\Substituting the action-angle variables, we have:
\begin{eqnarray}
\cJ_1 &=&\left.\sum_{1234}\right.^* \Big(\lambda_1+\frac{\mu_1}{2J_1}\Big)\sqrt{J_1 J_2 J_3 J_4}\cL^{+\sigma_2\sigma_3\sigma_4}_{1234}\Big\langle\psi_1^{-1}\psi_2^{\sigma_2}\psi_3^{\sigma_3}\psi_4^{\sigma_4}\prod_{\bk}\psi_{\bk}^{\mu_\bk}\Big\rangle_\psi \nonumber \\
&& \quad\;\;\;\btimes \Delta_T\left(-\omega_1+\sigma_2\omega_2+\sigma_3\omega_3+\sigma_4\omega_4\right)\delta_{\bk_1,\sigma_2\bk_2+\sigma_3\bk_3+\sigma_4\bk_4}
\label{J1b}\end{eqnarray}
where only the term within angular brackets depends on phases.
This term can be thought of as the sum of the contributions of all the possible closures ({\it Rule 3}) of the 
diagram in Fig. \ref{Fig.: figure4}. 
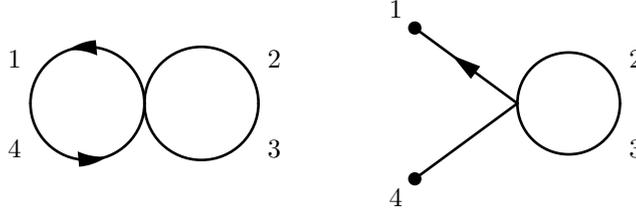
\begin{figure}[htbp]
	\begin{center}
	    \unitlength = 1mm
	    \begin{fmffile}{figure5a}
	        \begin{fmfgraph*}(30,20)
	            \fmfcurved
	            \fmfleft{i4,o4,i3,i1,i2}
	            \fmfright{i5,o3,i6,o2,i7}
	            \fmflabel{$1$}{i1}
	            \fmflabel{$2$}{o2}
	            \fmflabel{$3$}{o3}
	            \fmflabel{$4$}{o4}
	            \fmf{fermion,right}{v1,i3,v1}
	            \fmf{plain,left}{v1,i6,v1}
	        \end{fmfgraph*}
	    \end{fmffile}$\qquad\qquad$
	    \unitlength = 1mm
	    \begin{fmffile}{figure5b}
	        \begin{fmfgraph*}(30,20)
	            \fmfleft{o4,i1}
	            \fmfcurved
	            \fmfright{i4,o3,i3,o2,i2}
	            \fmfdot{i1,o4}
	            \fmflabel{$1$}{i1}
	            \fmflabel{$2$}{o2}
	            \fmflabel{$3$}{o3}
	            \fmflabel{$4$}{o4}
	            \fmf{fermion}{v1,i1}
	            \fmf{plain,left}{v1,i3,v1}
	            \fmf{plain}{v1,o4}
	        \end{fmfgraph*}
	    \end{fmffile}
	    \caption{Diagram 1 (type I, vanishing) and diagram 2 (type II)}
	    \label{Fig.: figure5}
	    \end{center}
\end{figure}
\begin{enumerate}
\item The contribution associated with diagram 1 in Fig. \ref{Fig.: figure5} may be directly written as
\be\sum_{\ul{\sigma}}{\sum_{\ul{\bk}}}^{*}\Big(\lambda_1+\frac{\mu_1}{2J_1}\Big)\sqrt{J_1 J_2 J_3 J_4}\cL^{+\sigma_2\sigma_2\sigma_4}_{1234} \prod_{m}\delta_{\mu_m,0}\quad\Delta_T\left(0\right)\delta_{\bk_1,\bk_4}\delta_{\bk_2,\bk_3} \ee
$$ \ul{\sigma}=\left( 1,\sigma_2,-\sigma_2,1\right)$$
The two Kronecker delta's of the internal couplings make the vertex delta redundant.
Applying {\it Rule 5}, one sees that this kind of graph is missing in the interaction. 
The physics of this diagram has already been included in the frequency renormalization and thus 
it must not been considered here. This implies that this is not a leading order term in $\cJ_1$.
\item For diagram 2 in Fig. \ref{Fig.: figure5}, one has the following contribution to $\cJ_1$:
\be
\sum_{\ul{\sigma}}{\sum_{\ul{\bk}}}^{'}\Big(\lambda_1+\frac{\mu_1}{2J_1}\Big)\sqrt{J_1 J_2 J_3 J_4}\cL^{+\sigma_2\sigma_2\sigma_4}_{1234}\delta_{\mu_1,1}\delta_{\mu_4,1}\delta_{\sigma_2,-\sigma_3}
\nonumber 
\ee
\be 
\;\;\;\quad\quad\btimes\prod_{m\neq1,-1}\delta_{\mu_m,0}\quad\Delta_T\left(-\omega_1-\omega_{-1}\right)\delta_{ \bk_2,\bk_3}\delta_{ \bk_1,-\bk_4}
\ee
Here $\ul{\sigma}=\left(1,\sigma_2,-\sigma_2,-1\right)$, because
the internal coupling between $2$ and $3$ needs $\sigma_2=-\sigma_3$ for the phase of $\bk_2$ to vanish. 
Then, $\bk_1=\sigma_4\bk_4=-\bk_4$, as $\sigma_4=-1$.
\begin{figure}[htbp]
	\begin{center}
	    \unitlength = 1mm
	    \begin{fmffile}{figure6a}
	        \begin{fmfgraph*}(30,20)
	            \fmfcurved
	            \fmfleft{i4,o4,i3,i1,i2}
	            \fmfright{o3,o2}
	            \fmfdot{o2,o3}
	            \fmflabel{$1$}{i1}
	            \fmflabel{$2$}{o2}
	            \fmflabel{$3$}{o3}
	            \fmflabel{$4$}{o4}
	            \fmf{plain}{v1,o2}
	            \fmf{plain}{v1,o3}
	            \fmf{fermion,right}{v1,i3,v1}
	        \end{fmfgraph*}
	    \end{fmffile}$\qquad\qquad$
	    \unitlength = 1mm
	    \begin{fmffile}{figure6b}
	        \begin{fmfgraph*}(30,20)
	            \fmfcurved
	            \fmfleft{i4,o4,i3,i1,i2}
	            \fmfright{i5,o3,i6,o2,i7}
	            \fmfdot{i6}
	            \fmflabel{$1$}{i1}
	            \fmflabel{$2$}{o2}
	            \fmflabel{$3$}{o3}
	            \fmflabel{$4$}{o4}
	            \fmf{fermion,right}{v1,i3,v1}
	            \fmf{plain,left}{v1,i6,v1}
	        \end{fmfgraph*}
	    \end{fmffile}
	    \caption{Diagram 3 (type II) and diagram 4 (type II, vanishing)}
	    \label{Fig.: figure6}
	    \end{center}
\end{figure}
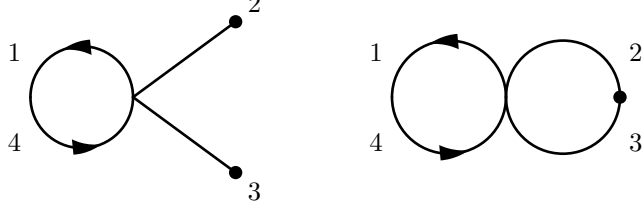
\item For diagram 3 in Fig. \ref{Fig.: figure6}, the contribution to $\cJ_1$ reads:
\be
\sum_{\ul{\sigma}}{\sum_{\ul{\bk}}}^{'}\Big(\lambda_1+\frac{\mu_1}{2J_1}\Big)\sqrt{J_1 J_2 J_3 J_4}\cL^{+\sigma_2\sigma_2\sigma_4}_{1234}\delta_{\mu_2,-\sigma_2}\delta_{\mu_3,-\sigma_2}\qquad\qquad\qquad
\nonumber 
\ee
\be 
\;\;\;\qquad\quad\quad\btimes\prod_{m\neq2,3}\delta_{\mu_m,0}\quad\Delta_T\left(\sigma_2\left(\omega_2+\omega_{-2}\right)\right)\delta_{\bk_2,-\bk_3}\delta_{\bk_1,\bk_4}
\ee
and $ \ul{\sigma}=\left( 1,\sigma_2,\sigma_2,1\right)$.
\item For diagram 4, the last Kronecker delta in (\ref{J1b}), which represents momentum 
conservation at the vertex, implies $\bk_2=\bk_3=0$. So this diagram does not represent an effective interaction. 
As a matter of fact, for spatially homogeneous WT fields there must be no coupling with the zero 
mode $\bk = \mathbf{0}$ because such coupling would violate momentum conservation, cf.\  
\cite{Nazarenko11,Choietal05b}.
If one of
the arguments of $\cL_{1234}$ vanishes, the matrix element is zero. That is to say that for any spatially
homogeneous WT system $\cL_{1234}$ is identically zero if one of 
$\bk_1$, $\bk_2$, $\bk_3$ or $\bk_4$ is zero. 
The situation is analogous for graphs obtained by permutations of the indices.
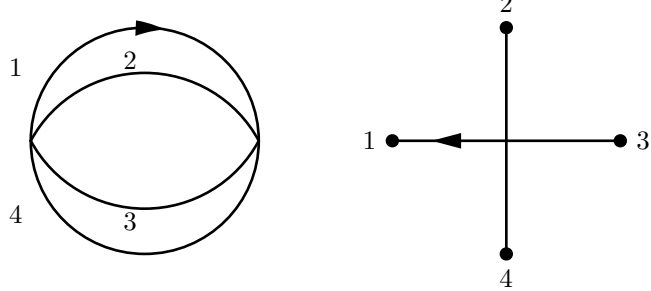
\begin{figure}[htbp]
	\begin{center}
	    \unitlength = 1mm
	    \begin{fmffile}{figure7a}
	        \begin{fmfgraph*}(30,30)
	            \fmfcurved
	            \fmfleft{l1,l2,i1,l3,l4}
	            \fmfright{i5,o4,i6,o1,i7}
	            \fmflabel{$1$}{l3}
	            \fmflabel{$4$}{l2}
	            \fmflabel{$2$}{v1}
	            \fmflabel{$3$}{v2}
	            \fmf{phantom}{i1,v1,i7}
	            \fmf{phantom}{i1,v2,i5}
	            \fmf{plain,left=0.6}{i1,i6}
	            \fmf{fermion,left,tension=0.5}{i1,i6}
	            \fmf{plain,right=0.6}{i1,i6}
	            \fmf{plain,right,tension=0.5}{i1,i6}
	        \end{fmfgraph*}
	    \end{fmffile}$\qquad\qquad$
	    \unitlength = 1mm
	    \begin{fmffile}{figure7b}
	        \begin{fmfgraph*}(30,30)
	            \fmfleft{i1}
	            \fmfright{o3}
	            \fmftop{o2}
	            \fmfbottom{o4}
	            \fmfdot{i1,o2,o3,o4}
	            \fmflabel{$1$}{i1}
	            \fmflabel{$2$}{o2}
	            \fmflabel{$3$}{o3}
	            \fmflabel{$4$}{o4}
	            \fmf{fermion}{v1,i1}
	            \fmf{plain}{v1,o2}
	            \fmf{plain}{v1,o3}
	            \fmf{plain}{v1,o4}
	        \end{fmfgraph*}
	    \caption{Diagram 5 (type II) and diagram 6 (type III)}
	    \end{fmffile}
	    \label{Fig.: figure7}
	    \end{center}
\end{figure}
\item Diagram 5 in Fig. \ref{Fig.: figure7} contributes as
\be\sum_{\ul{\sigma}}\sum_{\stackrel{\ul{\bk},}{\bk_1=\bk_2=\bk_3=\bk_4}}\Big(\lambda_1+\frac{\mu_1}{2J_1}\Big)\sqrt{J_1 J_2 J_3 J_4}\cL^{+\sigma_2\sigma_2\sigma_4}_{1234}\delta_{\sigma_2+\sigma_3+\sigma_4,1}\prod_{m}\delta_{\mu_m,0}\;\Delta_T\left(0\right) \ee
$$ 
\ul{\sigma}=\left( 1,\sigma_2,\sigma_3,\sigma_4\right)
$$
\item All other diagrams are type III (like {\em e.g.} diagram 6 in Fig. \ref{Fig.: figure7}) and give subleading contributions.
\end{enumerate}

\noindent
{\bf Normalization of amplitudes:} Let us introduce the change of variables
\be 
J_\bk=\left(\frac{2\pi}{L}\right)^d \tJ_\bk,\qquad \lambda_\bk=i\lambda(\bk) 
\label{subst}
\ee
This substitution implies that the characteristic function be 
expressed by:
\be 
\cZ_L[\lambda,\mu]\doteq \left \langle \exp \left( i\sum_{\bk \in \Lambda_L^*}\left(\frac{2\pi}{L}\right)^d\lambda(\bk) \tJ_\bk \right) \prod_{\bk \in \Lambda_L^*} \psi_\bk^{\mu_\bk}\right \rangle 
\label{eq: genfunct2}
\ee
where $\lambda(\bk)$ is a smooth test function and $\mu_\bk$ are integers.
Here we keep the time dependence implicit, for sake of notation. This characteristic function, after the transformation of the sum to an integral thanks to the large-$L$ limit, becomes a 
characteristic functional \cite{Eyink}.
The change of variables (\ref{subst}) 
is the key to a finite, well defined expression, in the thermodynamic limit.

\noindent
{\bf Main contributions to $\cJ_1$:}
Diagram 2 is the only type II diagram with mode $\bk_1$ pinned to an external blob, so that 
$\mu_1 \ne 0$. This graph contributes to $O(1)$, as it is order $O(L^d)$ 
(one free wavenumber, that is an unconstrained sum) multiplyed by order $O(L^{-d})$ 
(term proportional to $\mu_1$, see (\ref{J1})). 
A factor $3$ appears to account for the possible permutations of the indices.
There is no other leading order term. The other graphs contribute to 
order $O(L^{-d})$ and vanish in the $L\rightarrow \infty$ limit.
Summarizing, $\cJ_1$ may be written as:
\begin{eqnarray}
\cJ_1 &=&\frac{3}{2}\left(\frac{2\pi}{L}\right)^d\sum_{(1)}\Big[\sqrt{\frac{\tJ_2 \tJ_3 \tJ_4}{\tJ_1}}\cL^{+\sigma_2\sigma_3\sigma_4}_{1234}\delta_{\mu_1,1}\delta_{\mu_{-1},1}\prod_{m\neq\pm1}\delta_{\mu_m,0}\Delta_T\left(-(\omega_1+\omega_{-1})\right)\Big] \nonumber \\
&& \qquad\qquad+O\left(L^{-d}\right), \qquad
\mbox{where } ~~~~
\sum_{(1)}\doteq \sum_{\stackrel{\sigma_1=-\sigma_4=1,}{\sigma_2=-\sigma_3}}\sum_{\bk_2} 
\label{J1c}
\end{eqnarray}
\noindent
The contributions from the terms $\cJ_2,...,\cJ_5$, are given in \ref{appendix}.

\textit{Contribution of $\cJ_2$}
\begin{eqnarray}
\cJ_2 &=& \frac{i}{2}\left(\frac{2\pi}{L}\right)^{3d}\delta_{\mu,0} \Big\{9{\sum}_{(2)} \Big[ \lambda\left(\bk_1\right)\cL_{1234}^{+\sigma_2\sigma_3\sigma_4}\cL_{1567}^{-\sigma_5\sigma_6\sigma_7} \tJ_2 \tJ_4 \tJ_5 |\Delta_T\left(\tw_1+\tw_{-1}\right)|^2\Big] \nonumber\\
&& \quad\qquad+6{\sum}_{(3)} \Big[\lambda\left(\bk_1\right)|\cL_{1234}^{+\sigma_2\sigma_3\sigma_4}|^2 \tJ_2 \tJ_3 \tJ_4 |\Delta_T\left(\tw^1_{234}\right)|^2\Big]\Big\} + O\left(L^{-1}\right)\label{J2c}
\end{eqnarray}
\begin{align}
{\sum}_{(2)} \doteq& \sum_{\ul{\sigma}}{\sum_{\ul{\bk}}}'  \delta_{ \bk_4,-\bk_1}\delta_{ \bk_4,\bk_7}\delta_{ \bk_2,\bk_3}\delta_{ \bk_5,\bk_6},\;\ul{\sigma}=\left(1,\sigma_2,-\sigma_2,-1,\sigma_5,-\sigma_5,1\right)\\
{\sum}_{(3)} \doteq &\sum_{\ul{\sigma}}{\sum_{\ul{\bk}}}' \delta^1_{234}\delta_{\bk_2,\bk_5}\delta_{ \bk_4,\bk_7}\delta_{ \bk_3,\bk_6},\;\ul{\sigma}=\left(1,\sigma_2,\sigma_3,\sigma_4,-\sigma_2,-\sigma_3,-\sigma_4\right)
\end{align}

\textit{Contribution of $\cJ_3$}
\begin{eqnarray}
\cJ_3 &=&18i\left(\frac{2\pi}{L}\right)^{3d}\delta_{\mu,0} \nonumber \\
 &&\quad \btimes \Big\{{\sum}_{(4)} \Big[\lambda\left(\bk_1\right) \cL_{1234}^{+\sigma_2\sigma_3\sigma_4}\cL_{4567}^{\sigma_4\sigma_5\sigma_6\sigma_7} \tJ_1 \tJ_3 \tJ_5 E_T\left(0,\sigma_3\left(\omega_3+\omega_{-3}\right)\right)\Big] \nonumber\\
&& \;\qquad+\frac{1}{2}{\sum}_{(5)} \Big[\lambda\left(\bk_1\right) \cL_{1234}^{+\sigma_2\sigma_3\sigma_4}\cL_{4567}^{\sigma_4\sigma_5\sigma_6\sigma_7} \tJ_1 \tJ_3 \tJ_6  E_T\left(0,-\left(\tw_1+\tw_{-1}\right)\right)\Big] \nonumber\\
&& \;\qquad+{\sum}_{(6)} \Big[\lambda\left(\bk_1\right) \cL_{1234}^{+\sigma_2\sigma_3\sigma_4}\cL_{4567}^{\sigma_4\sigma_5\sigma_6\sigma_7} \tJ_1 \tJ_2 \tJ_3  E_T\left(0,\tw^1_{234}\right)\Big]\Big\} \nonumber\\
&&    +9\left(\frac{2\pi}{L}\right)^{2d}\Big\{{\sum}_{(7)} \Big[ \cL_{1234}^{+\sigma_2\sigma_3\sigma_4}\cL_{4567}^{\sigma_4\sigma_5\sigma_6\sigma_7}\sqrt{\frac{\tJ_{-1}}{\tJ_1}}\tJ_3 \tJ_5 \prod_{m\neq1,2}\delta_{\mu_m,0} \nonumber\\
&&\qquad \qquad\qquad \btimes  E_T\big(-\left(\tw_1+\tw_{-1}\right), -\left(\tw_1+\tw_{-1}\right)+\sigma_3\tw_3+\sigma_4 \tw_4\big)\Big]\nonumber
\end{eqnarray}
\begin{eqnarray}
&&  \quad \qquad\qquad+\frac{1}{2}{\sum}_{(8)} \Big[ \cL_{1234}^{+\sigma_2\sigma_3\sigma_4}\cL_{4567}^{\sigma_4\sigma_5\sigma_6\sigma_7}\sqrt{\frac{\tJ_{7}}{\tJ_1}}\tJ_3 \tJ_5 \prod_{m\neq1,7}\delta_{\mu_m,0} \nonumber\\
&&\qquad \qquad\qquad \btimes  E_T\big(-\left(\tw_1+\sigma_4\tw_4\right),-\left(\tw_1+\sigma_4\tw_4\right)\big)\Big] \nonumber\\
&&  \qquad \qquad\qquad+{\sum}_{(9)} \Big[ \cL_{1234}^{+\sigma_2\sigma_3\sigma_4}\cL_{4567}^{\sigma_4\sigma_5\sigma_6\sigma_7}\sqrt{\frac{\tJ_{-1}}{\tJ_1}}\tJ_3 \tJ_5 \prod_{m\neq1,6}\delta_{\mu_m,0} \nonumber \\
&&\;\;\qquad\qquad\qquad \btimes  E_T\big(-\left(\tw_1+\tw_{-1}\right),\tw^1_{234}\big)\Big]\Big\}+ O\left(L^{-1}\right) \label{J3c}
\end{eqnarray}
where
\begin{eqnarray}
{\sum}_{(4)} &\doteq &{\sum_{\ul{\sigma}, \ul{\bk}}}' \delta_{ \bk_4,-\bk_3}\delta_{ \bk_1,\bk_2}\delta_{ \bk_3,\bk_7}\delta_{ \bk_5,\bk_6},\;\ul{\sigma}=\left(1,1,\sigma_3,\sigma_3,\sigma_5,-\sigma_5,-\sigma_3\right)\nonumber\\
{\sum}_{(5)} &\doteq& {\sum_{\ul{\sigma}, \ul{\bk}}}' \delta_{ \bk_1,-\bk_4}\delta_{ \bk_1,\bk_5}\delta_{ \bk_3,\bk_2}\delta_{ \bk_6,\bk_7},\;\ul{\sigma}=\left(1,\sigma_2,-\sigma_2,-1,1,\sigma_6,-\sigma_6\right)\nonumber\\
{\sum}_{(6)} &\doteq& {\sum_{\ul{\sigma}, \ul{\bk}}}' \delta_{234}^1\delta_{ \bk_1,\bk_6}\delta_{ \bk_3,\bk_7}\delta_{ \bk_5,\bk_2},\;\ul{\sigma}=\left(1,\sigma_2,\sigma_3,\sigma_4,-\sigma_2,1,-\sigma_3\right)\nonumber\\
{\sum}_{(7)} &\doteq& {\sum_{\ul{\sigma}, \ul{\bk}}}' \delta_{ -\sigma_4\bk_4,\sigma_3\bk_3}\delta_{ \bk_1,-\bk_2}\delta_{ \bk_3,\bk_7}\delta_{ \bk_5,\bk_6},\; \ul{\sigma}=\left(1,-1,\sigma_3,\sigma_4,\sigma_5,-\sigma_5,-\sigma_3\right)\nonumber\\
{\sum}_{(8)}& \doteq &{\sum_{\ul{\sigma}, \ul{\bk}}}' \delta_{ \bk_1,\sigma_4\bk_4}\delta_{ \bk_1,\sigma_7\bk_7}\delta_{ \bk_3,\bk_2}\delta_{ \bk_5,\bk_6},\;\ul{\sigma}=\left(1,\sigma_2,-\sigma_2,\sigma_4,\sigma_5,-\sigma_5,\sigma_7\right)\nonumber\\
{\sum}_{(9)}& \doteq& {\sum_{\ul{\sigma}, \ul{\bk}}}' \delta_{234}^1\delta_{ \bk_1,-\bk_6}\delta_{ \bk_3,\bk_7}\delta_{ \bk_5,\bk_2},\;\ul{\sigma}=\left(1,\sigma_2,\sigma_3,\sigma_4,-\sigma_2,-1,-\sigma_3\right)
\end{eqnarray}

\textit{Contribution of $\cJ_4$}

One finds that $\cJ_4=O\left(L^{-d}\right)$, so it represents a subleading contribution.\\

\textit{Contribution of $\cJ_5$}
\begin{eqnarray}
\cJ_5&=&3\left(\frac{2\pi}{L}\right)^{2d} \sum_{(10)}\frac{\mu_1\mu_2}{4 J_1 J_2}\cL_{1345}^{+\sigma_3\sigma_4\sigma_5}\cL_{2678}^{+\sigma_6\sigma_7\sigma_8}\sqrt{J_1J_2J_3J_4J_5J_6J_7J_8} \nonumber\\
&&\qquad\btimes\delta_{\mu_1,1}\delta_{\mu_2,1} \prod_{m\neq1,2}\delta_{\mu_m,0}\, \Delta_T\big(\tw^1_{345}\big) \Delta_T\big(\tw^{(-1)345}\big)\nonumber\\
&& +9\left(\frac{2\pi}{L}\right)^{2d} \sum_{(11)}\frac{\mu_1\mu_2}{4 J_1 J_2}\cL_{1345}^{+\sigma_3\sigma_4\sigma_5}\cL_{2678}^{+\sigma_6\sigma_7\sigma_8}\sqrt{J_1J_2J_3J_4J_5J_6J_7J_8} \nonumber\\
&&\qquad\btimes\delta_{\mu_1,2}\delta_{\mu_2,2} \prod_{m\neq1,2}\delta_{\mu_m,0}\, \Delta_T\big(\tw^1_{(-1)44}\big) \Delta_T\big(\tw^{(-1)177}\big)+O\left(L^{-d}\right)\label{J5c}
\end{eqnarray}
\begin{align}
\sum_{(10)}&\doteq \sum_{\ul{\sigma}}{\sum_{\ul{\bk}}}'\delta_{\bk_1,-\bk_2}\delta_{\bk_3,\bk_6}\delta_{\bk_4,\bk_7}\delta_{\bk_5,\bk_8}\delta_{345}^1,\;\ul{\sigma}=\left(1,1,\sigma_3,\sigma_4,\sigma_5,-\sigma_3,-\sigma_4,-\sigma_5\right)\nonumber\\
\sum_{(11)}&\doteq \sum_{\ul{\sigma}}{\sum_{\ul{\bk}}}'\delta_{\bk_1,-\bk_2}\delta_{\bk_1,-\bk_3}\delta_{\bk_1,\bk_6}\delta_{\bk_4,\bk_5}\delta_{\bk_7,\bk_8},\;\ul{\sigma}=\left(1,1,-1,\sigma_4,-\sigma_4,-1,\sigma_7,-\sigma_7\right)
\end{align}

\subsection{Dynamical Multi-Mode Equation}
In this section we turn Eq.(\ref{eq: chi}) into a dynamical equation for the characteristic functional $Z$
taking the $L\longrightarrow\infty$ and $\epsilon\longrightarrow 0$ limits. The two limits do not commute:
the large-box limit must be taken first, the weak-nonlinearity limit after. 
The physical meaning of this operation is that there is a vast number of quasi-resonances (introduced 
by the large box limit, which leads to a continuous $\bk$-space sending $\Lambda_L^*\longrightarrow\Lambda^*$), 
each of which is as important as the exact resonances \cite{Nazarenko11}.

\subsubsection{Large-box limit}
Let us introduce the large-L asymptotics standard substitutions, and
\be 
\left(\frac{2\pi}{L}\right)^d\sum_\bk \Longrightarrow \int d^dk ~, \quad
 \left(\frac{L}{2\pi}\right)^d \delta_{\bk,\bk'} \Longrightarrow \delta^d(\bk-\bk')
\label{largeL} 
\ee
Recalling Eq.(\ref{eq: Zchi}), using (\ref{J1c}), (\ref{J2c}), (\ref{J3c}) and (\ref{J5c}), and
neglecting $O\left(L^{-d}\right)$ corrections, we can eventually write:
\begin{eqnarray*}
&&\Bigg\langle \exp\bigg[i\int d^dk\lambda(\bk)\tJ_\bk\bigg]\Bigg\{6i\epsilon \bigg(\sum_{\bk_1}\delta_{\mu_1,1}\delta_{\mu_{-1},1}\prod_{m\neq\pm1}\delta_{\mu_m,0}\bigg)  \qquad\qquad\qquad\qquad\\
&&\qquad\qquad\qquad\;\,\btimes \sum_{\sigma_2}\int d^d k_2\tJ_2\sqrt{\frac{\tJ_{-1}}{\tJ_1}}\cH_{1224}^{-\sigma_2(-\sigma_2)-}\Delta_T\big(-\left(\tw_1+\tw_{-1}\right)\big)+\\
&&+8i\epsilon^2  \delta_{\mu,0}\bigg[9\sum_{\sigma_2,\sigma_5}\int d^dk_1d^dk_2d^dk_5 \lambda\left(\bk_1\right)\cH_{122(-1)}^{-\sigma_2(-\sigma_2)-}\\
&& \quad\qquad\qquad\qquad\qquad\qquad\;\btimes\cH_{155(-1)}^{+\sigma_5(-\sigma_5)+}\tJ_{-1} \tJ_2 \tJ_5 |\Delta_T\left(\tw_1+\tw_{-1}\right)|^2\qquad\qquad\qquad\qquad \\
&& \quad\;\;+6\sum_{\sigma_2,\sigma_3,\sigma_4}\int d^dk_1d^dk_2d^dk_3d^dk_4 \lambda\left(\bk_1\right)|\cH_{1234}^{-\sigma_2\sigma_3\sigma_4}|^2 \tJ_2 \tJ_3 \tJ_4|\Delta_T\left(\tw^1_{234}\right)|^2 \delta^1_{234}\bigg]+\qquad\qquad\qquad\qquad\\
\end{eqnarray*}
\begin{eqnarray*}
&&+288i\epsilon^2 \delta_{\mu,0}\bigg[\sum_{\sigma_4,\sigma_5}\int d^dk_1d^dk_4d^dk_5 \left(-\sigma_4\right) \lambda\left(\bk_1\right)\cH_{11(-4)4}^{-+\sigma_4\sigma_4}\\
&&\quad\qquad\qquad\qquad\qquad\qquad \btimes \cH_{455(-4)}^{(-\sigma_4)\sigma_5(-\sigma_5)(-\sigma_4)}\tJ_1 \tJ_4 \tJ_5 E_T\left(0,\sigma_4\left(\omega_4+\omega_{-4}\right)\right)\qquad\qquad\qquad\qquad\\
&&\qquad\qquad\;+\frac{1}{2}\sum_{\sigma_2,\sigma_6}\int d^dk_1d^dk_2d^dk_6\lambda\left(\bk_1\right) \cH_{122(-1)}^{-\sigma_2(-\sigma_2)-} \\
&&\quad\qquad\qquad\qquad\qquad\qquad \btimes \cH_{(-1)166}^{++\sigma_6-(\sigma_6)} \tJ_1 \tJ_2 \tJ_6 E_T\left(0,-\left(\tw_1+\tw_{-1}\right)\right)\qquad\qquad\qquad\qquad\\
&&\qquad\qquad\; +\sum_{\sigma_2,\sigma_3,\sigma_4}\int d^dk_1d^dk_2d^dk_3d^dk_4 \left(-\sigma_4\right)\lambda\left(\bk_1\right)
|\cH_{1234}^{-\sigma_2\sigma_3\sigma_4}|^2 \tJ_1 \tJ_2 \tJ_3 E_T\left(0,\tw^1_{234}\right)\delta^1_{234}\bigg]+
\end{eqnarray*}
\begin{eqnarray*}
&&+144\epsilon^2 \bigg[\bigg(\sum_{\bk_1} \delta_{\mu_1,1}\delta_{\mu_{-1},1} \prod_{m\neq\pm1}\delta_{\mu_m,0}\bigg) 
\sum_{\sigma_3,\sigma_4,\sigma_5}\int d^dk_3d^dk_4d^dk_5\\
&&\qquad \qquad\quad \btimes\left(-\sigma_4\right)\cH_{1(-1)34}^{--\sigma_3\sigma_4}\cL_{4535}^{(-\sigma_4)\sigma_5(-\sigma_3)(-\sigma_5)}\sqrt{\frac{\tJ_{-1}}{\tJ_1}}\tJ_3 \tJ_5 \;\delta\left(\sigma_3\bk_3+\sigma_4\bk_4\right)\\
&&\qquad \qquad\quad \btimes E_T\big(-\left(\tw_1+\tw_{-1}\right), -\left(\tw_1+\tw_{-1}\right)+\sigma_3\tw_3+\sigma_4 \tw_{4}\big)\\
&&  \qquad +\frac{1}{2}\sum_{\sigma_2,\sigma_4,\sigma_5,\sigma_7}\bigg(\sum_{\bk_1} \delta_{\mu_1,1}\sum_{\bk_7}\delta_{\mu_{7},-\sigma_7} \prod_{m\neq1,7}\delta_{\mu_m,0}\bigg)  \\
&&\qquad\quad\btimes\int d^dk_2d^dk_4d^dk_5d^dk_7\left(-\sigma_4\right) \cH_{1224}^{-\sigma_2(-\sigma_2)\sigma_4}\cH_{4557}^{(-\sigma_4)\sigma_5(-\sigma_5)\sigma_7}\sqrt{\frac{\tJ_{7}}{\tJ_1}}\tJ_3 \tJ_5 \\
&&\qquad\quad\btimes  E_T\big(-\left(\tw_1+\sigma_4\tw_{4}\right),-\left(\tw_1+\sigma_4\tw_{4}\right)\big) \delta(\bk_1-\sigma_4\bk_4) \delta(\bk_1-\sigma_7\bk_7)\\
&& \qquad+\bigg(\sum_{\bk_1} \delta_{\mu_1,1}\delta_{\mu_{-1},1} \prod_{m\neq\pm1}\delta_{\mu_m,0}\bigg)  \sum_{\sigma_2,\sigma_3,\sigma_4}\int d^dk_2d^dk_3d^dk_4\\ 
&&\;\;\;\qquad\qquad \qquad\qquad \btimes\left(-\sigma_4\right)\cH_{1234}^{-\sigma_2\sigma_3\sigma_4}\cH_{42(-1)3}^{(-\sigma_4)(-\sigma_2)-(-\sigma_3)}\sqrt{\frac{\tJ_{-1}}{\tJ_1}}\tJ_2 \tJ_3 
E_T\big(-\left(\tw_1+\tw_{-1}\right),\tw^1_{234}\big)\delta^1_{234}\bigg]+\\
\end{eqnarray*}
\begin{eqnarray*}
&&-12\epsilon^2 \bigg(\sum_{\bk_1} \delta_{\mu_1,1}\delta_{\mu_{-1},1} \prod_{m\neq\pm1}\delta_{\mu_m,0}\bigg)  \sum_{\sigma_3,\sigma_4,\sigma_5}\qquad\qquad\\ 
&&\qquad \qquad\qquad \;\;\btimes \int d^dk_3d^dk_4d^dk_5\cH_{1345}^{-\sigma_3\sigma_4\sigma_5}\cH_{(-1)345}^{-(-\sigma_3)(-\sigma_4)(-\sigma_5)}
\frac{\tJ_3\tJ_4\tJ_5}{\sqrt{\tJ_1 \tJ_{-1}}}\Delta_T\big(\tw^1_{345}\big) \Delta_T\big(\tw^{(-1)345}\big) \delta^1_{345}
\end{eqnarray*}
\begin{eqnarray*}
&&-36\epsilon^2 \bigg(\sum_{\bk_1} \delta_{\mu_1,2}\delta_{\mu_{-1},2} \prod_{m\neq\pm1}\delta_{\mu_m,0}\bigg)  \sum_{\sigma_3,\sigma_4}
\int d^dk_3d^dk_4\cH_{1(-1)33}^{--\sigma_3(-\sigma_3)}\cH_{(-1)144}^{--\sigma_4(-\sigma_4)}\tJ_3\tJ_{4}
\end{eqnarray*}
\be 
\;\qquad\btimes \tJ_3\tJ_4 \Delta_T\big(\tw^1_{(-1)33}\big) \Delta_T\big(\tw^{(-1)144}\big)\Bigg\}\Bigg\rangle_J 
\label{Zchib}
\ee

\subsubsection{Weak-nonlinearity limit}\label{Sec: dineq}
Recall that in section \ref{Sec: expansion2}, we took 
$\frac{2\pi}{\tw_\bk}\ll T \ll \frac{2\pi}{\epsilon^2\tw_\bk}$, with $T$ between the wave period and the 
nonlinear time.
We can now take $T\sim\frac{2\pi}{\epsilon \tw_\bk}$, so that $\lim_{\epsilon\rightarrow 0}T=\infty$. 
Then, in (\ref{Zchib}) we must take the $T\rightarrow\infty$ limit, consistently 
with the large-$T$ asymptotics of $\Delta_T$ and $E_T$ \cite{Eyink,BenneyNewell69}:
\begin{eqnarray}
&& \Delta_T(x)\sim \wt{\Delta}(x)=\pi\delta(x)+iP\left(\frac{1}{x}\right),\,\,E_T(x;y)\sim\Delta_T(x)\Delta_T(y)\sim\wt{\Delta}(x)\wt{\Delta}(y), \cr
&& |\Delta_T(x)|^2\sim 2\pi T\delta(x)+2P\left(\frac{1}{x}\right)\frac{\partial}{\partial x},\,\,
E_T(x;0)\sim\wt{\Delta}(x)\left(T-i\frac{\partial}{\partial x}\right),\,\,
\label{smallep}\end{eqnarray}
Some considerations are in order.
\begin{enumerate}
\item 
in Eq.(\ref{Zchib}), only the terms containing $ |\Delta_T(x)|^2$, $E_T(x;0)$ or $E_T(0;y)$ give secular 
contributions (proportional to $T$); the non-secular contributions are irrelevant in the $T\rightarrow\infty$ ($\epsilon \rightarrow 0$) 
limit. Thus, only the terms with $\delta_{\mu,0}$ survive the weak-nonlinearity assumption, while 
those with $\delta_{\mu_1,1}$ etc. are subleading.
\item 
 The $\mu$-dependent part of $\cZ$ is constrained to be $1$ by $\delta_{\mu,0}$.
Then, using (\ref{subst}), switching to $i\lambda(\bk)$ and taking the large-box limit 
leads to the functional derivative
\be
\left(\frac{L}{2\pi}\right)^d \frac{\partial}{\partial\lambda_\bk} \Longrightarrow -i\frac{\delta}{\delta\lambda(\bk)}
\label{funcder}
\ee 
\item Replace $(\cZ[T]-\cZ[0])/T$ with the time derivative $\dot{\cZ}$.
This can be done (\cite{Nazarenko11},pg. 81) because time $T$ is small compared to the 
characteristic time of averaged quantities such as $\cZ$ (nonlinear time).
Indeed, the istantaneous time derivative can be of same order or even greater than the rate of change described by our substitution, 
but such rapid changes are oscillatory and they drop out.
\item We introduce a new time variable $\tau\doteq\epsilon^2T$.
\item Renaming indices, we split the integral with $4$ wavenumbers into identical contributions
\end{enumerate}
\begin{eqnarray*}
\frac{d\cZ[\lambda,\mu,\tau]}{d\tau}&=& -192\pi \delta_{\mu,0}\sum_{\ul{\sigma}=(1,\sigma_2,\sigma_3,\sigma_4)}\int d^dk_1d^dk_2d^dk_3d^dk_4 \lambda\left(\bk_1\right)|\cH_{1234}^{-\sigma_2\sigma_3\sigma_4}|^2 \\
&&\quad\delta\left(\tw^1_{234}\right) \delta^1_{234}\bigg(\frac{\delta^3 \cZ}{\delta\lambda(\bk_2)\delta\lambda(\bk_3)\delta\lambda(\bk_4)}-\sigma_2\frac{\delta^3 \cZ}{\delta\lambda(\bk_1)\delta\lambda(\bk_3)\delta\lambda(\bk_4)}\\
&&\qquad\quad\quad\;\;-\sigma_3\frac{\delta^3 \cZ}{\delta\lambda(\bk_1)\delta\lambda(\bk_2)\delta\lambda(\bk_4)}-\sigma_4\frac{\delta^3 \cZ}{\delta\lambda(\bk_1)\delta\lambda(\bk_2)\delta\lambda(\bk_3)}\bigg) \\
&& -288 \pi \delta_{\mu,0}\sum_{\ul{\sigma}=(1,\sigma_2,\sigma_3)}\int d^dk_1d^dk_2d^dk_3\lambda\left(\bk_1\right)\\
&&\quad\btimes\bigg[\cH_{1(-1)22}^{--\sigma_2(-\sigma_2)}\cH_{1(-1)33}^{++\sigma_3(-\sigma_3)} \delta\left(\tw_1+\tw_{-1}\right)  \\
&&\qquad\; \sum_{\sigma=\pm1}\frac{\delta^3 \cZ}{\delta\lambda(\sigma\bk_1)\delta\lambda(\bk_2)\delta\lambda(\bk_3)} -2\sigma_2 \cH_{2(-2)11}^{\sigma_2\sigma_2+-}
\end{eqnarray*}
\be  
\qquad\;\; \cH_{2(-2)33}^{(-\sigma_2)(-\sigma_2)\sigma_3(-\sigma_3)}\delta\left(\tw_2+\tw_{-2}\right)\frac{\delta^3 \cZ}{\delta\lambda(\bk_{1})\delta\lambda(\bk_2)\delta\lambda(\bk_3)}\bigg] 
\label{dineqZ}
\ee

\subsubsection{Resonance condition} \lb{Resonancecond}
Recall definition $\tw_\bk \doteq \omega_\bk+\Omega_\bk$. 
The definition of $\Omega_\bk$ (\ref{Omega}) and the thermodynamic limit imply:
\be 
\quad\Omega_1\stackrel{L\rightarrow\infty}{\longrightarrow} 24 \epsilon \int_{\Lambda^*} d^dk_2 
\mathcal{H}_{1122}^{+- +-} \tJ_2^{(0)} ~,
\qquad
\Omega_{-1} = 24 \epsilon \int_{\Lambda^*} d^dk_2 \mathcal{H}_{(-1)(-1)22}^{+- +-} \tJ_2^{(0)}
\ee
Note: each component of $\bk_2$, defined in the dual space $\Lambda^*$, ranges in the interval 
$\left[-k_{max},k_{max}\right]$ and space isotropy implies that our system is symmetric under 
the $\bk\rightarrow - \bk$ transformation. Therefore:
\be 
\Omega_{-1} = 24 \epsilon \int_{(-\Lambda^*)} d^d(-k_2) \mathcal{H}_{11(-2)(-2)}^{+- +-} \tJ_{-2}^{(0)}
\ee
where $(-\Lambda^*)$ means that we are integrating over each component of $\bk_2$ from $+k_{max}$ to $-k_{max}$ 
and not from $-k_{max}$ to $+k_{max}$ as it would be for $\Lambda^*$. 
However, the integration over $-\bk_2\in (-\Lambda^*)$ is equivalent to the integration over the variable 
$\bk_3\in\Lambda^*$, and this leads to:
\be 
\Omega_{-1} = 24 \epsilon \int_{\Lambda^*} d^d(k_3) \mathcal{H}_{1133}^{+- +-} \tJ_{3}^{(0)}\equiv\Omega_1 
\ee
Space isotropy also implies that $\omega_\bk=\omega_{-\bk}$ ($\omega_\bk$ is positive for all $\bk\in\Lambda^*$) and then: 
\be 
\tw_\bk+\tw_{-\bk}= \omega_\bk+\Omega_\bk+\omega_{-\bk}+\Omega_{-\bk}=2(\omega_\bk+\Omega_{\bk})
\ee
Thus, the condition to fulfill for the resonance in the second term (last four lines) of equation (\ref{dineqZ}) reads:
\be 
\omega_\bk+\Omega_\bk=0 \label{oO}
\ee
Also, $\tJ_2^{(0)}$ is positive, whereas the sign of $\cH_{1122}^{+-+-}$ implies that it is impossible to 
generalize without looking at the specific problem we want to describe.

If we take as a paradigmatic example a relatively simple, $4$-wave resonant system, namely the \textit{Nonlinear Klein Gordon} 
system, we easily notice that the Hamiltonian coefficients are strictly positive, see also~\cite{during2017wave}. 
A more accurate analysis is needed in other cases, such as the {deep water gravity 
waves},  whose effective coefficients have been derived in \cite{Zakharovetal92}. If the Hamiltonian coefficients are 
positive, then $\Omega_\bk$ is positive too.

Actually, many of the physical systems one usually considers have
positive Hamiltonian coefficients. Furthermore, 
this last condition is even not necessary to satisfy our weaker condition $\Omega_\bk \ge 0$, 
$\forall \bk \in \Lambda^*$. 
The reason to rely on such a condition is justified by the fact that those systems enjoy the property:
\be \lb{eq: pos-ren-freq}
\tw_\bk=\omega_\bk+\Omega_\bk \ge 0, \forall \bk\in\Lambda^* .
\ee
Thus, condition (\ref{oO}) is never fulfilled, implying that the arguments of the two Dirac delta's 
$\delta\left(\tw_1+\tw_{-1}\right)$ and $\delta\left(\tw_2+\tw_{-2}\right)$ in 
equation (\ref{dineqZ}) cannot vanish for any value of $\bk$ except from $\bk=0$, 
but in that case the Hamiltonian coefficients are identically null. As a consequence, 
for ``positive renormalized frequency'' systems ({\em{i.e.}} satisfying (\ref{eq: pos-ren-freq})) the last four lines of equation (\ref{dineqZ}) 
give zero identically and the dynamical multi-mode equation reduces to the really compact form (\ref{dineqZfin})
given below. 
Let us also note that the frequency 
$\Omega_\bk$, Eq.(\ref{Omega}), contains a factor $\epsilon$ and that the sum in (\ref{Omega}) is expected 
to converge if the energy of the system is finite, then $\Omega_\bk$ is of order $O(\epsilon)$.
Therefore, $\Omega_\bk \ll \omega_\bk$. 
As a matter of fact, even for a system where $\Omega_\bk$ can be negative, $\Omega_\bk$ does not 
nullify the frequency $\omega_\bk$. Thus, the relevant equation for 4-wave resonant systems is:
\begin{eqnarray}
\frac{d\cZ[\lambda,\mu,\tau]}{d\tau}&=& -192\pi \delta_{\mu,0}\sum_{\ul{\sigma}=(1,\sigma_2,\sigma_3,\sigma_4)}\int d^dk_1d^dk_2d^dk_3d^dk_4 \lambda\left(\bk_1\right)|\cH_{1234}^{-\sigma_2\sigma_3\sigma_4}|^2 \nonumber \\
&&\delta\left(\tw^1_{234}\right) \delta^1_{234}\bigg(\frac{\delta^3 \cZ}{\delta\lambda(\bk_2)\delta\lambda(\bk_3)\delta\lambda(\bk_4)}-\sigma_2\frac{\delta^3 \cZ}{\delta\lambda(\bk_1)\delta\lambda(\bk_3)\delta\lambda(\bk_4)} \nonumber \\
&&\quad\;\;-\sigma_3\frac{\delta^3 \cZ}{\delta\lambda(\bk_1)\delta\lambda(\bk_2)\delta\lambda(\bk_4)}-\sigma_4\frac{\delta^3 \cZ}{\delta\lambda(\bk_1)\delta\lambda(\bk_2)\delta\lambda(\bk_3)}\bigg) \label{dineqZfin}
\end{eqnarray}
which is a natural generalization to the $4$-wave case of Eq.(94) in \cite{Eyink}. 
It is worth emphasising that this equation has been obtained with the RP assumption but not with the RPA.

\subsection{Derivation of the spectral hierarchy}

We may now consider the characteristic functional of amplitudes only:\footnote{Then, we can consider equation (\ref{dineqZfin}) without the $\delta_{\mu,0}$ term.}
\be \cZ_L[\lambda,T]\doteq \left \langle \exp \bigg( \sum_{\bk \in \Lambda_L^*}\lambda_\bk J_\bk(T) \bigg)  \right\rangle \label{eq: genfunctbis}\ee

In analogy to \cite{Eyink}, from (\ref{dineqZfin}) we derive a hierarchy of evolution equations for the M-mode spectral correlation functions defined in (\ref{correl}), in the kinetic limit:
\be {\cN}^{(M)}(\bk_1,...,\bk_M,\tau)=\lim_{\epsilon\rightarrow 0}\lim_{L\rightarrow\infty} 
{\cN}^{(M)}_{L,\epsilon}(\bk_1,...,\bk_M,\epsilon^{-2}\tau). \ee
The hierarchy is easy to derive knowing the relation
\be\cN^{(M)}(\bk_1,...,\bk_M,\tau)=\left.(-i)^M\frac{\delta^M \cZ[\lambda,\tau]}{\delta\lambda(\bk_1)\cdot\cdot\cdot\delta\lambda(\bk_M)}\right|_{\lambda=0}. \ee
By taking $M$ functional derivatives of (\ref{dineqZfin}) and setting $\lambda\equiv 0$, one obtains:
\begin{align}
&\dot{\cN}^{(M)}(\bk_1,...,\bk_M,\tau)=192\pi\sum_{j=1}^M\sum_{\ul{\sigma}}
\int d^d\ol{k}_2 d^d\ol{k}_3d^d\ol{k}_4 \delta\left(\tw^1_{234}\right) \delta^1_{234}|H^{\ul{\sigma}}_{\ul{\bk}_j}|^2\nonumber\\
&\Big[\cN^{(M+2)}(\bk_1,...,\bk_{j-1},\bk_{j+1},...,\bk_M,\ol{\bk}_2,\ol{\bk}_3,\ol{\bk}_4,\tau)
-\sigma_2\cN^{(M+2)}(\bk_1,...,\bk_M,\ol{\bk}_3,\ol{\bk}_4,\tau)\nonumber\\
&-\sigma_3\cN^{(M+2)}(\bk_1,...,\bk_M,\ol{\bk}_2,\ol{\bk}_4,\tau)-\sigma_4\cN^{(M+2)}(\bk_1,...,\bk_M,\ol{\bk}_2,\ol{\bk}_3,\tau)\Big]
\lb{spect-hier} \end{align}
We shall refer to this set of equations as to the {\it spectral hierarchy} of kinetic wave turbulence. 
It is exactly analogous to the hierarchy
derived by Lanford from the BBGKY hierarchy in the low-density limit \cite{Lanford75,Lanford76}. If the spectral correlation functions
satisfy bounds on their growth for large orders $M$ that allow them to uniquely characterize the distribution of the empirical 
spectrum, then the spectral hierarchy (\ref{spect-hier}) is not only a consequence of the equation (\ref{dineqZfin}) but is in
fact equivalent to that equation. If the initial functional $\cZ[\lambda,0]$ is of exponential form (\ref{Zexp0}), 
as follows for an initial RP field with uncorrelated amplitudes, an exact solution of (\ref{dineqZfin}) is given by:
\be 
\cZ[\lambda,\tau]=\exp\Big(i\int d^d\bk\,\lambda(\bk)n(\bk,\tau)\Big), 
\lb{Zexp-sol} 
\ee
where $n(\bk,\tau)$ satisfies the standard wave kinetic equation with initial condition 
$n(\bk,0)=n(\bk).$
Equivalently, factorized $M$th-order correlation functions (\ref{spect-fac}) as initial data, entail a  
factorized solution of the spectral hierarchy (\ref{spect-hier}):
\be 
\cN^{(M)}(\bk_1,...,\bk_M,\tau)= \prod_{m=1}^M  n(\bk_m,\tau). 
\lb{spect-fac-sol} 
\ee
Under suitable conditions \cite{Eyink} this is analogous to the propagation of chaos of Boltzmann's 
{\it Stosszahlansatz} \cite{Lanford75,Lanford76}. The results above have an important implication. As follows from our discussion in section \ref{RPA}, the conditions 
(\ref{Zexp-sol}) or (\ref{spect-fac-sol}) imply a law of large numbers for  the empirical spectrum at positive times. That is,
with probability going to 1 in the kinetic limit (first $L\rightarrow\infty,$ then $\epsilon\rightarrow 0$), it follows that 
\be \hn_L(\bk,\epsilon^{-2}\tau) \simeq n(\bk,\tau), \,\,\,\, \tau>0 \ee
where $n(\bk,\tau)$ is the solution of the wave kinetic equation.  An interesting implication for laboratory 
and numerical experiments is that the wave kinetic equations hold for {\it typical} initial amplitudes and 
phases chosen from an RPA ensemble. 
Some technical comments are in order. 
As explained in section 2, the very definition of our generating function (\ref{genfuncintro}) entails that, 
in the thermodynamic limit, the solution has the form (\ref{Zexp0}) if the initial field is RPA. 
It is important to remark that this is an exact solution of Eq.(\ref{dineqZfin}), which has been derived asymptotycally under the sole RP assumption. Therefore, the result is not trivial, besides constituting 
a consistency check.

\section{Derivation of the PDF equation}

With respect to Section \ref{Sec: dinmulti}, we now consider a second possible limit involving only a fixed number of modes $\bk_m$, $m=1,2,...,M$, as the total number $N\rightarrow \infty$. As before, one must keep $\tJ_\bk=O(1)$ for all modes. We thus define the joint characteristic function:
\be \cZ_L^{(M)}(\lambda_1,...,\lambda_M,\mu_1,...,T;\bk_1,...)\doteq\left\langle\exp\left[i\sum_{m=1}^M\lambda_m\tJ_{\bk_M}(T)\right]\prod_{m=1}^M \psi_{\bk_m}^{\mu_m}(T)\right\rangle \label{ZPDF} \ee
This is the characteristic function (110) of \cite{Eyink}, which corresponds to the generating function (68) of \cite{Choietal05b} but with $\lambda_{\bk_m}=i\left(\frac{L}{2\pi}\right)^d\lambda_m,\quad J_{\bk_m}=\left(\frac{2\pi}{L}\right)^d\tJ_{\bk_m}, \quad m=1,...,M$, and for all the other modes $\lambda_\bk=0$.
It also corresponds to the generating functional (5.15) of \cite{Nazarenko11}, with same $\lambda_m$ and $\tJ_m$, but with an imaginary unit in the exponent, and with a finite number $M$ of nonzero arguments.

The reason why $\lambda_m$ is finite and $\lambda_{\bk_m}$ is not is that the exponent of 
(\ref{ZPDF}) contains finitely many terms $\lambda_m\tJ_{\bk_m}$, each of which 
is finite. Then, as $\tJ_{\bk_m}$ must be finite as $L\rightarrow\infty$, the same holds for $\lambda_m$.

We use the symbol $\cZ_L^{(M)}(\lambda,\mu,T)$ when there is no possibility of confusion, and we use the perturbation expansion in $\epsilon$ giving (\ref{eq: Zchi}) for the generating functions ($\forall M$), with the definitions (\ref{eq: chi}) of $\chi_L(\lambda,\mu,T)$ and (\ref{J1}) - (\ref{J5}) of the $\cJ$'s. As $\lambda_m$ is finite, different relations hold for the orders of the prefactors in the $\cJ$'s. In particular, for 
$\cJ_1$, $\cJ_2$, $\cJ_3$ we have:
\be 
\lambda_{\bk_1}+\frac{\mu_{\bk_1}}{2\tJ_{\bk_1}},\;\; \lambda_{\bk_1}+\lambda_{\bk_1}^2\tJ_{\bk_1}-\frac{\mu_{\bk_1}^2}{4\tJ_{\bk_1}} = O\left(L^d\right) 
\ee
and for $\cJ_4$, $\cJ_5$ we have:
\be 
\frac{1}{2} \lambda_{\bk_1}^2+\frac{\mu_{\bk_1}}{4\tJ_{\bk_1}^2}\left(\frac{\mu_{\bk_1}}{2}-1\right)+\frac{\lambda_{\bk_1}\mu_{\bk_1}}{2\tJ_{\bk_1}}, \;\; \lambda_{\bk_1} \lambda_{\bk_2}, \;\; \left(\lambda_{\bk_1}+\frac{\mu_{\bk_1}}{4\tJ_{\bk_1}}\right) \frac{\mu_{\bk_2}}{\tJ_{\bk_2}}= O\left(L^{2d}\right) 
\ee
To calculate the leading order contributions, one must note that some wavenumbers are discrete and take only $M$ values (mode $1$ for $\cJ_1$ - $\cJ_4$, modes $1$ and $2$ for $\cJ_5$), whereas all the others are continuous in the infinite-box limit. This is important to distinguish $O(L^d)$ from $O(M)$ terms.

\subsection{Derivation of the PDF hierarchy}
Collecting the contributions of the $\cJ_1-\cJ_5$ terms enumerated in \ref{pdfcontributions}, we
can neglect nonsecular terms. Furthermore, $\tw_\bk+\tw_{-\bk}=0$ is never fulfilled, so we ignore 
the non-resonant terms with a $\delta(\tw_\bk+\tw_{-\bk})$ contribution.
The two remaining contributions contain $\delta_{\mu,0}$, hence  we can write:
\be 
\quad \delta_{\mu,0}\langle \tJ_j e^{\sum_m i \lambda_m\tJ_m}\rangle_J =-i\delta_{\mu,0}\frac{\partial}{\partial\lambda_j}\cZ^{(M)}
\ee
for wavenumber $\bk_j, j=1,...,M$. Similarly, for mode $\bk \neq \bk_m, \forall m=0,...,M$ we have:
\be 
\delta_{\mu,0}i\langle \tJ_\bk e^{\sum_m i \lambda_m\tJ_m}\rangle_J =\delta_{\mu,0}\left.\frac{\partial}{\partial\lambda_\bk}\cZ^{(M+1)}\right|_{\lambda_\bk=0}
\ee
Subsequently, we consider an intermediate time between the linear time and the nonlinear time, cf.\ Section \ref{Sec: dineq}, $T\sim\frac{1}{\epsilon}$, and we take the $\epsilon\rightarrow0$ limit.
Because $\cZ^{(M)}(\lambda,\mu,T)=\chi^{(M)}(\lambda,\mu,T)+\chi^{(M)*}(-\lambda,-\mu,T)$, while
$ \chi^{(M)}(\lambda,\mu,T)=\chi^{(M)*}(-\lambda,-\mu,T)$ and $\cZ^{(M)}(\lambda,\mu,T)=2\chi^{(M)}(\lambda,\mu,T)$,
we get:
\be
\frac{\cZ^{(M)}(T)-\cZ^{(M)}(0)}{\epsilon^2 T}\sim \frac{\partial\cZ^{(M)}}{\epsilon^2\partial T}(\lambda,\mu,T)=\frac{\partial\cZ^{(M)}}{\partial \tau}(\lambda,\mu,\tau)
\ee
where $ \tau=\epsilon^2 T$ is the nonlinear time. This leads to:\footnote{The continuous quantities are
identified by a bar, and symmetrization is made in the three continuous modes $\overline{\bk}_2$, $\overline{\bk}_3$, $\overline{\bk}_4$.}
\begin{align} 
&\frac{\partial\cZ^{(M)}}{\partial \tau}(\lambda, \mu,\tau) = -192\pi  \delta_{\mu,0} \nonumber\\
&\;\;\btimes\sum_{j=1}^M \sum_{\left(1,\sigma_2,\sigma_3,\sigma_4\right)}
\int d^d\overline{k}_2 d^d\overline{k}_3 d^d\overline{k}_4 \delta^j_{234}\delta\left(\tw^j_{234}\right)\left|\cH^{-\sigma_2\sigma_3\sigma_4}_{j234}\right|^2 \nonumber\\
&\;\;\btimes \bigg\{\Big(\lambda_j+\lambda_j^2\frac{\partial}{\partial\lambda_j}\Big)\left.\frac{\partial^3\cZ^{(M+3)}}{\partial\overline{\lambda}_2\partial\overline{\lambda}_3\partial\overline{\lambda}_4} \right|_{\overline{\lambda}_2=\overline{\lambda}_3=\overline{\lambda}_4=0}- \sigma_2 \lambda_j \left.\frac{\partial^3\cZ^{(M+2)}}{\partial \lambda_j\partial\overline{\lambda}_3\partial\overline{\lambda}_4} \right|_{\overline{\lambda}_3=\overline{\lambda}_4=0}\nonumber\\
  &\qquad- \sigma_3 \lambda_j \left.\frac{\partial^3\cZ^{(M+2)}}{\partial \lambda_j\partial\overline{\lambda}_2\partial\overline{\lambda}_4} \right|_{\overline{\lambda}_2=\overline{\lambda}_4=0} 
  - \sigma_4 \lambda_j \left.\frac{\partial^3\cZ^{(M+2)}}{\partial \lambda_j\partial\overline{\lambda}_2\partial\overline{\lambda}_3} \right|_{\overline{\lambda}_2=\overline{\lambda}_3=0} \bigg\} \label{eqHIER}
\end{align}
for $M=1,2,3,...$.
An important fact is that $\delta_{\mu,0}$ implies that the RP property of the initial wavefield is preserved in
time. 
By Fourier transformation in the $\lambda$ variables, one can obtain an equivalent hierarchy of equations for the joint PDFs $\mathcal{P}^{(M)}(s_1,...,s_M,\tau;\bk_1,...,\bk_M)$, which appears more practical to implement boundary conditions on the amplitudes.

\subsection{The $M$-mode PDF equations}
From the definition of the joint characteristic function of amplitudes, one has:
\begin{eqnarray} \cZ^{(M)}(\lambda_1,...,\lambda_M)&=&\Big\langle e^{\sum_{m}i\lambda_m s_m}\Big\rangle_J \nonumber \\
&=&\int ds_1...ds_M e^{\sum_{m}i\lambda_m s_m} \mathcal{P}^{(M)}(s_1,...,s_M) \end{eqnarray}
$\mathcal{P}^{(M)}$ is the Fourier transform of $\cZ^{(M)}$, so that:
\be \mathcal{P}^{(M)}(s_1,...,s_M)=\frac{1}{2\pi}\int d\lambda_1...d\lambda_M e^{-\sum_{m}i\lambda_m s_m} \cZ^{(M)}(\lambda_1,...,\lambda_M) \ee
A straightforward Fourier transformation yields the following continuity equation:
\be \dot{\mathcal{P}}^{(M)}+\sum_{m=1}^M \frac{\partial}{\partial s_m}\mathcal{F}_m^{(M)}=0,\label{PDF3} \ee
\begin{align}
\mathcal{F}_m^{(M)}=&- 192\pi s_m \sum_{\ul{\sigma}=\left(1,\sigma_2,\sigma_3,\sigma_4\right)}
\int d^d\overline{k}_2 d^d\overline{k}_3 d^d\overline{k}_4  \delta^m_{234}\delta\left(\tw^m_{234}\right) \left|\cH^{-\sigma_2\sigma_3\sigma_4}_{m234}\right|^2 \nonumber\\
&\btimes \bigg[ \int d\overline{s}_2 d\overline{s}_3 d\overline{s}_4  \overline{s}_2\overline{s}_3\overline{s}_4 \frac{\partial \mathcal{P}^{(M+3)}}{\partial s_m}  +\sigma_2  \int  d\overline{s}_3 d\overline{s}_4 \overline{s}_3 \overline{s}_4
\mathcal{P}^{(M+2)} \nonumber\\
&\quad+\sigma_3  \int  d\overline{s}_2 d\overline{s}_4  \overline{s}_2 \overline{s}_4
\mathcal{P}^{(M+2)}+\sigma_4  \int  d\overline{s}_2 d\overline{s}_3 \overline{s}_2 \overline{s}_3
\mathcal{P}^{(M+2)}\bigg] \label{eqHIER2}
\end{align}
This is not a closed equation for $\mathcal{P}^{(M)}$, as it
contains $\mathcal{P}^{(M+2)}$ and $\mathcal{P}^{(M+3)}$, for $M=1,2,3,...$

\subsection{Relation with Peierls equation}
As recalled above, a similar diagrammatic calculation for the 3-wave case was performed in 
Ref.\cite{Choietal05b}. Starting from the same defintion of generating function adopted here, 
the authors derived the canonical Peierls equation, in their version of the thermodynamic 
limit \cite{Choietal05b,Nazarenko11}.
Later, it was shown that certain terms contributing to the Peierls equation are actually negligible, if
the variables are normalized so that the characteristic functional remains finite in the thermodynamic 
limit~\cite{Eyink}. Consequently, an equation that at the leading order differs from the Peierls 
equation was derived in Ref.\cite{Eyink}. \\ \indent
Because the PDF, rather than the generating function, is the object of physical interest, in this subsection 
we investigate the relation between the two asymptotic equations for the PDF, and we show that under two assumptions the Peierls equation reduces to the other leading order equation. We compare our 
PDF equation (\ref{PDF3},\ref{eqHIER2}), that follows from the leading-order equation ({\ref{eqHIER}}), 
with the 4-wave Peierls PDF equation, Eq.(6.120) of Ref.\cite{Nazarenko11}, that has been derived taking 
the Laplace transform of the generating function equation, obtained in the thermodynamic limit, in formal analogy with the 3-wave case.
Such a PDF equation, that takes the form:
\begin{eqnarray}
\dot{\mathcal{P}}=&\pi \epsilon ^4 \int \vert W_{nm}^{ij}\vert^2\delta(\omega^{ij}) \delta_{nm}^{ij} \left[\frac{\delta}{\delta s_j}+\frac{\delta}{\delta s_l}-\frac{\delta}{\delta s_m}-\frac{\delta}{\delta s_n} \right] \nonumber \\
& \times \left(s_j s_l s_m s_n \left[\frac{\delta}{\delta s_j}+\frac{\delta}{\delta s_l}-\frac{\delta}{\delta s_m}-\frac{\delta}{\delta s_n} \right] \mathcal{P} \right ) d{\bf k}_j d{\bf k}_l
d{\bf k}_m d{\bf k}_n \label{PeNa}
\end{eqnarray}
is meant to describe the behaviour of an infinite set of modes.
Unlike our case, there are no spins in Eq.(\ref{PeNa}), but this is irrelevant
for the following discussion. 
Equation (\ref{PeNa}) can also be written as a continuity equation, which reads:
\be \dot{\mathcal{P}}+\int
\frac{\partial}{\partial s_j}\mathcal{F}_j d{\bf k}_j =0, 
\ee
\be
\mathcal{F}_j=- 4\pi \epsilon ^4 s_j \int \vert W_{nm}^{lj}\vert^2 \delta(\omega^{lj}) \delta_{nm}^{lj}  s_j s_l s_m s_n \left[\frac{\delta}{\delta s_j}+\frac{\delta}{\delta s_l}-\frac{\delta}{\delta s_m}-\frac{\delta}{\delta s_n} \right] \mathcal{P} ~ d{\bf k}_l
d{\bf k}_m d{\bf k}_n ~.
\ee
To compare with our M-mode equation, let us assume that the Peierls equation holds with same form 
also in the case of large but finite N, so that we can write:
\be \dot{\mathcal{P}}^{(N)}+\sum_{j=1}^N
\frac{\partial}{\partial s_j}\mathcal{F}_j^{(N)} =0, 
\label{pdfnaz1}
\ee
\be
\mathcal{F}_j^{(N)}=- 4\pi \epsilon ^4 s_j \left (\frac{2 \pi}{L}\right)^{3d} \sum_{l,m,n=1}^N \vert W_{nm}^{lj}\vert^2 \delta(\omega^{lj}) \delta_{nm}^{lj}  s_j s_l s_m s_n \left[\frac{\delta}{\delta s_j}+\frac{\delta}{\delta s_l}-\frac{\delta}{\delta s_m}-\frac{\delta}{\delta s_n} \right] \mathcal{P}^{(N)}  ~.
\label{pdfnaz2}\ee
This is tantamount to commute the thermodynamic limit and the $T\sim 1/\epsilon\rightarrow \infty $ limit with 
the Laplace transform, if the $N \to \infty$ limit can be taken without further 
specifications.\footnote{Strictly speaking, here the derivative is only a finite difference.}
Now, one can integrate out $N-M$ variables, as in the standard BBGKY procedure, to obtain
\be 
\dot{\mathcal{P}}^{(M)}+\sum_{j=1}^M \frac{\partial}{\partial s_j}\mathcal{F}_j^{(M)}=0, \ee
with the flux given by
\begin{align}
\mathcal{F}_j^{(M)}=&
- 4\pi \epsilon ^4 \left (\frac{2 \pi}{L}\right)^{3d}  \times  \nonumber \\
& \left\{  \sum_{l,m,n=M+1}^N
\vert W_{nm}^{lj}\vert^2 \delta(\omega^{lj}) \delta_{nm}^{lj}  
\left(
\int s_l s_m s_n \frac{\partial \mathcal{P}^{M+3}}
{\partial s_j} ds_{l}  ds_{m}  ds_{n} 
 \right . \right . \nonumber \\
& \left . - \int s_m s_n  \mathcal{P}^{M+2}  ds_{m}  ds_{n} 
+2 \int s_m s_l  \mathcal{P}^{M+2}  ds_{m}  ds_{l} 
\right) \nonumber 
\end{align}
\begin{align}
& + \sum_{m,n=1}^M \sum_{l=M+1}^N
\vert W_{nm}^{lj}\vert^2 \delta(\omega^{lj}) \delta_{nm}^{lj} \left[ 
- s_j s_m s_n \left( \mathcal{P}^M +
\frac{\partial}{\partial s_j} \int ds_{l} s_{l} \mathcal{P}^{M+1} \right) \right] \nonumber \\
& + \left . \sum_{l=1}^M \sum_{m,n=M+1}^N 
\vert W_{nm}^{lj}\vert^2 \delta(\omega^{lj}) \delta_{nm}^{lj}  2  \left( \int  s_l s_m s_n \frac{\partial \mathcal{P}^{M+2}}{\partial s_j}  ds_{m}  ds_{n} 
+ \int s_l s_m  \mathcal{P}^{M+1} ds_{m}
\right) \right\}~.
\label{pdfnaz3}
\end{align}
Analogously to the analysis of Ref.\cite{Eyink} for the 3-wave case, we note that
this flux contains more terms than the leading order (\ref{PDF3})-(\ref{eqHIER2}). 
Nonetheless, taking $N\gg M$, and assuming that all the terms are individually of the 
same size, one may obtain the leading order only from the sum having all three indices
$l,m,n$ in $[N-M,N]$. The remaining terms can be neglected simply because they constitute
a negligibly small set compared to the others. Under this assumption, the last two lines 
of Eq.(\ref{pdfnaz3}) can be discarded, and the flux can be written as:
\begin{align}
\mathcal{F}_j^{(M)}=&
- 4\pi \epsilon ^4 \left (\frac{2 \pi}{L}\right)^{3d}  
\sum_{l,m,n=1}^N
\vert W_{nm}^{lj}\vert^2 \delta(\omega^{lj}) \delta_{nm}^{lj}  \left\{ 
\left(
\int s_l s_m s_n \frac{\partial \mathcal{P}^{M+3}}
{\partial s_j} ds_{l}  ds_{m}  ds_{n} 
\right . \right . \nonumber \\
& \left . - \int s_m s_n  \mathcal{P}^{M+2}  ds_{m}  ds_{n} 
+2 \int s_m s_l  \mathcal{P}^{M+2}  ds_{m}  ds_{l} 
\right) ~,
\label{pdfnaz3}
\end{align}
Then, taking the thermodynamic limit ($N,L\rightarrow\infty$) leads to our equation (\ref{eqHIER2}).\\ \indent
In summary, the procedure based on the Peierls equation leads to our same results, provided: (i)  the 
thermodynamic limit is not singular; (ii) the wave modes in the first M modes can be neglected, 
compared to all the others.

\subsection{The 1-mode PDF equation}

It is interesting to note that factorized initial conditions (which is equivalent to RPA property at the 
initial time) imply factorized solutions for Eq.(\ref{eqHIER2}), $\forall \tau \ge 0$:
\be \cZ^{(M)}(\lambda_1,...,\lambda_M,\tau;\bk_1,...,\bk_M)=\prod_{m=1}^M \cZ(\lambda_m,\tau;\bk_m), \quad \tau \ge 0 \ee
with each $\cZ(\lambda_\bk, \tau;\bk)$ satisfying
\be \frac{\partial \cZ(\lambda_\bk, \tau;\bk)}{\partial \tau}=i\eta_\bk \lambda_\bk \Big(1+\lambda_\bk \frac{\partial}{\partial \lambda_\bk}\Big)\cZ(\lambda_\bk, \tau;\bk) - \gamma_\bk \lambda_\bk\frac{\partial \cZ}{\partial \lambda_\bk}(\lambda_\bk, \tau;\bk) \ee
where
\begin{align}
\eta_\bk&\doteq192\pi\sum_{\ul{\sigma}}\int d^d{\bk}_2 d^d{\bk}_3 d^d{\bk}_4 \delta^\bk_{234} \delta\left(\tw^\bk_{234}\right) \left|\cH^{-\sigma_2\sigma_3\sigma_4}_{\bk234}\right|^2 n({\bk}_2) n({\bk}_3) n({\bk}_4) \ge 0, \label{eta} \\
\gamma_\bk&\doteq192\pi\sum_{\ul{\sigma}}\int d^d{\bk}_2 d^d{\bk}_3 d^d{\bk}_4 \delta^\bk_{234} \delta\left(\tw^\bk_{234}\right) \left|\cH^{-\sigma_2\sigma_3\sigma_4}_{\bk234}\right|^2 \nonumber \\
&\qquad\qquad\qquad\quad \btimes \Big[\sigma_2  n({\bk}_3) n({\bk}_4)+\sigma_3  n({\bk}_2) n({\bk}_3)+\sigma_4  n({\bk}_2) n({\bk}_3)\Big] \label{gamma}
\end{align}
For the PDF hierarchy an analogous result holds. Substituting a factorized solution 
into Eq.(\ref{PDF3}) we get the equation for the 1-mode PDF:
\be \mathcal{P}^{(M)}(s_1,...,s_M,\tau;\bk_1,...,\bk_M)=\prod_{m=1}^M P(s_m,\tau;\bk_m)\doteq \prod_{m=1}^M P_m, \quad \tau \ge 0 \ee
Equation (\ref{PDF3}) transforms into:
\begin{align}
\prod_{m\neq1}P_m &\frac{\partial{P}_{1}}{\partial\tau}+\prod_{m\neq2}P_m \frac{\partial{P}_{2}}{\partial\tau}...=\frac{\partial}{\partial s_1}\bigg\{ 192\pi s_1  \sum_{\ul{\sigma}}
\int d^d{k}_2 d^d{k}_3 d^d{k}_4  \delta^1_{234}\delta\left(\tw^1_{234}\right) \nonumber\\
&\qquad\btimes \left|\cH^{-\sigma_2\sigma_3\sigma_4}_{1234}\right|^2 \prod_{m\neq1}P_m  \bigg[ \frac{\partial\mathcal{P}_{1}}{\partial s_1} \int d\overline{s}_2 \overline{s}_2 P_2  \int d\overline{s}_3 \overline{s}_3 P_3  \int d\overline{s}_4 \overline{s}_4 P_4   \nonumber \\
&\quad  + \Big(\sigma_2 P_1  \int d\overline{s}_3 \overline{s}_3 P_3  \int d\overline{s}_4 \overline{s}_4 P_4 + (2 \leftrightarrow 3) +(2 \leftrightarrow 4)\Big)  \bigg]+ ... \bigg\} \label{PDF5}
\end{align}
Recall that $\int ds_i s_i P(s_i,\tau;\bk_i)= n(\bk_i,\tau)$, because of  the definition of the wave spectrum.
Equation (\ref{PDF5}) is made of $M$ independent parts, each of which can 
be written in the continuity equation form:
\be
\frac{\partial P}{\partial \tau}=\frac{\partial}{\partial s} \Big[ s \Big( \eta_\bk \frac{\partial P}{\partial s}+\gamma_\bk P\Big)\Big] \label{PDF4}
\ee
where $\eta_\bk$ and $\gamma_\bk$ are the same defined in (\ref{eta}) and (\ref{gamma}).
These are nonlinear Markov evolution equations in the sense of McKean, since 
the solutions satisfy the set of self-consistency conditions:
\be n(\bk,\tau)=\int ds \;s P(s,\tau;\bk) \label{consistency}\ee
where $n(\bk,\tau)$ is the same spectrum that appears in the formulas for the coefficients (\ref{eta}) and (\ref{gamma}).

These equations are the exact solutions of a model of ``self-consistent Langevin equations''. Here, the model equations take the form of the stochastic differential equations
\be ds_\bk=(\eta_\bk-\gamma_\bk s_\bk) d\tau + \sqrt{2\eta_\bk s_\bk}dW_\bk, \ee
interpreted in the Ito sense, with self-consistency determination of $n(\bk,\tau)$ via (\ref{consistency}).
This generalizes the 3-wave case of Ref.\cite{Eyink} where, $P(s,\tau;\bk)$
relaxes to
\be 
Q(s,\tau;\bk) = \frac{1}{n(\bk,\tau)} \exp(-s/n(\bk,\tau)), 
\lb{rayleigh}  
\ee
which corresponds to a Gaussian distribution of the canonical variable $\tilde{b}=\left(\frac{2\pi}{L}\right)^{d/2} b$.
For any solution $n(\bk,\tau)$ of the wave kinetic equation, 
$Q(s,\tau;\bk)$ solves the 1-mode PDF equation (\ref{PDF4}). Also, the 
relaxation of a general solution $P$ to $Q$ is indicated by an $H$-theorem for the {\it relative entropy}
\be H(P|Q)=\int ds\,\, P(s)\ln\left(\frac{P(s)}{Q(s)}\right)
 = \int ds\,\, P(s)\ln P(s) +\ln n+1. \ee
 This is a convex function of $P,$ non-negative, and vanishing only for $P=Q$. Taking the time-derivative using (\ref{PDF4}), it is straightforward to derive
\be \frac{d}{d\tau} H(P(\tau)|Q(\tau))= -\eta \int ds\,\,\frac{s|\partial_s P(s,\tau)|^2}{P(s,\tau)}
   +\frac{\eta}{n(\tau)},\ee
where 
\be \int -s\partial_s P(s,\tau)\,ds=\int P(s,\tau)\,ds=1 \lb{pdfid} \ee 
is used to cancel terms involving $\gamma$. The self-consistency condition 
$n(\tau)=\int s\,P(s,\tau)\,ds$ implies 
\be \frac{d}{d\tau} H(P(\tau)|Q(\tau))= -\eta \left(
\int ds\,\,\frac{s|\partial_s P(s,\tau)|^2}{P(s,\tau)}
   -\frac{1}{\int s\,P(s,\tau)\,ds}\right)\leq 0. \ee
The inequality follows from the Cauchy-Schwartz inequality applied to (\ref{pdfid})
\be 1=\int \sqrt{sP}\cdot \sqrt{\frac{s}{P}}(-\partial_s P)\,ds 
      \leq \sqrt{\int sP(s)\,ds\cdot\int \frac{s|\partial_s P|^2}{P}\,ds}. \ee   
Equality holds and relative entropy production vanishes if and only if 
$\sqrt{sP}=c \sqrt{\frac{s}{P}}(-\partial_s P)$, or $P=-c\partial_sP$ for some $c$. 
The solution of this latter equation
gives $P=Q$ with $n=c.$ Then, $P(\tau)$  relaxes to $Q(\tau)$
as $\tau$ increases, assuming that kinetic theory holds over the entire amplitudes range 
$s\in (0,\infty)$.

\section{Conclusions}
\begin{enumerate}
\item We have worked within the framework of WWT. We 
have considered a \textit{Hamiltonian system} in $d$ dimensions, 
with a quartic small perturbation implying \textit{4-wave interactions}. 
From Hamilton equations, we have derived the equations of motion expressed in canonical normal variables.
\item To reach a closure for the problem,
we have assumed that the canonical wavefield has the  
RP property at the initial time, allowing a statistical description of the field through its modes. 
We have averaged over phases using a method based on the \textit{Feynman-Wyld diagrams}.
\item For the large-box limit, we have normalized the amplitudes to keep
the wave spectrum finite, which is crucial for a correct evaluation 
of the contributions of the different diagrams \cite{Eyink}. 
The result differs significantly from the previous approach of Ref.\cite{Nazarenko11}, 
but it has been shown that the approach of Ref.\cite{Nazarenko11}
is equivalent to ours, under two technical assumptions.
\item We have formally taken the large-box 
(\textit{thermodynamic}) limit, followed by the small nonlinearity limit, obtaining
the following closed equation:
\begin{eqnarray}
\frac{d\cZ[\lambda,\mu,\tau]}{d\tau}&=& -192\pi \delta_{\mu,0}\sum_{\ul{\sigma}}\int d^dk_1d^dk_2d^dk_3d^dk_4 \lambda\left(\bk_1\right)|\cH_{1234}^{-\sigma_2\sigma_3\sigma_4}|^2\delta(\tw^1_{234}) \nonumber \\
&& \btimes\delta^1_{234}\bigg(\frac{\delta^3 \cZ}{\delta\lambda(\bk_2)\delta\lambda(\bk_3)\delta\lambda(\bk_4)}-\sigma_2\frac{\delta^3 \cZ}{\delta\lambda(\bk_1)\delta\lambda(\bk_3)\delta\lambda(\bk_4)}+ \nonumber \\
&&\quad-\sigma_3\frac{\delta^3 \cZ}{\delta\lambda(\bk_1)\delta\lambda(\bk_2)\delta\lambda(\bk_4)}-\sigma_4\frac{\delta^3 \cZ}{\delta\lambda(\bk_1)\delta\lambda(\bk_2)\delta\lambda(\bk_3)}\bigg) \label{concl1}
\end{eqnarray}
where $\tau$ is the nonlinear time. Note that:
\begin{itemize}
\item Due to the $\delta_{\mu,0}$ factor, the RP property of the initial field is preserved as time goes on. This fact is crucial as it ensures the validity of the equation itself at $\tau>0$.
\item The stricter initial RPA property for the wavefield, and thus a factorized form 
for $\cZ[\lambda,\mu,0]$, entails a solution preserving the factorized form in time.
\item Differentiating the characteristic functional in the variables $\lambda_\bk$'s, 
one obtains the \textit{spectral hierarchy}, which is analogous to the BBGKY hierarchy of Kinetic Theory. 
Assuming RPA for the initial field, the hierarchy is closed obtaining the kinetic wave equation for the
spectrum. This connects our work, that gives for the first time the general derivation for the 4-wave case,
with the existing literature.
\end{itemize}
\item We have defined a different \textit{characteristic function}, for a \textit{finite number of modes} $M$, and derived a hierarchy of equations for its time evolution for any value of $M$, which reads:
\begin{align} &\frac{\partial\cZ^{(M)}}{\partial \tau}(\lambda, \mu,\tau) = -192\pi  \delta_{\mu,0} \nonumber\\
&\;\;\btimes\sum_{j=1}^M \sum_{\left(1,\sigma_2,\sigma_3,\sigma_4\right)}
\int d^d\overline{k}_2 d^d\overline{k}_3 d^d\overline{k}_4 \delta^j_{234}\delta\left(\tw^j_{234}\right)\left|\cH^{-\sigma_2\sigma_3\sigma_4}_{j234}\right|^2 \nonumber\\
&\;\;\btimes \bigg\{\Big(\lambda_j+\lambda_j^2\frac{\partial}{\partial\lambda_j}\Big)\left.\frac{\partial^3\cZ^{(M+3)}}{\partial\overline{\lambda}_2\partial\overline{\lambda}_3\partial\overline{\lambda}_4} \right|_{\overline{\lambda}_2=\overline{\lambda}_3=\overline{\lambda}_4=0}- \sigma_2 \lambda_j \left.\frac{\partial^3\cZ^{(M+2)}}{\partial \lambda_j\partial\overline{\lambda}_3\partial\overline{\lambda}_4} \right|_{\overline{\lambda}_3=\overline{\lambda}_4=0}\nonumber\\
  &\qquad- \sigma_3 \lambda_j \left.\frac{\partial^3\cZ^{(M+2)}}{\partial \lambda_j\partial\overline{\lambda}_2\partial\overline{\lambda}_4} \right|_{\overline{\lambda}_2=\overline{\lambda}_4=0} 
  - \sigma_4 \lambda_j \left.\frac{\partial^3\cZ^{(M+2)}}{\partial \lambda_j\partial\overline{\lambda}_2\partial\overline{\lambda}_3} \right|_{\overline{\lambda}_2=\overline{\lambda}_3=0} \bigg\} \lb{concl2}
\end{align}
\item By taking the Fourier transform of equation (\ref{concl2}), we have derived a hierarchy for the $M$-\textit{mode joint PDFs}, which can be written in \textit{continuity equation} form:
\be
\dot{\mathcal{P}}^{(M)}+\sum_{m=1}^M \frac{\partial}{\partial s_m}\mathcal{F}_m^{(M)}=0, \label{concl3}
\ee
\begin{align*}
\mathcal{F}_m^{(M)}=&- 192\pi s_m \sum_{\ul{\sigma}=\left(1,\sigma_2,\sigma_3,\sigma_4\right)}
\int d^d\overline{k}_2 d^d\overline{k}_3 d^d\overline{k}_4  \delta^m_{234}\delta\left(\tw^m_{234}\right) \left|\cH^{-\sigma_2\sigma_3\sigma_4}_{m234}\right|^2 \\
&\btimes \bigg[ \int d\overline{s}_2 d\overline{s}_3 d\overline{s}_4  \overline{s}_2\overline{s}_3\overline{s}_4 \frac{\partial \mathcal{P}^{(M+3)}}{\partial s_m}  +\sigma_2  \int  d\overline{s}_3 d\overline{s}_4 \overline{s}_3 \overline{s}_4
\mathcal{P}^{(M+2)} \\
&\quad+\sigma_3  \int  d\overline{s}_2 d\overline{s}_4  \overline{s}_2 \overline{s}_4
\mathcal{P}^{(M+2)}+\sigma_4  \int  d\overline{s}_2 d\overline{s}_3 \overline{s}_2 \overline{s}_3
\mathcal{P}^{(M+2)}\bigg]
\end{align*}
where $\mathcal{F}_m^{(M)}$ is the flux for one of the $M$ modes. As in the case of 
Eq.(\ref{concl1}) we have:
\begin{itemize}
\item RP property for the wavefield at $\tau=0$ remains fullfilled for the field at $\tau >0$. So, equation (\ref{concl2}) is valid for any nonlinear time $\tau\ge0$.
\item An initial RPA field remains RPA under the evolution of Eq.(\ref{concl3}). 
\item Under RPA, the hierarchy (\ref{concl3}) can be closed to yield the
the 1-mode PDF equation:
\be
\frac{\partial P}{\partial \tau}=\frac{\partial}{\partial s} 
\Big[ s \Big( \eta_\bk \frac{\partial P}{\partial s}+\gamma_\bk P\Big)\Big] 
\label{concl4},
\ee
\begin{align}
\eta_\bk&\doteq192\pi\sum_{\ul{\sigma}}\int d^d{\bk}_2 d^d{\bk}_3 d^d{\bk}_4 \delta^\bk_{234} \delta\left(\tw^\bk_{234}\right) \left|\cH^{-\sigma_2\sigma_3\sigma_4}_{\bk234}\right|^2 \nonumber \\
&\qquad\qquad\qquad\quad \btimes n({\bk}_2) n({\bk}_3) n({\bk}_4) \ge 0, \label{etaconcl}\\
\gamma_\bk&\doteq192\pi\sum_{\ul{\sigma}}\int d^d{\bk}_2 d^d{\bk}_3 d^d{\bk}_4 \delta^\bk_{234} \delta\left(\tw^\bk_{234}\right) \left|\cH^{-\sigma_2\sigma_3\sigma_4}_{\bk234}\right|^2 \nonumber \\
&\qquad\qquad\qquad\quad \btimes \Big[\sigma_2  n({\bk}_3) n({\bk}_4)+\sigma_3  n({\bk}_2) n({\bk}_3)+\sigma_4  n({\bk}_2) n({\bk}_3)\Big] \label{gammaconcl}
\end{align}
that can be efficiently treated numerically. 
The spectrum $n(\bk)$ in $\eta_\bk$ and $\gamma_\bk$ can be determined using the kinetic equation.
\item Our Eq.(\ref{concl4}) is more general than Eq.(6.51) of \cite{Nazarenko11}, as it
contains all interactions, not only the ``2 waves $\rightarrow$ 2 waves'' interactions.\footnote{Remarkably, see the system of vibrating elastic plates treated in \cite{During06}.}
\item An important solution to (\ref{concl5}) is represented by the Rayleigh distribution:
\be 
Q(s,\tau;\bk)=\frac{1}{n(\bk,\tau)} e^{-s/n(\bk,\tau)} 
\lb{concl5}
\ee
corresponding to equilibrium. In absence of forcing and damping, $P$
tends to the Rayleigh form (\ref{concl5}) for any typical initial condition. This was tested numerically in \cite{chibbaroetal2017}. 
\end{itemize}
\item
In the most general case, Eqs.(\ref{concl1},\ref{concl2},\ref{concl3}) would have some supplementary terms (see equation (\ref{dineqZ})). However, as argued in Section \ref{Resonancecond}, we think they 
are irrelevant for the known physical systems of wave turbulence, since the resonant condition is never fulfilled.
\item For any system where the leading nonlinear phenomena are $N$-wave resonances, our results suggest the
conjecture that the coefficient preceding the right-hand side in Eq.(\ref{concl1}) equals $12 i^{2-N}A_N$,
where $A_N$ is a number and $A_3=3, A_4=16$. Integration over the $N$ wavenumber variables, on which also 
the Hamiltonian coefficients and the two delta's depend yields:
\begin{align}
&\bigg(\frac{\delta^{N-1} \cZ}{\delta\lambda(\bk_2)\delta\lambda(\bk_3)...\delta\lambda(\bk_N)}-\sum_{i=2}^N \sigma_i\frac{\delta^{N-1} \cZ}{\prod_{j\neq i} \delta\lambda(\bk_i)}\bigg)
\end{align}
\item We conclude noting that our derivation of the wave kinetics is not mathematically rigorous, as is common in the
specialized literature. In particular, analogously to Ref.\cite{Eyink} for 3-wave systems, we have not shown that $O(\epsilon^3)$ terms are negligible in the perturbation expansion (\ref{eq: powerseriesb}). The kinetic limit consists indeed of a delicate combination of large box and small nonlinearity limits~\cite{spohn2006phonon}, whereas rigorous proofs based on asymptotic methods are problematic and presently limited to particular systems, see e.g. Ref.\cite{lukkarinen2011weakly}. At the same time, treating a discrete system in a finite
volume and successively taking a suitable large system limit makes physical sense: the various quantities are well
defined, classical mechanics issues are naturally cast in a discrete formalism, and splitting schemes can be mathematically
justified in a variety of circumstances, including kinetic equations \cite{temam1968stabilite,desvillettes1996splitting,preziosi1999conservative}. Furthermore, this
approach allows us to identify and test the properties of the leading order equations of the 4-wave dynamics. As a matter of fact, the recent work~\cite{chibbaroetal2017} has demonstrated the agreement of part of our results with the kinetic equation in~\cite{During06},
that had been derived through asymptotic methods~\cite{during2017wave}. Reference~\cite{chibbaroetal2017} also shows the agreement of the predictions of the PDF equation with direct numerical simulations of relevant 4-wave systems. This further vindicates the approach developed in the present paper.
\end{enumerate}

\section*{Acknowledgements}
The authors want to thank Gregory Eyink, Miguel Onorato and Umberto Giuriato for enlightening correspondence
and discussions. The authors are grateful to Sergey Nazarenko for extensive, detailed and accurate remarks.

\appendix

\section{Consistency with Choi et al. 2005 \cite{Choietal05a}}
Let us assess the consistency of Eq.(\ref{eq: b1}) with Eq.(3) in \cite{Choietal05a}, 
which only concerns ``$\text{2 waves } \rightarrow \text{2 waves}$'' interactions. If Eq.(\ref{eq: b1}) is to describe 
that particular case, the following factor has to be added:
\be 
\left(\delta_{\sigma_2,1} \delta_{\sigma_3,1}\delta_{\sigma_4,-1} + \delta_{\sigma_2,1} \delta_{\sigma_3,-1}\delta_{\sigma_4,1} + \delta_{\sigma_2,-1} \delta_{\sigma_3,1}\delta_{\sigma_4,1}\right) 
\ee
turning (\ref{eq: b1}) into:\footnote{In this paragraph we omit the superscript $(0)$ to simplify the notation.}
\begin {eqnarray}
b_1^{(1)+}(T)&=& \left.\sum_{\bk_2\bk_3\bk_4}\right.^*  \cL_{1234}^{+++-}b_2^+b_3^+b_4^- \Delta_T(\tw^{12}_{34}) \delta^{12}_{34} + (2\leftrightarrow4) + (3\leftrightarrow 4) \nonumber\\
&=& 12i \sum_{\bk_2\bk_3\bk_4}  \cH_{1234}^{-++-}b_2^+b_3^+b_4^- \Delta_T(\tw^{12}_{34}) \delta^{12}_{34} \nonumber \\
&&- 6 \sum_{\bk_3}\cL_{1133}^{+++-} |b_3|^2 \Delta_T b_1^+ \label{long}
\end{eqnarray}
We used Eq.(\ref{star1}) and definition (\ref{LH}). Then, definition (\ref{Omega}) yields:
\be
i b_1^{(1)+}(T) = -12 \sum_{\bk_2\bk_3\bk_4}  \cH_{1234}^{--++}b_2^-b_3^+b_4^+ \Delta_T(\tw^{12}_{34}) \delta^{12}_{34} + \frac{\Omega_1}{\epsilon}b_1^+ T 
\label{compare1}
\ee
This can be compared with Eq.(3) of \cite{Choietal05a}, which in our notation writes:
\be 
i b_1^{(1)+}(T) = \sum_{\bk_2\bk_3\bk_4}  \mathcal{W}^{12}_{34}b_2^-b_3^+b_4^+ \Delta_T(\tw^{34}_{12}) \delta^{34}_{12} - \frac{\Omega_1}{\epsilon}b_1^+ T 
\label{compare2}
\ee
\begin{enumerate}
\item The two equations differ by a sign. This is due to a different definition of the initial field 
$A_\bk\doteq \frac{1}{\sqrt{2}}\left(p_\bk+i q_\bk \right)$, defined as $\frac{1}{\sqrt{2}}\left(q_\bk+i p_\bk \right)$ in \cite{Choietal05a}. 
This also explains why in equation (\ref{compare2}) there are the factors $\Delta_T(\tw^{34}_{12})\; \delta^{34}_{12}$ instead of our $\Delta_T(\tw^{12}_{34})\; \delta^{12}_{34}$ in (\ref{compare1}).
\item Recall that $\cH_{1234}^{--++}=\frac{1}{24} \mathcal{W}^{12}_{34}$, where $\frac{\mathcal{W}}{4}$ instead 
of $\frac{\mathcal{W}}{2}$ \cite{Choietal05a}. So, writing
\be 
\cH_{1234}^{--++} \doteq \frac{1}{2}\frac{1}{3}\frac{1}{2}\mathcal{W}^{12}_{34} = \frac{1}{12}\mathcal{W}^{12}_{34}
\ee
one obtains  
\be 12\; \cH_{1234}^{--++}=\mathcal{W}^{12}_{34} \label{HW}\ee
Therefore, the first two terms in the right handside of equations (\ref{compare1}) and (\ref{compare2}) are consistent.
\item In \cite{Choietal05a} the linear terms inside the interaction term are grouped in $\Omega_1$ this way\footnote{We recall that at time $t=0$ the fields $A_\bk$, $a_\bk$ and $b_\bk$ are equal, because they differ by an exponential factor with $t$ in the exponent, which gives $1$ at the initial time.}:
\be 
\Omega_1 \doteq 2\epsilon \sum_{\bk_2} \mathcal{W}^{12}_{12}|A_2(0)|^2= 2\epsilon \sum_{\bk_2} \mathcal{W}^{12}_{12}|b_2|^2 
\label{Omega1}
\ee
Then, from definition (\ref{Omega}) and using (\ref{LH}) again, our $\Omega_1$ obeys:
\be \Omega_1 \doteq 24\epsilon \sum_{\bk_3} \mathcal{H}_{1133}^{-++-}|b_3|^2\ee
Replacing $12\;\mathcal{H}_{1133}^{-++-}=12\;\mathcal{H}_{1313}^{--++}$ with $\mathcal{W}^{13}_{13}$ using (\ref{HW}) again, we obtain:
\be \Omega_1 = 2\epsilon \sum_{\bk_3} \mathcal{W}^{13}_{13}|b_3|^2 \ee
which is identical to (\ref{Omega1}).
\end{enumerate}
We can thus state that equations (\ref{compare1}) and (\ref{compare2}) are consistent, where (\ref{compare1}) 
results from Eq.(\ref{eq: b1}) for a Hamiltonian like the one of Ref.\cite{Choietal05a}, containing only 
($2\rightarrow2$) interactions. In other words, Eq.(3) of Ref.\cite{Choietal05a} is obtained as a particular case of our evolution equation (\ref{eq: b1}) for the first-order term.
This result is a check meant to assess the agreement of the formalism used in this paper with how 4-wave systems have already been treated in the past.

\section{Proof of the \textit{Lemma} in section \ref{sec: rules}}\label{theorem1}
\begin{itemize}
\item \textit{Role of couplings}\\
The initial wavefield is an RP field and let us assume the $k_i$'s are distinct. Then:
$$
\Big\langle\prod_i\psi_i^{p_i}\Big\rangle_\psi=\prod_i \delta_{p_i,0}
$$
So, the phase average
$$\Big\langle\prod_l\psi_l^{\mu_l}\psi_1\cdot\cdot\cdot\psi_p\psi_{p+1}^*\cdot\cdot\cdot\psi_{q}^*\Big\rangle$$
is the sum of products of Kronecker delta's. Before phase-averaging each $\bk_i \in \Lambda^*_L$, $i=1,...,q$ carries 
an independent degree of freedom. Each delta other than $\delta_{\mu_l,0}$ would cause the degrees of freedom to drop by one. The phase average is zero, unless either $\psi_k$'s and $\psi^*_m$'s  cancel each other (internal coupling) or they cancel with $\psi^{\mu_l}_l$ (external coupling).\\
Every external coupling concerning $\bk_i$ would lead to a Kronecker delta $\delta_{\mu_i,-1}$ or $\delta_{\mu_i,+1}$, which would cause the degrees of freedom to drop by one.
Every internal coupling of pair $\bk_i$, $\bk_j$ (with $\sigma_i=+1, \sigma_j=-1$), would lead to a Kronecker delta $\delta_{\bk_i,\bk_j}$ which would cause the degrees of freedom to drop by one.\\
Suppose a term has $m$ internal couplings, so that $2m$ wavenumbers are internally coupled (each internal coupling concerns a pair of wavenumbers). Then, the term is non-zero only if the remaining $q-2m$ wavenumbers are externally coupled. Thus, the term totally drops by $m+(q-2m)=q-m$ degrees of freedom, that is the non-null contribution of the term is given by a product of $q-m$ Kronecker delta's different from $\delta_{\mu_l,0}$.\\
Therefore, a diagram has $m$ free wavenumbers, where $m$ is the number of internal couplings.
\item \textit{Role of vertices}\\
Consider the role of the momentum conservation delta's at the vertices and how they act with respect to the degrees of freedom.
\begin{itemize}
\item \textit{Diagrams with one vertex}\\
The Kronecker delta of momentum conservation does not lower the degrees of freedom of the graph:
\begin{enumerate}
\item if there are four external couplings there are no more degrees of freedom to reduce;
\item if there are one internal coupling and two external couplings there is one degree of freedom, but in the momentum conservation delta the two internally coupled terms delete each other and thus one wavenumber remains free;
\item if there are two internal couplings, two pairs of two 
wavenumbers delete each other in the momentum conservation delta, and then two free wavenumbers are preserved.
\end{enumerate}
\item \textit{Diagrams with two vertices}\\
The bridge has initially one degree of freedom. Let us consider the three wavenumbers on one side of the bridge.
\begin{enumerate}
\item \textit{No cross-internal couplings}
\begin{itemize}
\item If those three wavenumbers are pinned to a blob, the momentum conservation delta fixes the bridge to a value so that it does not bring any degree of freedom to the diagram;
\item if one is pinned to a blob and two are in-internally coupled, the momentum conservation delta fixes the bridge to the value of the pinned wavenumber, because in the condition of the delta the two in-internally coupled wavenumbers delete each other.
\end{itemize}
\item \textit{One cross-internal coupling}\\
If only one wavenumber is cross-internally coupled with a wavenumber on the other side of the bridge, the two momentum conservation delta's (one is redundant) reduce from two to one the degrees of freedom of the bridge and the cross-internal coupling.
\item \textit{Two cross-internal couplings}\\
If two wavenumbers on one side are internally coupled with two wavenumbers on the other side (so the other two wavenumbers, one for each side, are externally coupled), the two delta's, giving the same condition, reduce the degrees of freedom of the bridge and the two cross-internal couplings from three to two.
\item \textit{Three cross-internal couplings}\\
If all the three wavenumbers on one side are cross-internally coupled with the other side, the two delta's reduce the total degrees of freedom from four to three.
\end{enumerate}
In these four cases, with the only purpose of counting the degrees of freedom, it is as if the bridge never brings any degree of freedom to the diagram and each cross-internal coupling brings one, as a normal internal coupling.
\end{itemize}
We conclude, both for one vertex and two vertices diagrams, that the momentum conservation delta's only delete one degree of freedom when a bridge is present.
\end{itemize}
Collecting the contributions of the couplings and of the vertices, the lemma is proved.

\section{Phase averaging with Feynman-Wyld diagrams}\label{appendix}
\setcounter{equation}{0}
The expression of $\cJ_2$ in action-angle variables, changing variables in (\ref{J2}), is the following:
\begin{eqnarray}
\cJ_2&=&\frac{1}{2}\sum_1{\sum_{234}}^*{\sum_{567}}^*\Big(\lambda_1+\lambda_1^2 J_1-\frac{\mu_1^2}{4 J_1}\Big)\cL_{1234}^{+\sigma_2\sigma_3\sigma_4}\cL_{1567}^{-\sigma_5\sigma_6\sigma_7}\sqrt{J_2 J_3 J_4 J_5 J_6 J_7}\nonumber\\
&&\btimes \Big\langle\psi_2^{\sigma_2}\psi_3^{\sigma_3}\psi_4^{\sigma_4}\psi_5^{\sigma_5}\psi_6^{\sigma_6}\psi_7^{\sigma_7}\prod_{\bk}\psi_{\bk}^{\mu_\bk}\Big\rangle_\psi \Delta_T\left(\tw_{234}^1\right) \Delta_T\left(\tw_{1567}\right) \delta_{234}^1 \delta_{1567} \nonumber \\
\end{eqnarray}
\begin{figure}[htbp]
	\begin{center}
	    	    \unitlength = 1mm
	    \begin{fmffile}{figure13}
	        \begin{fmfgraph*}(45,35)
	            \fmfleft{i1}
	            \fmfright{o6}
	            \fmftop{i2,o2,i4,o5,i5}
	            \fmfbottom{i3,o3,i4b,o7,i7}
	            \fmflabel{$3$}{i1}
	            \fmflabel{$2$}{o2}
	            \fmflabel{$4$}{o3}
	            \fmflabel{$1$}{o4}
	            \fmflabel{$5$}{o5}
	            \fmflabel{$6$}{o6}
	            \fmflabel{$7$}{o7}
	            \fmf{plain}{i1,v1}
	            \fmf{phantom}{i2,o2}	            
	            \fmf{plain}{o2,v1}
	            \fmf{phantom}{v1,i4,v2}
	            \fmf{phantom}{v1,i4b,v2}
	            \fmf{phantom}{i3,o3}
	            \fmf{plain}{o3,v1}
	            \fmf{dashes_arrow}{v1,o4}
	            \fmf{dashes}{o4,v2}
	            \fmf{plain}{v2,o5}
	            \fmf{phantom}{o5,i5}
	            \fmf{plain}{v2,o7}
	            \fmf{phantom}{o7,i7}
	            \fmf{plain}{v2,o6}
	        \end{fmfgraph*}
	    \end{fmffile}
	    \caption{Diagram associated to $\cJ_2$ before phase averaging}
	    \label{Fig.: figure13}
	    \end{center}
\end{figure}
\begin{itemize}
\item Type $0$ diagrams:
\begin{figure}[htbp]
	\begin{center}
	    \unitlength = 1mm
	    \begin{fmffile}{figure14a}
	        \begin{fmfgraph*}(45,15)
	            \fmfleft{l1,l2,i1,l3,l4}
	            \fmfright{r1,r2,o6,r3,r4}
	            \fmftop{i2,o2,i4,o5,i5}
	            \fmfbottom{i3,o3,i4b,o7,i7}
	            \fmflabel{$3$}{o3}
	            \fmflabel{$2$}{o2}
	            \fmflabel{$4\qquad7$}{i4b}
	            \fmflabel{$1$}{o4}
	            \fmflabel{$5$}{o5}
	            \fmflabel{$6$}{o7}
	            \fmf{phantom}{i2,o2}	            
	            \fmf{plain,right}{v1,i1,v1}
	            \fmf{phantom}{v1,i4,v2}
	            \fmf{phantom}{v1,i4b,v2}
	            \fmf{phantom}{i3,o3}
	            \fmf{dashes}{v1,o4}
	            \fmf{dashes_arrow}{o4,v2}
	            \fmf{phantom}{o5,i5}
	            \fmf{plain,right,tension=0.2}{v1,v2}
	            \fmf{plain,right}{v2,o6,v2}
	            \fmf{phantom}{o7,i7}
	        \end{fmfgraph*}
	    \end{fmffile} $\qquad\qquad$
	    	    \unitlength = 1mm
	    \begin{fmffile}{figure14b}
	        \begin{fmfgraph*}(30,28)
	            \fmfcurved
	            \fmfleft{l1,l2,i1,l3,l4}
	            \fmfright{i5,o4,i6,o1,i7}
	            \fmflabel{$1$}{v1}
	            \fmflabel{$\quad3\qquad 6$}{v2}
	            \fmflabel{$\;\;2$}{l3}
	            \fmflabel{$\;\;4$}{l2}
	            \fmflabel{$5$}{o1}
	            \fmflabel{$7$}{o4}
	            \fmf{phantom}{i1,v1,i7}
      	            \fmf{phantom}{i1,v2,i5}
	            \fmf{dashes_arrow,left=0.6}{i1,i6}
	            \fmf{plain,left,tension=0.5}{i1,i6}
	            \fmf{plain,right=0.6}{i1,i6}
	            \fmf{plain,right,tension=0.5}{i1,i6}
	        \end{fmfgraph*}
	    \end{fmffile}
	    \caption{Diagrams 1 and 2 (type 0 diagrams)}
	    \label{Fig.: figure14}
	    \end{center}
\end{figure}
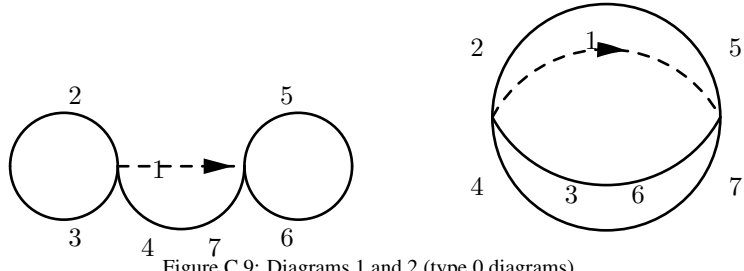
having three free wavenumbers, these are the best candidates for the leading contributions to $\cJ_2$.
\begin{enumerate}
\item The contribution associated with diagram 1 in Fig. \ref{Fig.: figure14} is the following:
\be \frac{1}{2}\sum_1{\sum_{234}}^{*'}{\sum_{567}}^{*'}\Big(\lambda_1+\lambda_1^2 J_1-\frac{\mu_1^2}{4 J_1}\Big)\cL_{1234}^{+\sigma_2\sigma_3\sigma_4}\cL_{1567}^{-\sigma_5\sigma_6\sigma_7}\sqrt{J_2 J_3 J_4 J_5 J_6 J_7}\nonumber\ee
\be \btimes\prod_{m}\delta_{\mu_m,0}\Delta_T\left(\sigma_4\tw_4-\tw_1\right)\Delta_T\left(\tw_1-\sigma_4\tw_4\right)\delta_{ \bk_1,\sigma_4 \bk_4}\delta_{ \bk_2,\bk_3}\delta_{ \bk_4,\bk_7}\delta_{ \bk_5,\bk_6}\ee
$$\ul{\sigma}=\left(1,\sigma_2,-\sigma_2,\sigma_4,\sigma_5,-\sigma_5,-\sigma_4\right)$$
The internal coupling between $2$ and $3$ and between $5$ and $6$ leads to $\Delta_T$'s and vertices $\delta$'s with 
just two terms each. Then, the $\delta_{ \bk_1,\sigma_4 \bk_4}$ term (left vertex) is the the only independent vertex condition, due to the internal coupling between $4$ and $7$. In turn, the $\prod_{m}\delta_{\mu_m,0}$ factor is due to the absence of external couplings.
However, this graph does not entirely contribute to $\cJ_2$. Applying {\it Rule 5} one finds two cases
in graph 1 (Fig. \ref{Fig.: figure14}):
\begin{itemize}
\item $\bk_1=\bk_4 \Longrightarrow \sigma_4=1\quad$ 
This term has been excluded from the interaction when frequency renormalization was done.
\item $\bk_1=-\bk_4 \Longrightarrow \sigma_4=-1\quad$
Now, neither the left vertex nor the right one ($\bk_7=-\bk_1$) are in the configurations giving linear terms. We may conclude that this second case is effectively contributing to $\cJ_2$, so that diagram $1.$ finally gives:
\end{itemize}
\be \frac{1}{2}\sum_{\ul{\sigma}}{\sum_{\ul{\bk}}}'\Big(\lambda_1+\lambda_1^2 J_1-\frac{\mu_1^2}{4 J_1}\Big)\cL_{1234}^{+\sigma_2\sigma_3\sigma_4}\cL_{1567}^{-\sigma_5\sigma_6\sigma_7}\sqrt{J_2 J_3 J_4 J_5 J_6 J_7}\;\delta_{\mu,0}\nonumber\ee
\be \btimes\Delta_T\left(-\tw_1-\tw_{-1}\right)\Delta_T\left(\tw_1+\tw_{-1}\right)\delta_{ \bk_4,-\bk_1}\delta_{ \bk_4,\bk_7}\delta_{ \bk_2,\bk_3}\delta_{ \bk_5,\bk_6}\ee
$$\ul{\sigma}=\left(1,\sigma_2,-\sigma_2,-1,\sigma_5,-\sigma_5,1\right)$$
$$\ul{\bk}=\left(\bk_1,\bk_2,\bk_3,\bk_4,\bk_5,\bk_6,\bk_7\right)$$
\item The contribution to $\cJ_2$ of diagram 2 in Fig. \ref{Fig.: figure14} is the following:
\begin{align}
\frac{1}{2}\sum_{\ul{\sigma}}&{\sum_{\ul{\bk}}}'\Big(\lambda_1+\lambda_1^2 J_1-\frac{\mu_1^2}{4 J_1}\Big)\cL_{1234}^{+\sigma_2\sigma_3\sigma_4}\cL_{1567}^{-\sigma_5\sigma_6\sigma_7}\sqrt{J_2 J_3 J_4 J_5 J_6 J_7}\nonumber\\ &\btimes\prod_{m}\delta_{\mu_m,0}\Delta_T\left(\tw^1_{234}\right)\Delta_T\left(\tw_1^{234}\right)\delta^1_{234}\delta_{\bk_2,\bk_5}\delta_{ \bk_4,\bk_7}\delta_{ \bk_3,\bk_6}
\end{align}
$$\ul{\sigma}=\left(1,\sigma_2,\sigma_3,\sigma_4,-\sigma_2,-\sigma_3,-\sigma_4\right)$$
The Kronecker $\delta$ of the right vertex is redundant, the conditions of {\it Rule 5} for a graph to 
contribute are met, so there are no more simplifications in this case (we omit $^*$ on the sums).
\end{enumerate}
\item Type I diagrams:
\begin{figure}[htbp]
	\begin{center}
	    \unitlength = 1mm
	    \begin{fmffile}{figure15a}
	        \begin{fmfgraph*}(45,15)
	            \fmfleft{l1,l2,i1,l3,l4}
	            \fmfright{r1,r2,o6,r3,r4}
	            \fmftop{i2,o2,i4,o5,i5}
	            \fmfbottom{i3,o3,i4b,o7,i7}
	            \fmflabel{$3$}{o3}
	            \fmflabel{$2$}{o2}
	            \fmflabel{$4\qquad7$}{i4b}
	            \fmflabel{$1$}{o4}
	            \fmflabel{$5$}{o5}
	            \fmflabel{$6$}{o7}
	            \fmfdot{r1,r4}
	            \fmf{phantom}{i2,o2}	            
	            \fmf{plain,right}{v1,i1,v1}
	            \fmf{phantom}{v1,i4,v2}
	            \fmf{phantom}{v1,i4b,v2}
	            \fmf{phantom}{i3,o3}
	            \fmf{dashes}{v1,o4}
	            \fmf{dashes_arrow}{o4,v2}
	            \fmf{phantom}{o5,i5}
	            \fmf{plain,right,tension=0.2}{v1,v2}
	            \fmf{plain}{v2,r4}
	            \fmf{plain}{v2,r1}
	            \fmf{phantom}{o7,i7}
	        \end{fmfgraph*}
	    \end{fmffile} $\qquad\qquad$
	    	    \unitlength = 1mm
	    \begin{fmffile}{figure15b}
	        \begin{fmfgraph*}(45,15)
	            \fmfleft{l1,l2,i1,l3,l4}
	            \fmfright{r1,r2,o6,r3,r4}
	            \fmftop{i2,o2,i4,o5,i5}
	            \fmfbottom{i3,o3,b1,i4b,b2,o7,i7}
	            \fmflabel{$3$}{i3}
	            \fmflabel{$2$}{o2}
	            \fmflabel{$4\qquad7$}{i4b}
	            \fmflabel{$1$}{o4}
	            \fmflabel{$5$}{o5}
	            \fmflabel{$6$}{i7}
	            \fmfdot{b1,b2}
	            \fmf{phantom}{i2,o2}	            
	            \fmf{plain,right}{v1,i1,v1}
	            \fmf{phantom}{v1,i4,v2}
	            \fmf{phantom}{v1,i4b,v2}
	            \fmf{phantom}{i3,o3}
	            \fmf{dashes}{v1,o4}
	            \fmf{dashes_arrow}{o4,v2}
	            \fmf{phantom}{o5,i5}
	            \fmf{plain}{v1,b1}
	            \fmf{plain}{v2,b2}
	            \fmf{plain,right}{v2,o6,v2}
	            \fmf{phantom}{o7,i7}
	        \end{fmfgraph*}
	    \end{fmffile}
	    \caption{Diagrams 3 and 4 (type I diagrams)}
	    \label{Fig.: figure15}
	    \end{center}
\end{figure}
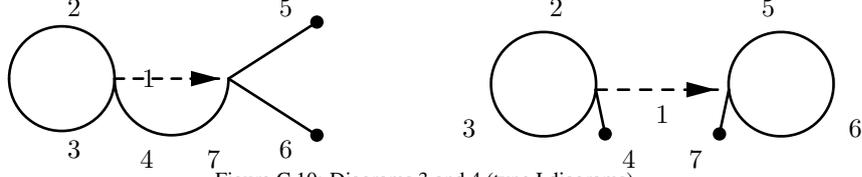
diagrams such as 3 and 4 in Fig.\ref{Fig.: figure15} have two free wavenumbers, hence two unconstrained 
sums and, besided type 0 graphs, they represent the second choice for the leading 
order terms in $\cJ_2$. Together with those with permuted indices, they represent all type 
I diagrams.
\end{itemize}
\textbf{Main contributions to $\cJ_2$.}
If present, the terms proportional to $\mu_1^2$ carry a factor $O(L^{-2d})$ and are greater in order than the terms proportional to $\lambda(\bk_1)$, which have a factor $O(L^{-3d})$. 
\begin{itemize}
\item Type 0 diagrams: the 
contributions of diagrams 1 and 2 are of order $O\left(L^{-3d}\right)$ because $\bk_1$ is not pinned and so the average over phases gives a factor $\delta_{\mu,0}$, implying $\mu_1$ is identically zero. These diagrams have three free summations, so they are of order $O\left(L^{-3d}\right)O\left(L^{3d}\right)=O(1)$.
In total, there are $9$ graphs similar to 1, because as far as the left vertex is concerned, the role of $\bk_4$ can be played by $\bk_2$, $\bk_3$ and $\bk_4$, and, as far as the right vertex is concerned, the role of $\bk_7$ can be played by $\bk_5$, $\bk_6$ and $\bk_7$. So there are $3\btimes 3$ possibilities.
There are $6$ graphs similar to 2: one configuration has $\bk_2$ coupled with one of the three different wavenumbers on the right side. In turn, $\bk_3$ can be coupled with one of the two remaining wavenumbers and $\bk_4$ has no more freedom to make a choice. So, there are $3\btimes 2\btimes 1$ different ways to ``close'' $\cJ_2$ with the shape of diagram 2.
\item Type I diagrams: note that neither 3 nor 4 (our type I graphs) have $\bk_1$ pinned. Thanks to an argument similar to the one used above, $\mu_1$ is null because of the presence of the $\delta_{\mu,0}$. Thus, the two free summations of these graphs make their total contribution at most of order $O\left(L^{-3d}\right)O\left(L^{2d}\right)=O\left(L^{-d}\right)$, which is subleading with respect to the $O(1)$ terms.
\item All the other types of diagrams represent subleading contributions.
\end{itemize}
Just keeping the terms of order $O(1)$, $\cJ_2$ takes the form of equation (\ref{J2c}).\footnote{Remember that these summations have to be intended with $\bk_i\neq\bk_j$, $i\neq j$, except when there is explicitly one $\delta_{\bk_i,\bk_j}$ term.}\\
\textbf{Calculation of $\cJ_3$.}
Substituting the action-angle variables into the expression of $\cJ_3$ (\ref{J3}), we obtain
\be
\cJ_3=\sum_1\Big[{\sum_{234567}}^*\Big(\lambda_1+\frac{\mu_1}{2 J_1}\Big)\cL_{1234}^{+\sigma_2\sigma_3\sigma_4}\cL_{4567}^{\sigma_4\sigma_5\sigma_6\sigma_7}\sqrt{J_1 J_2 J_3  J_5 J_6 J_7}
\ee
\begin{eqnarray}
&&\qquad\quad\btimes \Big\langle\psi_1^-\psi_2^{\sigma_2}\psi_3^{\sigma_3}\psi_5^{\sigma_5}\psi_6^{\sigma_6}\psi_7^{\sigma_7}\prod_{\bk}\psi_{\bk}^{\mu_\bk}\Big\rangle_\psi E_T\left(\tw_{23567}^1,\tw^1_{234}\right) \delta_{234}^1 \delta_{567}^4 \nonumber \\
&& +(4\longleftrightarrow2) + (4\longleftrightarrow3) + \int_0^T\cD_1b_1dt\Big] \label{J3b}
\end{eqnarray}
We start by taking into account the diagrams associated to $\cJ_3$, without the last term containing $\cD_1$. Later, we will separately consider this term and its contribution.
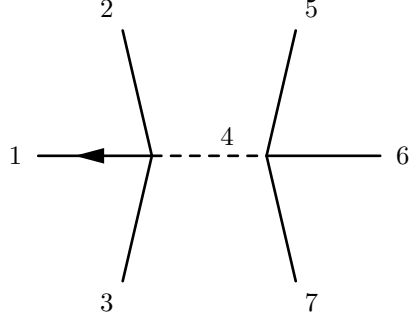
\begin{figure}[htbp]
	\begin{center}
	    	    \unitlength = 1mm
	    \begin{fmffile}{figure16}
	        \begin{fmfgraph*}(45,35)
	            \fmfleft{i1}
	            \fmfright{o6}
	            \fmftop{i2,o2,i4,o5,i5}
	            \fmfbottom{i3,o3,i4b,o7,i7}
	            \fmflabel{$1$}{i1}
	            \fmflabel{$2$}{o2}
	            \fmflabel{$3$}{o3}
	            \fmflabel{$4$}{o4}
	            \fmflabel{$5$}{o5}
	            \fmflabel{$6$}{o6}
	            \fmflabel{$7$}{o7}
	            \fmf{fermion}{v1,i1}
	            \fmf{phantom}{i2,o2}	            
	            \fmf{plain}{o2,v1}
	            \fmf{phantom}{v1,i4,v2}
	            \fmf{phantom}{v1,i4b,v2}
	            \fmf{phantom}{i3,o3}
	            \fmf{plain}{o3,v1}
	            \fmf{dashes}{v1,o4}
	            \fmf{dashes}{o4,v2}
	            \fmf{plain}{v2,o5}
	            \fmf{phantom}{o5,i5}
	            \fmf{plain}{v2,o7}
	            \fmf{phantom}{o7,i7}
	            \fmf{plain}{v2,o6}
	        \end{fmfgraph*}
	    \end{fmffile}
	    \caption{Diagram associated to $\cJ_3$ before phase averaging}
	    \label{Fig.: figure16}
	    \end{center}
\end{figure}
The first diagrams we consider are the type $0$ diagrams ($3$ free wavenumbers).\\
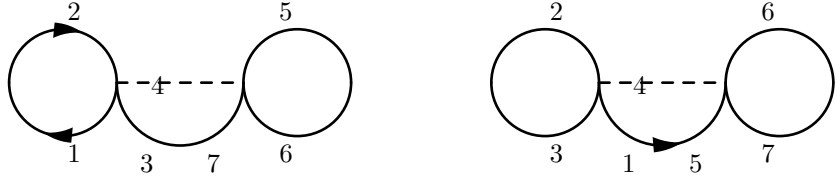
\begin{figure}[htbp]
	\begin{center}
	    \unitlength = 1mm
	    \begin{fmffile}{figure17a}
	        \begin{fmfgraph*}(45,15)
	            \fmfleft{l1,l2,i1,l3,l4}
	            \fmfright{r1,r2,o6,r3,r4}
	            \fmftop{i2,o2,i4,o5,i5}
	            \fmfbottom{i3,o3,i4b,o7,i7}
	            \fmflabel{$1$}{o3}
	            \fmflabel{$2$}{o2}
	            \fmflabel{$3\qquad7$}{i4b}
	            \fmflabel{$4$}{o4}
	            \fmflabel{$5$}{o5}
	            \fmflabel{$6$}{o7}
	            \fmf{phantom}{i2,o2}	            
	            \fmf{fermion,left}{v1,i1,v1}
	            \fmf{phantom}{v1,i4,v2}
	            \fmf{phantom}{v1,i4b,v2}
	            \fmf{phantom}{i3,o3}
	            \fmf{dashes}{v1,o4}
	            \fmf{dashes}{o4,v2}
	            \fmf{phantom}{o5,i5}
	            \fmf{plain,right,tension=0.2}{v1,v2}
	            \fmf{plain,right}{v2,o6,v2}
	            \fmf{phantom}{o7,i7}
	        \end{fmfgraph*}
	    \end{fmffile} $\qquad\qquad$
	    	    \unitlength = 1mm
	    \begin{fmffile}{figure17b}
	        \begin{fmfgraph*}(45,15)
	            \fmfleft{l1,l2,i1,l3,l4}
	            \fmfright{r1,r2,o6,r3,r4}
	            \fmftop{i2,o2,i4,o5,i5}
	            \fmfbottom{i3,o3,i4b,o7,i7}
	            \fmflabel{$3$}{o3}
	            \fmflabel{$2$}{o2}
	            \fmflabel{$1\qquad5$}{i4b}
	            \fmflabel{$4$}{o4}
	            \fmflabel{$6$}{o5}
	            \fmflabel{$7$}{o7}
	            \fmf{phantom}{i2,o2}	            
	            \fmf{plain,left}{v1,i1,v1}
	            \fmf{phantom}{v1,i4,v2}
	            \fmf{phantom}{v1,i4b,v2}
	            \fmf{phantom}{i3,o3}
	            \fmf{dashes}{v1,o4}
	            \fmf{dashes}{o4,v2}
	            \fmf{phantom}{o5,i5}
	            \fmf{fermion,right,tension=0.2}{v1,v2}
	            \fmf{plain,right}{v2,o6,v2}
	            \fmf{phantom}{o7,i7}
	        \end{fmfgraph*}
	    \end{fmffile}
	    \caption{Diagrams 1 and 2 (type 0 diagrams)}
	    \label{Fig.: figure17}
	    \end{center}
\end{figure}
\begin{enumerate}
\item Diagram 1 in Fig. \ref{Fig.: figure17} contributes to $\cJ_3$ as:
\begin{align}
\sum_{1}{\sum_{234567}}^{*'}&\Big(\lambda_1+\frac{\mu_1}{2 J_1}\Big)\cL_{1234}^{+\sigma_2\sigma_3\sigma_4}\cL_{4567}^{\sigma_4\sigma_5\sigma_6\sigma_7}\sqrt{J_1 J_2 J_3  J_5 J_6 J_7}\;\prod_{m}\delta_{\mu_m,0}\nonumber\\
& \btimes E_T\left(0,\sigma_3\tw_3+\sigma_4\tw_4\right) \delta_{\sigma_4\bk_4,-\sigma_3\bk_3} \delta_{\bk_1,\bk_2}\delta_{\bk_3,\bk_7}\delta_{\bk_5,\bk_6}
\end{align}
$$\ul{\sigma}=\left(1,1,\sigma_3,\sigma_4,\sigma_5,-\sigma_5,-\sigma_3\right)$$
The internal couplings between $1$ and $2$, between $5$ and $6$ and between $3$, $4$ and $7$ result into a great simplification inside the arguments of $E_T$ and the vertices delta's (both of which give the same condition). The term $\prod_{m}\delta_{\mu_m,0}$ is due to the fact that there are no external couplings.
Similarly to the case of graph 1 of $\cJ_2$ (Fig. \ref{Fig.: figure14}), this graph is not entirely contributing to $\cJ_3$. Applying {\it Rule 5} of our phase-averaging method, in graph 1 there are two cases:
\begin{itemize}
\item $\bk_3=\bk_4=\bk_7 \Longrightarrow \sigma_3=-\sigma_4=-\sigma_7, \quad$ non contributing;
\item $\bk_4=-\bk_3=-\bk_7 \Longrightarrow \sigma_4=\sigma_3=-\sigma_7, \quad$ contributing and giving:
\end{itemize}
\begin{align}
18\sum_{\ul{\sigma}}{\sum_{\ul{\bk}}}'&\Big(\lambda_1+\frac{\mu_1}{2 J_1}\Big)\cL_{1234}^{+\sigma_2\sigma_3\sigma_4}\cL_{4567}^{\sigma_4\sigma_5\sigma_6\sigma_7}\sqrt{J_1 J_2 J_3  J_5 J_6 J_7}\;\prod_{m}\delta_{\mu_m,0}\nonumber\\
& \btimes E_T\left(0,\sigma_3\left(\omega_3+\omega_{-3}\right)\right) \delta_{\bk_4,-\bk_3} \delta_{\bk_1,\bk_2}\delta_{\bk_3,\bk_7}\delta_{\bk_5,\bk_6}
\end{align}
$$\ul{\sigma}=\left(1,1,\sigma_3,\sigma_3,\sigma_5,-\sigma_5,-\sigma_3\right)$$
$$\ul{\bk}=\left(\bk_1,\bk_2,\bk_3,\bk_4,\bk_5,\bk_6,\bk_7\right)$$
As a matter of fact, there are $18$ graphs similar to 1, because the role of $\bk_4$ can be played by $\bk_2$, $\bk_3$ 
and $\bk_4$, and they all give equivalent results.
\item The contribution to $\cJ_3$ of diagram 2 in Fig. \ref{Fig.: figure17} is the following:
\begin{align}
\sum_{1}{\sum_{234567}}^{*'}&\Big(\lambda_1+\frac{\mu_1}{2 J_1}\Big)\cL_{1234}^{+\sigma_2\sigma_3\sigma_4}\cL_{4567}^{\sigma_4\sigma_5\sigma_6\sigma_7}\sqrt{J_1 J_2 J_3  J_5 J_6 J_7}\;\prod_{m}\delta_{\mu_m,0}\nonumber\\
& \btimes E_T\left(0,-\tw_1+\sigma_4\tw_4\right) \delta_{\bk_1,\sigma_4\bk_4} \delta_{\bk_2,\bk_3}\delta_{\bk_1,\bk_5}\delta_{\bk_6,\bk_7}
\end{align}
$$\ul{\sigma}=\left(1,\sigma_2,-\sigma_2,\sigma_4,1,\sigma_6,-\sigma_6\right)$$
The proof of the above expression is anologous to diagram 1, with two cases:
\begin{itemize}
\item $\bk_1=\bk_4=\bk_5 \Longrightarrow \sigma_4=1=\sigma_5, \quad$ non contributing;
\item $\bk_4=-\bk_1=-\bk_5 \Longrightarrow \sigma_4=-1=-\sigma_5, \quad$ contributing and giving:
\begin{align}
9\sum_{\ul{\sigma}}{\sum_{\ul{\bk}}}'&\Big(\lambda_1+\frac{\mu_1}{2 J_1}\Big)\cL_{1234}^{+\sigma_2\sigma_3\sigma_4}\cL_{4567}^{\sigma_4\sigma_5\sigma_6\sigma_7}\sqrt{J_1 J_2 J_3  J_5 J_6 J_7}\;\prod_{m}\delta_{\mu_m,0}\nonumber\\
& \btimes E_T\left(0,-\left(\tw_1+\tw_{-1}\right)\right) \delta_{\bk_1,-\bk_4} \delta_{\bk_2,\bk_3}\delta_{\bk_1,\bk_5}\delta_{\bk_6,\bk_7}
\end{align}
$$\ul{\sigma}=\left(1,\sigma_2,-\sigma_2,-1,1,\sigma_6,- \sigma_6\right)$$
\end{itemize}
\begin{figure}[htbp]
	\begin{center}
	    \unitlength = 1mm
	    \begin{fmffile}{figure18}
	        \begin{fmfgraph*}(30,28)
	            \fmfcurved
	            \fmfleft{l1,l2,i1,l3,l4}
	            \fmfright{i5,o4,i6,o1,i7}
	            \fmflabel{$4$}{v1}
	            \fmflabel{$\quad1\qquad 6$}{v2}
	            \fmflabel{$\;\;2$}{l3}
	            \fmflabel{$\;\;3$}{l2}
	            \fmflabel{$5$}{o1}
	            \fmflabel{$7$}{o4}
	            \fmf{phantom}{i1,v1,i7}
      	            \fmf{phantom}{i1,v2,i5}
	            \fmf{dashes,left=0.6}{i1,i6}
	            \fmf{plain,left,tension=0.5}{i1,i6}
	            \fmf{fermion,right=0.6}{i1,i6}
	            \fmf{plain,right,tension=0.5}{i1,i6}
	        \end{fmfgraph*}
	    \end{fmffile}
	    \caption{Diagram 3 (type $0$)}
	    \label{Fig.: figure18}
	    \end{center}
\end{figure}
\item The contribution to $\cJ_3$ of diagram 3 in Fig. \ref{Fig.: figure18} is the following:
\begin{align}
18\sum_{\ul{\sigma}}{\sum_{\ul{\bk}}}'&\Big(\lambda_1+\frac{\mu_1}{2 J_1}\Big)\cL_{1234}^{+\sigma_2\sigma_3\sigma_4}\cL_{4567}^{\sigma_4\sigma_5\sigma_6\sigma_7}\sqrt{J_1 J_2 J_3  J_5 J_6 J_7}\;\prod_{m}\delta_{\mu_m,0}\nonumber\\
& \btimes E_T\left(0,\tw^1_{234}\right) \delta_{234}^1 \delta_{\bk_2,\bk_5}\delta_{\bk_1,\bk_6}\delta_{\bk_3,\bk_7}
\end{align}
$$\ul{\sigma}=\left(1,\sigma_2,\sigma_3,\sigma_4,-\sigma_2,1,-\sigma_3\right)$$
In this case, the pinning between wavenumbers of the same vertex is not done. 
So, with the condition that the wavenumbers of each vertex are different from each other (${\sum}'$), this graph 
does not vanish because of {\it Rule 5}. Multiplicity is 18 also here.
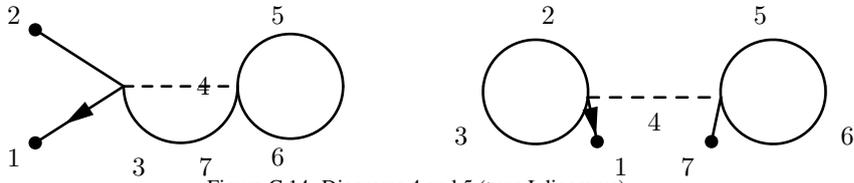
\begin{figure}[htbp]
	\begin{center}
	    \unitlength = 1mm
	    \begin{fmffile}{figure19a}
	        \begin{fmfgraph*}(45,15)
	            \fmfleft{l1,l2,i1,l3,l4}
	            \fmfright{r1,r2,o6,r3,r4}
	            \fmftop{i2,o2,i4,o5,i5}
	            \fmfbottom{i3,o3,i4b,o7,i7}
	            \fmflabel{$4$}{o4}
	            \fmflabel{$2$}{l4}
	            \fmflabel{$3\qquad7$}{i4b}
	            \fmflabel{$1$}{l1}
	            \fmflabel{$5$}{o5}
	            \fmflabel{$6$}{o7}
	            \fmfdot{l1,l4}
	            \fmf{phantom}{i2,o2}	            
	            \fmf{plain}{v1,l4}
	            \fmf{fermion}{v1,l1}
	            \fmf{phantom}{v1,i4,v2}
	            \fmf{phantom}{v1,i4b,v2}
	            \fmf{phantom}{i3,o3}
	            \fmf{dashes}{v1,o4}
	            \fmf{dashes}{o4,v2}
	            \fmf{phantom}{o5,i5}
	            \fmf{plain,right,tension=0.2}{v1,v2}
	            \fmf{plain,right}{v2,o6,v2}
	            \fmf{phantom}{o7,i7}
	        \end{fmfgraph*}
	    \end{fmffile} $\qquad\qquad$
	    	    \unitlength = 1mm
	    \begin{fmffile}{figure19b}
	        \begin{fmfgraph*}(45,15)
	            \fmfleft{l1,l2,i1,l3,l4}
	            \fmfright{r1,r2,o6,r3,r4}
	            \fmftop{i2,o2,i4,o5,i5}
	            \fmfbottom{i3,o3,b1,i4b,b2,o7,i7}
	            \fmflabel{$3$}{i3}
	            \fmflabel{$2$}{o2}
	            \fmflabel{$1\qquad7$}{i4b}
	            \fmflabel{$4$}{o4}
	            \fmflabel{$5$}{o5}
	            \fmflabel{$6$}{i7}
	            \fmfdot{b1,b2}
	            \fmf{phantom}{i2,o2}	            
	            \fmf{plain,right}{v1,i1,v1}
	            \fmf{phantom}{v1,i4,v2}
	            \fmf{phantom}{v1,i4b,v2}
	            \fmf{phantom}{i3,o3}
	            \fmf{dashes}{v1,o4}
	            \fmf{dashes}{o4,v2}
	            \fmf{phantom}{o5,i5}
	            \fmf{fermion}{v1,b1}
	            \fmf{plain}{v2,b2}
	            \fmf{plain,right}{v2,o6,v2}
	            \fmf{phantom}{o7,i7}
	        \end{fmfgraph*}
	    \end{fmffile}
	    \caption{Diagrams 4 and 5 (type I diagrams)}
	    \label{Fig.: figure19}
	    \end{center}
\end{figure}
\item From graph 4 (Fig.\ref{Fig.: figure19}), we analyze type I diagrams,\footnote{We only draw the type I diagrams with $\bk_1$ pinned, knowing that they give a leading contribution as $L\rightarrow\infty$, since they allow $\mu_1\neq0$ and so they get an extra factor $L^d$ with respect to the other graphs.}
to write the contribution to $\cJ_3$.
The right-vertex $\delta$ becomes $\delta_{\sigma_4\bk_4,\sigma_7\bk_7}\Longrightarrow \delta_{\sigma_4\bk_4,-\sigma_3\bk_3}$. In the left-vertex condition, this yields $\delta_{\bk_1,\sigma_2\bk_2}$; but $\bk_1\neq\bk_2$ in this graph, so $\bk_1=-\bk_2$ and $\sigma_2=-1$. Considering the number of equivalent diagrams obtained by permutation of the indices, 
this contribution writes:
\begin{align}
18\sum_{\ul{\sigma}}{\sum_{\ul{\bk}}}'&\Big(\lambda_1+\frac{\mu_1}{2 J_1}\Big)\cL_{1234}^{+\sigma_2\sigma_3\sigma_4}\cL_{4567}^{\sigma_4\sigma_5\sigma_6\sigma_7}\sqrt{J_1 J_2 J_3  J_5 J_6 J_7}\;\delta_{\mu_1,1}\delta_{\mu_2,-\sigma_2}\nonumber\\
&\quad\btimes \prod_{m\neq1,2}\delta_{\mu_m,0}E_T\big(-\left(\tw_1+\tw_{-1}\right), -\left(\tw_1+\tw_{-1}\right)+\sigma_3\tw_3+\sigma_4 \tw_4\big)\nonumber\\
& \qquad \btimes \delta_{\bk_1,-\bk_2}\delta_{-\sigma_3\bk_3,\sigma_4\bk_4}\delta_{\bk_3,\bk_7}\delta_{\bk_5,\bk_6}
\end{align}
$$ \ul{\sigma}=\left(1,-1,\sigma_3,\sigma_4,\sigma_5,-\sigma_5,-\sigma_3\right)$$
\item Diagram 5 in Fig. \ref{Fig.: figure19} gives this contribution to $\cJ_3$ (proof similar to above):
\begin{align}
&9\sum_{\ul{\sigma}}{\sum_{\ul{\bk}}}'\Big(\lambda_1+\frac{\mu_1}{2 J_1}\Big)\cL_{1234}^{+\sigma_2\sigma_3\sigma_4}\cL_{4567}^{\sigma_4\sigma_5\sigma_6\sigma_7}\sqrt{J_1 J_2 J_3  J_5 J_6 J_7}\;\delta_{\mu_1,1}\delta_{\mu_7,-\sigma_7}\nonumber \\
& \btimes \prod_{m\neq1,7}\delta_{\mu_m,0}E_T\big(-\left(\tw_1+\sigma_4\tw_4\right),-\left(\tw_1+\sigma_4\tw_4\right)\big)\delta_{\bk_1,\sigma_4\bk_4}\delta_{\bk_1,\sigma_7\bk_7}\delta_{\bk_2,\bk_3}\delta_{\bk_5,\bk_6}\nonumber\\
& \qquad\qquad\qquad\qquad\ul{\sigma}=\left(1,\sigma_2,-\sigma_2,\sigma_4,\sigma_5,-\sigma_5,\sigma_7\right)
\end{align}
\begin{figure}[htbp]
	\begin{center}
	    \unitlength = 1mm
	    \begin{fmffile}{figure20}
	        \begin{fmfgraph*}(45,13)
	            \fmfleft{l1,l2,i1,l3,l4}
	            \fmfright{r1,r2,o6,r3,r4}
	            \fmftop{i2,o2,i4,o5,i5}
	            \fmfbottom{i3,o3,i4b,o7,i7}
	            \fmflabel{$4$}{o4}
	            \fmflabel{$3\qquad7$}{i4b}
	            \fmflabel{$1$}{i1}
	            \fmflabel{$6$}{o6}
	            \fmflabel{$2\qquad5$}{i4}
	            \fmfdot{i1,o6}
	            \fmf{phantom}{v1,i2}	            
	            \fmf{phantom}{v1,i3}
	            \fmf{fermion}{v1,i1}
	            \fmf{phantom}{v1,i4,v2}
	            \fmf{phantom}{v1,i4b,v2}
	            \fmf{dashes}{v1,o4}
	            \fmf{dashes}{o4,v2}
	            \fmf{phantom}{v2,i5}
	            \fmf{phantom}{v2,i7}
	            \fmf{plain,right}{v1,v2,v1}
	            \fmf{plain}{v2,o6}
	        \end{fmfgraph*}
	    \end{fmffile}
	    \caption{Diagram 6 (type I)}
	    \label{Fig.: figure20}
	    \end{center}
\end{figure}
\item The contribution to $\cJ_3$ of diagram 6 (Fig. \ref{Fig.: figure20}) is:
\begin{align}
18\sum_{\ul{\sigma}}{\sum_{\ul{\bk}}}'&\Big(\lambda_1+\frac{\mu_1}{2 J_1}\Big)\cL_{1234}^{+\sigma_2\sigma_3\sigma_4}\cL_{4567}^{\sigma_4\sigma_5\sigma_6\sigma_7}\sqrt{J_1 J_2 J_3  J_5 J_6 J_7}\;\delta_{\mu_1,1}\delta_{\mu_6,-\sigma_6}\nonumber \\
&\prod_{m\neq1,6}\delta_{\mu_m,0}
E_T\big(-\left(\tw_1+\tw_{-1}\right),-\tw_1+\sigma_2\tw_2+\sigma_3\tw_3+\sigma_4\tw_4\big)\nonumber\\
& \qquad \btimes \delta_{\bk_1,\sigma_2\bk_2+\sigma_3\bk_3+\sigma_4\bk_4}\delta_{\bk_1,-\bk_6}\delta_{\bk_2,\bk_5}\delta_{\bk_3,\bk_7}
\end{align}
$$ \ul{\sigma}=\left(1,\sigma_2,\sigma_3,\sigma_4,-\sigma_2,-1,-\sigma_3\right)$$
\end{enumerate}
\textbf{Calculation of diagrams corresponding to $\int_0^T\cD_1 b_1 dt$.}
Starting from definition (\ref{D1}), such term can be written in the equivalent form:
\begin{eqnarray}
\int_0^T\cD_1 b_1 dt&=& 3\sum_{\sigma_3=\pm1}\sum_{\bk_2\bk_3\bk_4}\left.\sum_{567}\right.^*\Big(\lambda_1+\frac{\mu_1}{2 J_1}\Big)\mathcal{L}_{1234}^{++\sigma_3-\sigma_3} \mathcal{L}_{3567}^{\sigma_3\sigma_5\sigma_6\sigma_7} \nonumber\\
&&\quad\btimes b_2^{+}b_4^{-\sigma_3}b_5^{\sigma_5}b_6^{\sigma_6}b_7^{\sigma_7} \Delta_T\left(\tw_{567}^3\right) \delta_{567}^3 \delta_{\bk_1,\bk_2}\delta_{\bk_3,\bk_4}\label{D1b}
\end{eqnarray}
The diagram associated to this expression before phase averaging is given in Fig.\ref{Fig.: figure21}, with the  constraint $\bk_3=\bk_4$ to keep in mind, because it cannot be represented in the graph.
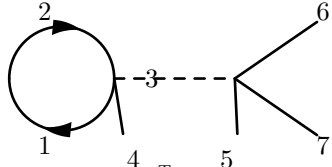
\begin{figure}[htbp]
	\begin{center}
	    \unitlength = 1mm
	    \begin{fmffile}{figure21}
	        \begin{fmfgraph*}(45,15)
	            \fmfleft{l1,l2,i1,l3,l4}
	            \fmfright{r1,r2,o6,r3,r4}
	            \fmftop{i2,o2,t1,i4,t2,o5,i5}
	            \fmfbottom{i3,o3,b1,i4b,b2,o7,i7}
	            \fmflabel{$1$}{o3}
	            \fmflabel{$2$}{o2}
	            \fmflabel{$4\qquad\quad5$}{i4b}
	            \fmflabel{$3$}{o4}
	            \fmflabel{$\;6$}{o5}
	            \fmflabel{$\;7$}{o7}
	            \fmf{phantom}{i2,o2}	            
	            \fmf{fermion,left}{v1,i1,v1}
	            \fmf{phantom}{v1,i4,v2}
	            \fmf{phantom}{v1,i4b,v2}
	            \fmf{phantom}{i3,o3}
	            \fmf{dashes}{v1,o4}
	            \fmf{dashes}{o4,v2}
	            \fmf{phantom}{o5,i5}
	            \fmf{plain}{v1,b1}
	            \fmf{plain}{v2,b2}
	            \fmf{phantom}{v1,t1}
	            \fmf{phantom}{v2,t2}
	            \fmf{plain}{v2,r4}
	            \fmf{plain}{v2,r1}
	            \fmf{phantom}{o7,i7}
	        \end{fmfgraph*}
	    \end{fmffile}
	    \caption{Diagram associated to $\int_0^T\cD_1 b_1 dt$, with the condition that $\bk_3=\bk_4$}
	    \label{Fig.: figure21}
	    \end{center}
\end{figure}
\begin{figure}[htbp]
	\begin{center}
	    \unitlength = 1mm
	    \begin{fmffile}{figure22}
	        \begin{fmfgraph*}(45,15)
	            \fmfleft{l1,l2,i1,l3,l4}
	            \fmfright{r1,r2,o6,r3,r4}
	            \fmftop{i2,o2,i4,o5,i5}
	            \fmfbottom{i3,o3,i4b,o7,i7}
	            \fmflabel{$1$}{o3}
	            \fmflabel{$2$}{o2}
	            \fmflabel{$4\qquad5$}{i4b}
	            \fmflabel{$3$}{o4}
	            \fmflabel{$6$}{o5}
	            \fmflabel{$7$}{o7}
	            \fmf{phantom}{i2,o2}	            
	            \fmf{fermion,left}{v1,i1,v1}
	            \fmf{phantom}{v1,i4,v2}
	            \fmf{phantom}{v1,i4b,v2}
	            \fmf{phantom}{i3,o3}
	            \fmf{dashes}{v1,o4}
	            \fmf{dashes}{o4,v2}
	            \fmf{phantom}{o5,i5}
	            \fmf{plain,right,tension=0.2}{v1,v2}
	            \fmf{plain,right}{v2,o6,v2}
	            \fmf{phantom}{o7,i7}
	        \end{fmfgraph*}
	    \end{fmffile}
	    \caption{Diagram 7 (type 0 diagram, vanishing)}
	    \label{Fig.: figure22}
	    \end{center}
\end{figure}
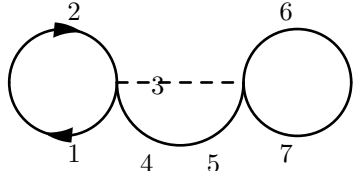
\begin{itemize}
\item[7.]
The graph in Fig. \ref{Fig.: figure21} is closed preserving the maximum number of free wavenumbers as 
in diagram 7 in Fig. \ref{Fig.: figure22}. However, $\delta_{\bk_5,\bk_3}$ and $\delta_{\bk_6,\bk_7}$ 
imply that such diagram does not contribute.
\end{itemize}
Other two type I graphs (8 and 9, Fig. \ref{Fig.: figure23}) are obtained closing the diagram in Fig. \ref{Fig.: figure21} with two external pinnings. These diagrams have two free wavenumbers, but $\bk_1$ is not pinned so 
they vanish identically because of the $\delta_{\mu_1,0}$.
Then, the term with $\cD_1$ inside $\cJ_3$ (equation (\ref{J3b})) can be neglected.
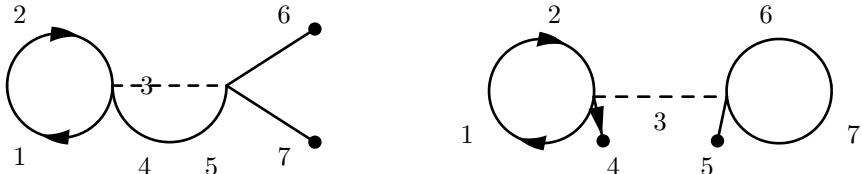
\begin{figure}[htbp]
	\begin{center}
	    \unitlength = 1mm
	    \begin{fmffile}{figure23a}
	        \begin{fmfgraph*}(45,15)
	            \fmfleft{l1,l2,i1,l3,l4}
	            \fmfright{r1,r2,o6,r3,r4}
	            \fmftop{i2,o2,i4,o5,i5}
	            \fmfbottom{i3,o3,i4b,o7,i7}
	            \fmflabel{$3$}{o4}
	            \fmflabel{$2$}{l4}
	            \fmflabel{$4\qquad5$}{i4b}
	            \fmflabel{$1$}{l1}
	            \fmflabel{$6$}{o5}
	            \fmflabel{$7$}{o7}
	            \fmfdot{r1,r4}
	            \fmf{phantom}{i2,o2}	            
	            \fmf{fermion,left}{v1,i1,v1}
	            \fmf{phantom}{v1,i4,v2}
	            \fmf{phantom}{v1,i4b,v2}
	            \fmf{phantom}{i3,o3}
	            \fmf{dashes}{v1,o4}
	            \fmf{dashes}{o4,v2}
	            \fmf{phantom}{o5,i5}
	            \fmf{plain,right,tension=0.2}{v1,v2}
	            \fmf{plain}{v2,r1}
	            \fmf{plain}{v2,r4}
	            \fmf{phantom}{o7,i7}
	        \end{fmfgraph*}
	    \end{fmffile} $\qquad\qquad$
	    	    \unitlength = 1mm
	    \begin{fmffile}{figure23b}
	        \begin{fmfgraph*}(45,15)
	            \fmfleft{l1,l2,i1,l3,l4}
	            \fmfright{r1,r2,o6,r3,r4}
	            \fmftop{i2,o2,i4,o5,i5}
	            \fmfbottom{i3,o3,b1,i4b,b2,o7,i7}
	            \fmflabel{$1$}{i3}
	            \fmflabel{$2$}{o2}
	            \fmflabel{$4\quad\qquad5$}{i4b}
	            \fmflabel{$3$}{o4}
	            \fmflabel{$6$}{o5}
	            \fmflabel{$7$}{i7}
	            \fmfdot{b1,b2}
	            \fmf{phantom}{i2,o2}	            
	            \fmf{fermion,left}{v1,i1,v1}
	            \fmf{phantom}{v1,i4,v2}
	            \fmf{phantom}{v1,i4b,v2}
	            \fmf{phantom}{i3,o3}
	            \fmf{dashes}{v1,o4}
	            \fmf{dashes}{o4,v2}
	            \fmf{phantom}{o5,i5}
	            \fmf{fermion}{v1,b1}
	            \fmf{plain}{v2,b2}
	            \fmf{plain,right}{v2,o6,v2}
	            \fmf{phantom}{o7,i7}
	        \end{fmfgraph*}
	    \end{fmffile}
	    \caption{Diagrams 8 and 9 (type I diagrams), but without $\bk_1$ pinned to an external blob}
	    \label{Fig.: figure23}
	    \end{center}
\end{figure}

\noindent
\textbf{Main contributions to $\cJ_3$.}
If present, the terms proportional to $\mu_1$ carry a factor $O(L^{-2d})$ and are greater in order than the terms proportional to $\lambda(\bk_1)$, which are of order $O(L^{-3d})$. As usual, we must be careful when a $\delta_{\mu,0}$ is present, since that implies the terms in $\mu_1$ to identically vanish.
Among type 0 diagrams, the contributions of diagrams 1, 2 and 3 take with them an $O\left(L^{-3d}\right)$ because $\bk_1$ is not pinned and so the average over phases gives a factor $\delta_{\mu,0}$, implying $\mu_1$ to be identically zero. These diagrams have 3 free summations, so the resulting terms turn out to be of order $O\left(L^{-3d}\right)O\left(L^{3d}\right)=O(1)$. As far as type I diagrams are concerned, thanks to the external coupling of $\bk_1$, diagrams 4, 5 and 6 have $\mu_1$ constrained to the value $1$. Their leading contribution brings an $O\left(L^{-2d}\right)$ and therefore their two free summations are sufficient to make the total contribution $O(1)$.
The other type I diagrams contribute with an $O\left(L^{-d}\right)$ and so are neglected in our work. All the other diagrams (type II and type III) represent subleading contributions.\\
\textbf{Calculation of $\cJ_4$}
\begin{eqnarray}
\cJ_4&=&\sum_1\Big(\frac{1}{2}\lambda_1^2+\frac{\mu_1}{4 J_1^2}\Big(\frac{\mu_1}{2}-1\Big)+\frac{\lambda_1\mu_1}{2 J_1}\Big){\sum_{234}}^*{\sum_{567}}^*\cL_{1234}^{+\sigma_2\sigma_3\sigma_4}\cL_{1567}^{+\sigma_5\sigma_6\sigma_7}\nonumber\\
&&\qquad\quad\btimes J_1\sqrt{J_2 J_3 J_4 J_5 J_6 J_7} \Big\langle\psi_1^{-2}\psi_2^{\sigma_2}\psi_3^{\sigma_3}\psi_4^{\sigma_4}\psi_5^{\sigma_5}\psi_6^{\sigma_6}\psi_7^{\sigma_7}\prod_{\bk}\psi_{\bk}^{\mu_\bk}\Big\rangle_\psi \nonumber\\
&&\qquad\quad\btimes \Delta_T\left(\tw_{234}^1\right)\Delta_T\left(\tw_{567}^1\right) \delta_{234}^1 \delta_{567}^1 \label{J4b}
\end{eqnarray}
\begin{figure}[htbp]
	\begin{center}
	    	    \unitlength = 1mm
	    \begin{fmffile}{figure24}
	        \begin{fmfgraph*}(45,35)
	            \fmfleft{i1}
	            \fmfright{o6}
	            \fmftop{i2,o2,i4,o5,i5}
	            \fmfbottom{i3,o3,i4b,o7,i7}
	            \fmflabel{$3$}{i1}
	            \fmflabel{$2$}{o2}
	            \fmflabel{$4$}{o3}
	            \fmflabel{$1$}{o4}
	            \fmflabel{$5$}{o5}
	            \fmflabel{$6$}{o6}
	            \fmflabel{$7$}{o7}
	            \fmfdot{o4}
	            \fmf{plain}{i1,v1}
	            \fmf{phantom}{i2,o2}	            
	            \fmf{plain}{o2,v1}
	            \fmf{phantom}{v1,i4,v2}
	            \fmf{phantom}{v1,i4b,v2}
	            \fmf{phantom}{i3,o3}
	            \fmf{plain}{o3,v1}
	            \fmf{fermion}{v1,o4}
	            \fmf{fermion}{v2,o4}
	            \fmf{plain}{v2,o5}
	            \fmf{phantom}{o5,i5}
	            \fmf{plain}{v2,o7}
	            \fmf{phantom}{o7,i7}
	            \fmf{plain}{v2,o6}
	        \end{fmfgraph*}
	    \end{fmffile}
	    \caption{Diagram associated to $\cJ_4$ before phase averaging}
	    \label{Fig.: figure24}
	    \end{center}
\end{figure}
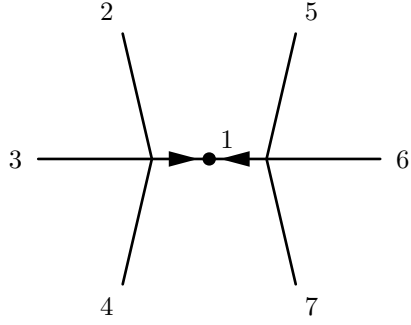
$\bk_1$ must be pinned to a blob so that $\mu_1-2=0$. Let us begin with
type I diagrams, with two free wavenumbers.
\begin{figure}[htbp]
	\begin{center}
	    \unitlength = 1mm
	    	    \begin{fmffile}{figure25a}
	        \begin{fmfgraph*}(30,28)
	            \fmfcurved
	            \fmfleft{l1,l2,l5,i1,l3,l6,l4}
	            \fmfright{i5,o4,r1,i6,o1,r2,i7}
	            \fmflabel{$1$}{v3}
	            \fmflabel{$\quad3\qquad 6$}{v2}
	            \fmflabel{$\;\;2$}{l3}
	            \fmflabel{$\;\;4$}{l2}
	            \fmflabel{$5$}{o1}
	            \fmflabel{$7$}{o4}
	            \fmfdot{v3}
	            \fmf{phantom}{i1,v1,i7}
      	            \fmf{phantom}{i1,v2,i5}
	            \fmf{plain,left=0.6}{i1,i6}
	            \fmf{phantom,left,tension=0.5}{i1,i6}
	            \fmf{fermion}{i1,v3}
	            \fmf{fermion}{i6,v3}
	            \fmf{plain,right=0.6}{i1,i6}
	            \fmf{plain,right,tension=0.5}{i1,i6}
	        \end{fmfgraph*}
	    \end{fmffile}$\qquad\qquad$
	    	    \unitlength = 1mm
	    \begin{fmffile}{figure25b}
	        \begin{fmfgraph*}(45,15)
	            \fmfleft{l1,l2,i1,l3,l4}
	            \fmfright{r1,r2,o6,r3,r4}
	            \fmftop{i2,o2,i4,o5,i5}
	            \fmfbottom{i3,o3,i4b,o7,i7}
	            \fmflabel{$3$}{o3}
	            \fmflabel{$2$}{o2}
	            \fmflabel{$4\qquad7$}{i4b}
	            \fmflabel{$1$}{o4}
	            \fmflabel{$5$}{o5}
	            \fmflabel{$6$}{o7}
	            \fmfdot{o4}
	            \fmf{phantom}{i2,o2}	            
	            \fmf{plain,right}{v1,i1,v1}
	            \fmf{phantom}{v1,i4,v2}
	            \fmf{phantom}{v1,i4b,v2}
	            \fmf{phantom}{i3,o3}
	            \fmf{fermion}{v1,o4}
	            \fmf{fermion}{v2,o4}
	            \fmf{phantom}{o5,i5}
	            \fmf{plain,right,tension=0.2}{v1,v2}
	            \fmf{plain,right}{v2,o6,v2}
	            \fmf{phantom}{o7,i7}
	        \end{fmfgraph*}
	    \end{fmffile}
	    \caption{Diagrams 1 and 2 (type I diagrams)}
	    \label{Fig.: figure25}
	    \end{center}
\end{figure}
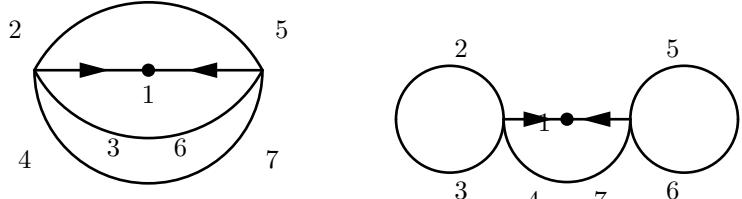
\begin{enumerate}
\item (Fig. \ref{Fig.: figure25})
The two conditions at the vertices imply $\bk_1=\bk_{-1}=0$, so the contribution to $\cJ_4$ is null.
Due to the internal coupling between wavenumbers, we have:
\be
\bk_1 = \sigma_2\bk_2+\sigma_3\bk_3+\sigma_4\bk_4 ~, \quad 
\bk_1 = -\sigma_2\bk_2-\sigma_3\bk_3-\sigma_4\bk_4
\ee
\item (Fig. \ref{Fig.: figure25})
As for the previous graph, the two conditions at the vertices imply $\bk_1=\bk_{-1}=0$, so also 
this contribution to $\cJ_4$ is null. Due to the internal coupling between wavenumbers, we have
\begin{eqnarray*}
\bk_1 = \sigma_4\bk_4 ~, \quad \bk_1 = -\sigma_4\bk_4
\end{eqnarray*}
and $\bk_1=0$ follows.\\\\
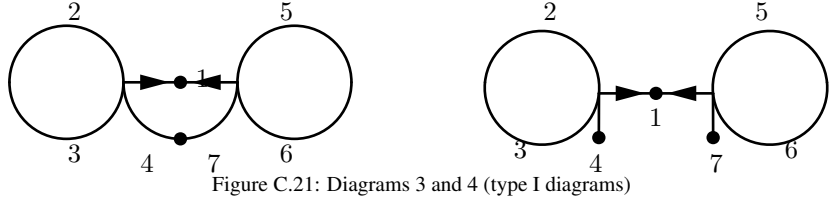
\begin{figure}[htbp]
	\begin{center}
	    \unitlength = 1mm
	    	    \begin{fmffile}{figure26a}
	        \begin{fmfgraph*}(45,15)
	            \fmfleft{l1,l2,i1,l3,l4}
	            \fmfright{r1,r2,o6,r3,r4}
	            \fmftop{i2,o2,i4,o5,i5}
	            \fmfbottom{i3,o3,i4b,o7,i7}
	            \fmflabel{$3$}{o3}
	            \fmflabel{$2$}{o2}
	            \fmflabel{$4\qquad7$}{i4b}
	            \fmflabel{$1$}{o4}
	            \fmflabel{$5$}{o5}
	            \fmflabel{$6$}{o7}
	            \fmfdot{i4b,o4}
	            \fmf{phantom}{i2,o2}	            
	            \fmf{plain,left}{v1,i1,v1}
	            \fmf{phantom}{v1,i4,v2}
	            \fmf{phantom}{v1,i4b,v2}
	            \fmf{phantom}{i3,o3}
	            \fmf{fermion}{v1,o4}
	            \fmf{fermion}{v2,o4}
	            \fmf{phantom}{o5,i5}
	            \fmf{plain,right,tension=0.5}{v1,v2}
	            \fmf{plain,right}{v2,o6,v2}
	            \fmf{phantom}{o7,i7}
	        \end{fmfgraph*}
	    \end{fmffile}$\qquad\qquad$
	    	    \unitlength = 1mm
	    \begin{fmffile}{figure26b}
	        \begin{fmfgraph*}(45,15)
	            \fmfleft{l1,l2,i1,l3,l4}
	            \fmfright{r1,r2,o6,r3,r4}
	            \fmftop{i2,o2,i4,o5,i5}
	            \fmfbottom{i3,o3,b1,i4b,b2,o7,i7}
	            \fmflabel{$3$}{o3}
	            \fmflabel{$2$}{o2}
	            \fmflabel{$4\qquad\qquad7$}{i4b}
	            \fmflabel{$1$}{o4}
	            \fmflabel{$5$}{o5}
	            \fmflabel{$6$}{o7}
	            \fmfdot{b1,b2,o4}
	            \fmf{phantom}{i2,o2}	            
	            \fmf{plain,left}{v1,i1,v1}
	            \fmf{phantom}{v1,i4,v2}
	            \fmf{phantom}{v1,i4b,v2}
	            \fmf{phantom}{i3,o3}
	            \fmf{fermion}{v1,o4}
	            \fmf{fermion}{v2,o4}
	            \fmf{phantom}{o5,i5}
	            \fmf{plain}{v1,b1}
	            \fmf{phantom,right,tension=0.5}{v1,v2}
	            \fmf{plain}{v2,b2}
	            \fmf{plain,right}{v2,o6,v2}
	            \fmf{phantom}{o7,i7}
	        \end{fmfgraph*}
	    \end{fmffile}
	    \caption{Diagrams 3 and 4 (type I diagrams)}
	    \label{Fig.: figure26}
	    \end{center}
\end{figure}
\item  (Fig. \ref{Fig.: figure26})
This diagram gives one of the non-zero leading contributions to $\cJ_4$ and we keep in mind that it has two free wavenumbers implying two free sums.
\be
\bk_1 = \sigma_4\bk_4 ~, \bk_1=\sigma_7\bk_7 ~, \quad
\Longrightarrow \sigma_4=-\sigma_7,\quad \bk_4=-\bk_7
\ee
because $\bk_4$ must be different from $\bk_7$.
\item  (Fig. \ref{Fig.: figure26})
Diagram 4 gives the other non-zero leading contribution to $\cJ_4$ and it implies two free sums too,
$ \sigma_4=\sigma_7,\quad \bk_4=\bk_7.$
\end{enumerate}

\noindent
\textbf{Main contributions to $\cJ_4$.}
The term $\frac{1}{J_1^2}$ has a $\Big( \frac{\mu_1}{2}-1\Big)$ factor which vanishes with $\mu_1$ pinned to the 
value $\mu_1=2$.
The term $\frac{1}{J_1}$ gives the leading contribution, $O(L^{-3d})$, coming from the $J_i$'s. This can at most 
reach order $O(L^{-d})$ thanks to the two free sums of graphs 3 and 4. Type II and type III graphs cannot give 
larger contributions, having at most one free summation. Finally, $\cJ_4$ is of order $O(L^{-d})$ for large L,
and hence negligible with respect to $\cJ_2$ and $\cJ_3$.

\noindent
\textbf{Calculation of $\cJ_5$.}
Before starting, here, we remark that $\bk_2$ has to be treated like $\bk_1$;\footnote{We can see this looking back to (\ref{J5}).} so, as for $\bk_1$ there is no $\sigma_2$ degeneration in the following formulas. Let us rewrite (\ref{J5}) in the form:
\be
\cJ_5=\frac{1}{2}\sum_{\bk_1\neq \bk_2}\Big[\lambda_1\lambda_2(B_1+ B_2)+\Big(\frac{\lambda_1\mu_2}{J_2}+\frac{\mu_1\mu_2}{4J_1 J_2}\Big)(B_1-B_3)\Big]
\label{J5b} 
\ee
where, substituting the action-angle variables, (\ref{J5}) gives
\begin{eqnarray}
B_1&=&{\sum_{345}}^*{\sum_{678}}^*\cL_{1345}^{+\sigma_3\sigma_4\sigma_5}\cL_{2678}^{+\sigma_6\sigma_7\sigma_8}\sqrt{J_1J_2J_3J_4J_5J_6J_7J_8}\Big\langle \psi_1^-\psi_2^-\psi_3^{\sigma_3}\psi_4^{\sigma_4} \nonumber\\
&&\qquad\psi_5^{\sigma_5}\psi_6^{\sigma_6}\psi_7^{\sigma_7}\psi_8^{\sigma_8}\prod_{\bk}\psi_{\bk}^{\mu_\bk}\Big\rangle_\psi \Delta_T\left(\tw^1_{345}\right)\Delta_T\left(\tw^2_{678}\right)\delta^1_{345}\delta^2_{678}\\
B_2&=&{\sum_{345}}^*{\sum_{678}}^*\cL_{1345}^{-\sigma_3\sigma_4\sigma_5}\cL_{2678}^{+\sigma_6\sigma_7\sigma_8}\sqrt{J_1J_2J_3J_4J_5J_6J_7J_8}\Big\langle \psi_1^+\psi_2^-\psi_3^{\sigma_3}\psi_4^{\sigma_4} \nonumber\\
&&\qquad\psi_5^{\sigma_5}\psi_6^{\sigma_6}\psi_7^{\sigma_7}\psi_8^{\sigma_8}\prod_{\bk}\psi_{\bk}^{\mu_\bk}\Big\rangle_\psi \Delta_T\left(\tw_{1345}\right)\Delta_T\left(\tw^2_{678}\right)\delta_{1345}\delta^2_{678}\\
B_3&=&{\sum_{345}}^*{\sum_{678}}^*\cL_{1345}^{+\sigma_3\sigma_4\sigma_5}\cL_{2678}^{-\sigma_6\sigma_7\sigma_8}\sqrt{J_1J_2J_3J_4J_5J_6J_7J_8}\Big\langle \psi_1^-\psi_2^+\psi_3^{\sigma_3}\psi_4^{\sigma_4} \nonumber\\
&&\qquad\psi_5^{\sigma_5}\psi_6^{\sigma_6}\psi_7^{\sigma_7}\psi_8^{\sigma_8}\prod_{\bk}\psi_{\bk}^{\mu_\bk}\Big\rangle_\psi \Delta_T\left(\tw^1_{345}\right)\Delta_T\left(\tw_{2678}\right)\delta^1_{345}\delta_{2678}
\end{eqnarray}

\noindent
The three $B_i$'s can be drawn like in Figs. \ref{Fig.: figure27}, \ref{Fig.: figure28} and \ref{Fig.: figure29}. 
Then, $B_2$ and $B_3$ have the same diagram under permutation $(1,3,4,5) \leftrightarrow (2,6,7,8)$.
\begin{figure}[htbp]
	\begin{center}
	    \unitlength = 1mm
	    	    \begin{fmffile}{figure27}
	        \begin{fmfgraph*}(65,35)
	            \fmfleft{i4}
	            \fmfright{i7}
	            \fmftop{o2,i3,o4,i6,o7}
	            \fmfbottom{b2,i5,b4,i8,b7}
	            \fmflabel{$3$}{i3}
	            \fmflabel{$2$}{i2}
	            \fmflabel{$4$}{i4}
	            \fmflabel{$1$}{i1}
	            \fmflabel{$5$}{i5}
	            \fmflabel{$6$}{i6}
	            \fmflabel{$7$}{i7}
	            \fmflabel{$8$}{i8}
	            \fmf{phantom}{i1,i2}	            
	            \fmf{plain}{i4,v1}
	            \fmf{plain}{i3,v1}
	            \fmf{plain}{i5,v1}
	            \fmf{plain}{i7,v2}
	            \fmf{plain}{i6,v2}
	            \fmf{plain}{i8,v2}
	            \fmf{fermion}{v1,i1}
	            \fmf{fermion}{v2,i2}
	            \fmf{phantom}{i1,i2}
	        \end{fmfgraph*}
	    \end{fmffile}
	    \caption{Diagram associated to $B_1$}
	    \label{Fig.: figure27}
	    \end{center}
\end{figure}
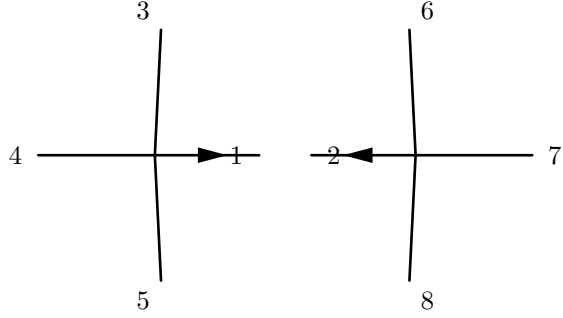
\begin{figure}[htbp]
	\begin{center}
	    \unitlength = 1mm
	    	    \begin{fmffile}{figure28}
	        \begin{fmfgraph*}(65,35)
	            \fmfleft{i4}
	            \fmfright{i7}
	            \fmftop{o2,i3,o4,i6,o7}
	            \fmfbottom{b2,i5,b4,i8,b7}
	            \fmflabel{$3$}{i3}
	            \fmflabel{$2$}{i2}
	            \fmflabel{$4$}{i4}
	            \fmflabel{$1$}{i1}
	            \fmflabel{$5$}{i5}
	            \fmflabel{$6$}{i6}
	            \fmflabel{$7$}{i7}
	            \fmflabel{$8$}{i8}
	            \fmf{phantom}{i1,i2}	            
	            \fmf{plain}{i4,v1}
	            \fmf{plain}{i3,v1}
	            \fmf{plain}{i5,v1}
	            \fmf{plain}{i7,v2}
	            \fmf{plain}{i6,v2}
	            \fmf{plain}{i8,v2}
	            \fmf{fermion}{i1,v1}
	            \fmf{fermion}{v2,i2}
	            \fmf{phantom}{i1,i2}
	        \end{fmfgraph*}
	    \end{fmffile}
	    \caption{Diagram associated to $B_2$}
	    \label{Fig.: figure28}
	    \end{center}
\end{figure}
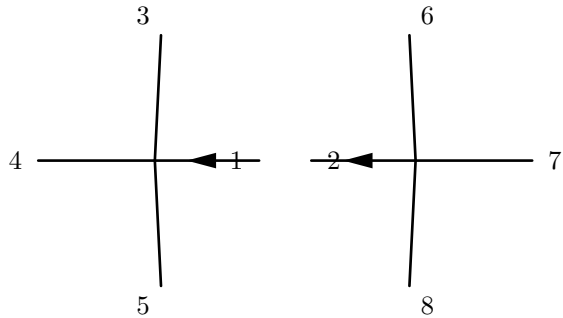
\begin{figure}[htbp]
	\begin{center}
	    \unitlength = 1mm
	    	    \begin{fmffile}{figure29}
	        \begin{fmfgraph*}(65,35)
	            \fmfleft{i4}
	            \fmfright{i7}
	            \fmftop{o2,i3,o4,i6,o7}
	            \fmfbottom{b2,i5,b4,i8,b7}
	            \fmflabel{$3$}{i3}
	            \fmflabel{$2$}{i2}
	            \fmflabel{$4$}{i4}
	            \fmflabel{$1$}{i1}
	            \fmflabel{$5$}{i5}
	            \fmflabel{$6$}{i6}
	            \fmflabel{$7$}{i7}
	            \fmflabel{$8$}{i8}
	            \fmf{phantom}{i1,i2}	            
	            \fmf{plain}{i4,v1}
	            \fmf{plain}{i3,v1}
	            \fmf{plain}{i5,v1}
	            \fmf{plain}{i7,v2}
	            \fmf{plain}{i6,v2}
	            \fmf{plain}{i8,v2}
	            \fmf{fermion}{v1,i1}
	            \fmf{fermion}{i2,v2}
	            \fmf{phantom}{i1,i2}
	        \end{fmfgraph*}
	    \end{fmffile}
	    \caption{Diagram associated to $B_3$}
	    \label{Fig.: figure29}
	    \end{center}
\end{figure}
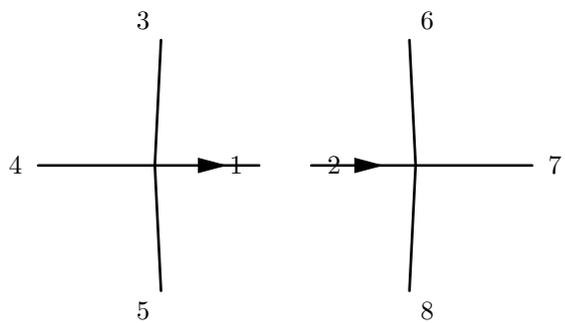
Consider type 0 diagrams with $3$ free wavenumbers. 
\begin{figure}[htbp]
	\begin{center}
	    \unitlength = 1mm
	    	    \begin{fmffile}{figure30a}
	        \begin{fmfgraph*}(30,28)
	            \fmfcurved
	            \fmfleft{l1,l2,l5,i1,l3,l6,l4}
	            \fmfright{i5,o4,r1,i6,o1,r2,i7}
	            \fmflabel{$\quad5\qquad 8$}{v2}
	            \fmflabel{$\;\;3$}{l3}
	            \fmflabel{$\;\;4$}{l2}
	            \fmflabel{$1\quad\qquad 6$}{v3}
	            \fmflabel{$2$}{o1}
	            \fmflabel{$7$}{o4}
	            \fmf{phantom}{i1,v3,i6}
	            \fmf{phantom}{i1,v1,i7}
      	            \fmf{phantom}{i1,v2,i5}
	            \fmf{fermion,right=0.6}{i6,i1}
	            \fmf{phantom,left,tension=0.5}{i1,i6}
	            \fmf{fermion}{i1,i6}
	            \fmf{plain,right=0.6}{i1,i6}
	            \fmf{plain,right,tension=0.5}{i1,i6}
	        \end{fmfgraph*}
	    \end{fmffile}$\qquad\qquad$
	    	    \unitlength = 1mm
	    \begin{fmffile}{figure30b}
	        \begin{fmfgraph*}(30,28)
	            \fmfcurved
	            \fmfleft{l1,l2,l5,i1,l3,l6,l4}
	            \fmfright{i5,o4,r1,i6,o1,r2,i7}
	            \fmflabel{$\quad5\qquad 8$}{v2}
	            \fmflabel{$\;\;3$}{l3}
	            \fmflabel{$\;\;4$}{l2}
	            \fmflabel{$1\quad\qquad 2$}{v3}
	            \fmflabel{$6$}{o1}
	            \fmflabel{$7$}{o4}
	            \fmf{phantom}{i1,v3,i6}
	            \fmf{phantom}{i1,v1,i7}
      	            \fmf{phantom}{i1,v2,i5}
	            \fmf{plain,right=0.6}{i6,i1}
	            \fmf{phantom,left,tension=0.5}{i1,i6}
	            \fmf{fermion}{i6,i1}
	            \fmf{plain,right=0.6}{i1,i6}
	            \fmf{plain,right,tension=0.5}{i1,i6}
	        \end{fmfgraph*}
	    \end{fmffile}
	    \caption{Diagram 1 (for $B_1$) and diagram 2 (for $B_2$); both are type 0 diagrams}
	    \label{Fig.: figure30}
	    \end{center}
\end{figure}
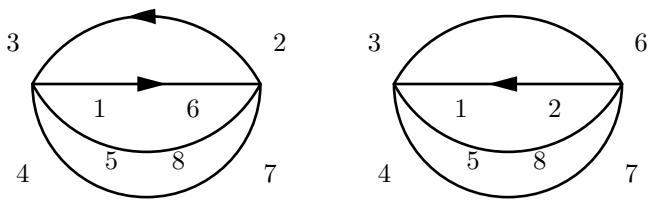
\begin{enumerate}
\item (Fig. \ref{Fig.: figure30})
This graph and similar ones for $B_2$ and $B_3$ (with $\bk_1$ coupled with $\bk_6$ too) contribute, but 
they imply $\mu_1=\mu_2=0$, so they are subleading.
\item (Fig. \ref{Fig.: figure30})
This graph and a similar one for $B_3$ (with $\bk_1$ coupled with $\bk_2$) are not allowed due to the condition 
$\bk_1\neq\bk_2$ in (\ref{J5b}). Such a diagram for $B_1$ is not allowed because the internal coupling 
between $\bk_1$ and $\bk_2$ would lead to $\sigma_1=-\sigma_2$, but $B_1$ already has $\sigma_1=\sigma_2$.
\end{enumerate}
Consider type I diagrams, illustrated in Figs. \ref{Fig.: figure31} and \ref{Fig.: figure32}.
\begin{figure}[htbp]
	\begin{center}
	    \unitlength = 1mm
	    	    \begin{fmffile}{figure31a}
	        \begin{fmfgraph*}(30,28)
	            \fmfcurved
	            \fmfleft{l1,l2,l5,i1,l3,l6,l4}
	            \fmfright{i5,o4,r1,i6,o1,r2,i7}
	            \fmftop{t1,t2,t3,t4,t5}
	            \fmflabel{$\quad5\qquad 8$}{v2}
	            \fmflabel{$\;\;4$}{l2}
	            \fmflabel{$3\quad\qquad 6$}{v3}
	            \fmflabel{$1$}{t2}
	            \fmflabel{$2$}{t4}
	            \fmflabel{$7$}{o4}
	            \fmfdot{t2,t4}
	            \fmf{phantom}{i1,v3,i6}
	            \fmf{phantom}{i1,v1,i7}
      	            \fmf{phantom}{i1,v2,i5}
	            \fmf{fermion}{i1,t2}
	            \fmf{fermion}{i6,t4}
	            \fmf{phantom,left,tension=0.5}{i1,i6}
	            \fmf{plain}{i1,i6}
	            \fmf{plain,right=0.6}{i1,i6}
	            \fmf{plain,right,tension=0.5}{i1,i6}
	        \end{fmfgraph*}
	    \end{fmffile}$\qquad\qquad$
	    	    \unitlength = 1mm
	    \begin{fmffile}{figure31b}
	        \begin{fmfgraph*}(45,15)
	            \fmfleft{l1,l2,i1,l3,l4}
	            \fmfright{r1,r2,o6,r3,r4}
	            \fmftop{i2,o2,i4,o5,i5}
	            \fmfbottom{i3,o3,b1,i4b,b2,o7,i7}
	            \fmflabel{$4$}{i3}
	            \fmflabel{$5$}{o2}
	            \fmflabel{$1\quad\qquad2$}{i4b}
	            \fmflabel{$3\quad\,\,6$}{o4}
	            \fmflabel{$8$}{o5}
	            \fmflabel{$7$}{i7}
	            \fmfdot{b1,b2}
	            \fmf{phantom}{i2,o2}	            
	            \fmf{plain,left}{v1,i1,v1}
	            \fmf{phantom}{v1,i4,v2}
	            \fmf{phantom}{v1,i4b,v2}
	            \fmf{phantom}{i3,o3}
	            \fmf{plain}{v1,o4}
	            \fmf{plain}{o4,v2}
	            \fmf{phantom}{o5,i5}
	            \fmf{fermion}{v1,b1}
	            \fmf{fermion}{v2,b2}
	            \fmf{plain,right}{v2,o6,v2}
	            \fmf{phantom}{o7,i7}
	        \end{fmfgraph*}
	    \end{fmffile}
	    \caption{Diagram 3 and diagram 4 (both for $B_1$ and both type I)}
	    \label{Fig.: figure31}
	    \end{center}
\end{figure}
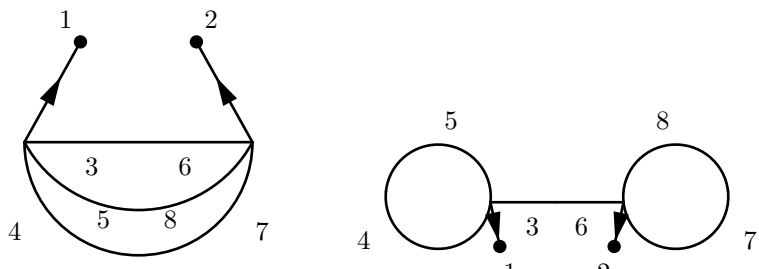
\begin{itemize}
\item[3.] (Fig. \ref{Fig.: figure31})
The internal couplings yield:
$ \bk_3=\bk_6$, $\bk_4=\bk_7$, $\bk_5=\bk_8$, $\sigma_3=-\sigma_6$, 
$\sigma_4=-\sigma_7$, $\sigma_5=-\sigma_8$.
The two deltas at the vertices imply the conditions
\begin{eqnarray}
\bk_1&=&\sigma_3\bk_3+\sigma_4\bk_4+\sigma_5\bk_5 ~; \quad
\bk_2 = \sigma_6\bk_6+\sigma_7\bk_7+\sigma_8\bk_8\nonumber\\
\Longrightarrow \bk_2&=&-\sigma_3\bk_3-\sigma_4\bk_4-\sigma_5\bk_5 \quad
\text{and so } \quad\bk_2=-\bk_1\nonumber
\end{eqnarray}
This holds for $B_1$, because for $B_2$ and $B_3$ the $\delta$'s at the vertices would lead to $\bk_1=\bk_2$,
which is forbidden by (\ref{J5b}).
The contribution of diagram 3 to $\cJ_5$ (only due to $\frac{1}{2}\sum_{1\neq2}(...)B_1$) reads:
\begin{align}
\frac{1}{2}&\sum_{\ul{\sigma}}{\sum_{\ul{\bk}}}'(...)\cL_{1345}^{+\sigma_3\sigma_4\sigma_5}\cL_{2678}^{+\sigma_6\sigma_7\sigma_8}\sqrt{J_1J_2J_3J_4J_5J_6J_7J_8}\;\delta_{\mu_1,1}\delta_{\mu_2,1}\nonumber\\
& \btimes \prod_{m\neq1,2}\delta_{\mu_m,0}\,\delta_{\bk_1,-\bk_2}\delta_{\bk_3,\bk_6}\delta_{\bk_4,\bk_7}\delta_{\bk_5,\bk_8}\delta_{345}^1 \Delta_T\big(\tw^1_{345}\big) \Delta_T\big(\tw^{(-1)345}\big)
\end{align}
$$ \ul{\sigma}=\left(1,1,\sigma_3,\sigma_4,\sigma_5,-\sigma_3,-\sigma_4,-\sigma_5\right) $$
\item[4.] (Fig. \ref{Fig.: figure31})
For the same reason as above, this kind of contribution is only present for $B_1$.
The left-vertex condition gives $\bk_1=\sigma_3 \bk_3$; the right-vertex condition gives $\bk_2=-\sigma_3 \bk_3$. So, $\bk_2=-\bk_1$.
Due to {\it Rule 5}, the right-vertex here is null and then the contribution vanishes.
So, finally diagram 4 does not contribute to $\cJ_5$.\\
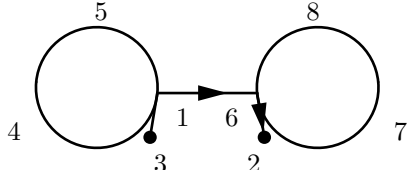
\begin{figure}[htbp]
	\begin{center}
	    	    \unitlength = 1mm
	    \begin{fmffile}{figure32}
	        \begin{fmfgraph*}(45,15)
	            \fmfleft{l1,l2,i1,l3,l4}
	            \fmfright{r1,r2,o6,r3,r4}
	            \fmftop{i2,o2,i4,o5,i5}
	            \fmfbottom{i3,o3,b1,i4b,b2,o7,i7}
	            \fmflabel{$4$}{i3}
	            \fmflabel{$5$}{o2}
	            \fmflabel{$3\quad\qquad2$}{i4b}
	            \fmflabel{$1\quad\,\,6$}{o4}
	            \fmflabel{$8$}{o5}
	            \fmflabel{$7$}{i7}
	            \fmfdot{b1,b2}
	            \fmf{phantom}{i2,o2}	            
	            \fmf{plain,left}{v1,i1,v1}
	            \fmf{phantom}{v1,i4,v2}
	            \fmf{phantom}{v1,i4b,v2}
	            \fmf{phantom}{i3,o3}
	            \fmf{phantom}{v1,o4}
	            \fmf{phantom}{o4,v2}
	            \fmf{phantom}{o5,i5}
	            \fmf{plain}{v1,b1}
	            \fmf{fermion}{v2,b2}
	            \fmf{plain,right}{v2,o6,v2}
	            \fmf{phantom}{o7,i7}
	            \fmf{fermion}{v1,v2}
	        \end{fmfgraph*}
	    \end{fmffile}
	    \caption{Diagram $5$ (for $B_1$); type I diagram}
	    \label{Fig.: figure32}
	    \end{center}
\end{figure}
\item[5.] (Fig. \ref{Fig.: figure32})
An analogous kind of graph also exists for $B_2$ and $B_3$.
There are two free wavenumbers; however, $\mu_1=0$ identically and then, as we are going to see, the contribution to $\cJ_5$ is not leading order.
Similar graphs with $\bk_2$ internally coupled are possible, but with $\mu_2=0$ their contributions to $\cJ_5$ are even smaller.
\item[6.] (Fig. \ref{Fig.: figure33})
One last contribution comes from the type I graph referred to $B_1$.
$$ \sigma_6=-1,\quad \sigma_3=-1, \quad \bk_1=-\bk_3, \quad \bk_1= \bk_6, \quad \bk_2=\sigma_6\bk_6=-\bk_1$$
\begin{figure}[htbp]
	\begin{center}
	    \unitlength = 1mm
	    	    \begin{fmffile}{figure33}
	        \begin{fmfgraph*}(45,15)
	            \fmfleft{l1,l2,i1,l3,l4}
	            \fmfright{r1,r2,o6,r3,r4}
	            \fmftop{i2,o2,i4,o5,i5}
	            \fmfbottom{i3,o3,i4b,o7,i7}
	            \fmflabel{$5$}{o3}
	            \fmflabel{$4$}{o2}
	            \fmflabel{$3\qquad2$}{i4b}
	            \fmflabel{$6$}{v2}
	            \fmflabel{$\quad\quad 1$}{v1}
	            \fmflabel{$7$}{o5}
	            \fmflabel{$8$}{o7}
	            \fmfdot{i4b,o4}
	            \fmf{phantom}{i2,o2}	            
	            \fmf{plain,left}{v1,i1,v1}
	            \fmf{phantom}{v1,i4,v2}
	            \fmf{phantom}{v1,i4b,v2}
	            \fmf{phantom}{i3,o3}
	            \fmf{fermion}{v1,o4}
	            \fmf{fermion}{v2,o4}
	            \fmf{phantom}{o5,i5}
	            \fmf{plain,right,tension=0.5}{v1,v2}
	            \fmf{plain,right}{v2,o6,v2}
	            \fmf{phantom}{o7,i7}
	        \end{fmfgraph*}
	    \end{fmffile}
	    \caption{Diagram 6 (for $B_1$); type I diagram}
	    \label{Fig.: figure33}
	    \end{center}
\end{figure}
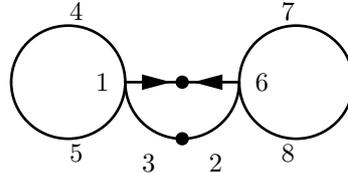
The contribution of diagram 6 to $\cJ_5$ (only due to $\frac{1}{2}\sum_{1\neq2}(...)B_1$) is the following:
\begin{align}
\frac{1}{2}\sum_{\ul{\sigma}}{\sum_{\ul{\bk}}}'&(...)\cL_{1345}^{+\sigma_3\sigma_4\sigma_5}\cL_{2678}^{+\sigma_6\sigma_7\sigma_8}\sqrt{J_1J_2J_3J_4J_5J_6J_7J_8}\;\delta_{\mu_1,2}\delta_{\mu_2,2}\prod_{m\neq1,2}\delta_{\mu_m,0}\nonumber \\
& \btimes \delta_{\bk_1,-\bk_2}\delta_{\bk_1,-\bk_3}\delta_{\bk_1,\bk_6}\delta_{\bk_4,\bk_5}\delta_{\bk_7,\bk_8} \Delta_T\big(\tw^1_{(-1)44}\big) \Delta_T\big(\tw^{(-1)177}\big)
\end{align}
$$ \ul{\sigma}=\left(1,1,-1,\sigma_4,-\sigma_4,-1,\sigma_7,-\sigma_7\right) $$
\end{itemize}
Type II diagrams (with one free wavenumber) and type III (with no free wavenumbers) certainly give subleading contributions and so it is not worth looking at them in detail.\\

\noindent
\textbf{Main contributions to $\cJ_5$.}
Consider the order of the various terms for growing $L$, using the variables 
$\tJ_\bk$ and $\lambda(\bk)$, cf.\ Eq.(\ref{subst}).
\begin{itemize}
\item For $\frac{1}{2}\sum_{1\neq2}\lambda_1\lambda_2 \left(B_1+B_2\right)$, the leading contribution is 
of order $O\left(L^{-d}\right)=O\left(L^{3d}\right)O\left(L^{-4d}\right)$ ($3$ free sums and $8$ factors $\sqrt{J}$).
\item For $\frac{1}{2}\sum_{1\neq2}\frac{\lambda_1\mu_2}{J_2} \left(B_1-B_3\right)$, the leading contribution is 
of order $O\left(L^{-d}\right)=O\left(L^{2d}\right)O\left(L^{-4d}\right)O\left(L^{d}\right)$ ($2$ free sums, 
$8$ factors $\sqrt{J}$ and $J_2$ at denominator).
\item For $\frac{1}{2}\sum_{1\neq2}\frac{\mu_1\mu_2}{4J_1 J_2} B_1$, graph 3 contribution (Fig. \ref{Fig.: figure31})
is of order $O\left(L^{2d}\right)O\left(L^{-4d}\right)O\left(L^{2d}\right)$ ($2$ free sums, 
$8$ factors $\sqrt{J}$ and $J_1 J_2$ at denominator). This graph has multiplicity $6$.
Graph 6 in Fig.(\ref{Fig.: figure33}) and permutations contribute to same order. The multiplicity is $18$.
\item For $\frac{1}{2}\sum_{1\neq2}\frac{\mu_1\mu_2}{4 J_1 J_2}B_3$, the leading contribution is given by type II diagrams (type I do not exist for $B_3$, cf.\ diagram 3, Fig. \ref{Fig.: figure30}) and is of order 
$O\left(L^{-d}\right)=$ $O\left(L^{d}\right)O\left(L^{-4d}\right)O\left(L^{2d}\right)$ ($1$ free sum, $8$ 
factors $\sqrt{J}$ and $J_1 J_2$ at denominator).
\end{itemize}
The final expression of $\cJ_5$ is the one given in equation (\ref{J5c}).

\section{The PDF hierarchy}

\subsection{The five contributions ${\mathcal J}_1-{\mathcal J}_5$}\label{pdfcontributions}
\begin{itemize}
\item $\mathbf{\cJ_1}\qquad$ As seen in Section \ref{Sec: Feynman}, the leading order graph is diagram 2 (Fig. \ref{Fig.: figure5}), a type II diagram with one free wavenumber. This free wavenumber is continuous, as resulting from the internal coupling of $\bk_2$ and $\bk_3$; so, it brings an $O\left(L^d\right)$ factor. Then, we have an $O\left(L^d\right)$ from the prefactor and an $O\left(L^{-2d}\right)$ from the factor $\sqrt{J_1...J_4}$. Thus, the contribution is $O(1)$ and it looks like (\ref{J1c}):
\begin{align}
\Big\langle e^{\sum_\bk i \lambda_\bk J_\bk}\cJ_1 \Big\rangle_J &\sim\left(\frac{2\pi}{L}\right)^d\sum_{(1)}\Big\langle\Big[i\lambda_j\sqrt{\tJ_j}+\frac{1}{2\sqrt{\tJ_j}}\Big] \sqrt{\tJ_2 \tJ_3 \tJ_4}e^{\sum_{m=1}^M i \lambda_m \tJ_m}\Big\rangle_J\nonumber \\
& \btimes \cL^{+\sigma_2\sigma_2-}_{j22(-j)}\delta_{\mu_j,1}\delta_{\mu_{-j},1} \prod_{m\neq\pm j}\delta_{\mu_m,0}\Delta_T\left(-2(\tw_j+\tw_{-j})\right)\Big] 
\end{align}
\be 
\text{with }\;\;\sum_{(1)}\doteq \sum_{j=1}^M\sum_{\sigma_2=\pm1}\sum_{\bk_2} 
\ee
\item $\mathbf{\cJ_2} \qquad$ The prefactor $O\left(L^d\right)$ and the $\sqrt{J_1...}$ term ($O\left(L^{-3d}\right)$) give $O\left(L^{-2d}\right)$. The leading order graphs are diagrams 1 and 2 in Fig. \ref{Fig.: figure14} (type 0).
\begin{itemize}
\item In diagram 1 we have three free wavenumbers, but the one given by $\bk_1$ (and consequently by $\bk_4$ and $\bk_7$, dependent on $\bk_1$) is discrete. So, the total contribution is $O(1)$.
\item In diagram 2 the situation is similar: $\bk_1$ is discrete and the $\delta^1_{234}$ allows other two wavenumbers to be continuous, so it is $O(1)$ too.
\item Diagrams 3 and 4 (type I) in Fig.\ref{Fig.: figure15} only have one continuous free wavenumber and thus they are subleading (of order $O\left(L^{-d}\right)$).
\end{itemize}
The total contribution to $\cJ_2$ reads:
\begin{align}
\Big\langle e^{\sum_\bk i \lambda_\bk J_\bk}\cJ_2 \Big\rangle_J\sim& \frac{1}{2}\delta_{\mu,0}\bigg\{ 9 \sum_{(2)} \left(\frac{2\pi}{L}\right)^{2d}\Big\langle \Big[i\lambda_j-\lambda_j^2\tJ_j\Big]\tJ_2\tJ_5\tJ_{-j}e^{\sum_{m} i \lambda_m \tJ_m}\Big\rangle_J \nonumber \\
&\quad\qquad \cL^{+\sigma_2(-\sigma_2)-}_{j22(-j)}\cL^{-\sigma_5(-\sigma_5)+}_{j55(-j)} \left| \Delta_T\left(\tw_1+\tw_{-1}\right)\right|^2+ \nonumber \\
&\qquad+6 \sum_{(3)} \left(\frac{2\pi}{L}\right)^{2d}\Big\langle \Big[i\lambda_j-\lambda_j^2\tJ_j\Big]\tJ_2\tJ_3\tJ_4 e^{\sum_{m} i \lambda_m \tJ_m}\Big\rangle_J \nonumber \\
&\quad\qquad \cL^{+\sigma_2\sigma_3\sigma_4}_{j234}\cL^{-(-\sigma_2)(-\sigma_3)(-\sigma_4)}_{j234} \left| \Delta_T\left(\tw^1_{234}\right)\right|^2\bigg\}
\end{align}
where similar definitions to those given in Section \ref{Sec: Feynman} hold:
\begin{align}
{\sum}_{(2)} \doteq& \sum_{j=1}^M \sum_{\ul{\sigma}=\left(1,\sigma_2,-\sigma_2,-1,\sigma_5,-\sigma_5,1\right)} {\sum_{\bk_2...\bk_7}}'  \delta_{ \bk_4,-\bk_1}\delta_{ \bk_4,\bk_7}\delta_{ \bk_2,\bk_3}\delta_{ \bk_5,\bk_6}\\
{\sum}_{(3)} \doteq & \sum_{j=1}^M   \sum_{\ul{\sigma}=\left(1,\sigma_2,\sigma_3,\sigma_4,-\sigma_2,-\sigma_3,-\sigma_4\right)} {\sum_{\bk_2...\bk_7}}' \delta^1_{234}\delta_{\bk_2,\bk_5}\delta_{ \bk_4,\bk_7}\delta_{ \bk_3,\bk_6}
\end{align}

\item $\mathbf{\cJ_3}\qquad$ The prefactor is $O(L^d)$ and the $\sqrt{J_1...}$ term is $O\left(L^{-3d}\right)$. The leading order graphs are diagrams 1, 2, 3, 4, 5 and 6.
\begin{itemize}
\item In diagrams 1, 2 and 3 we have three free wavenumbers, but $\bk_1$ is discrete and so there are two free wavenumbers, making the contributions $O(1)$.
\item In diagrams 4, 5 and 6 $\bk_1$ is discrete too but it is pinned. Two degrees of freedom remain, due to two free continuous wavenumbers.
\item All the other graphs are subleading.
\end{itemize}
The following contributions for $\cJ_3$ result:
\begin{align}
\Big\langle &e^{\sum_\bk i \lambda_\bk J_\bk}\cJ_3 \Big\rangle_J \sim \left(\frac{2\pi}{L}\right)^{2d}\delta_{\mu,0}\nonumber\\
\btimes&\bigg\{18 \sum_{(4)}\Big\langle i\lambda_j\tJ_j\tJ_3\tJ_5 e^{\sum...}\Big\rangle_J \cL^{++\sigma_3\sigma_3}_{jj3(-3)}\cL^{\sigma_3\sigma_5(-\sigma_5)(-\sigma_3)}_{(-3)553} E_T\left(0,\sigma_3(\tw_3+\tw_{-3}\right) \nonumber\\
&+9 \sum_{(5)}\Big\langle i\lambda_j\tJ_j\tJ_3\tJ_6 e^{\sum...}\Big\rangle_J \cL^{+\sigma_2(-\sigma_2)-}_{j22(-j)}\cL^{-+\sigma_6(-\sigma_6)}_{(-j)j66} E_T\left(0,-(\tw_j+\tw_{-j}\right)\nonumber\\
&+18 \sum_{(6)}\Big\langle i\lambda_j\tJ_j\tJ_2\tJ_3 e^{\sum...}\Big\rangle_J \cL^{+\sigma_2\sigma_3\sigma_4}_{j234}\cL^{\sigma_4(-\sigma_2)+(-\sigma_3)}_{42j3} E_T\left(0,\tw^j_{234}\right)\bigg\}\nonumber\\
+&\left(\frac{2\pi}{L}\right)^{2d} \bigg\{18\sum_{(7)}\Big\langle \Big[i\lambda_j+\frac{1}{2\tJ_j}\Big] \sqrt{\tJ_j\tJ_{-j}}\tJ_3\tJ_5 e^{\sum...}\Big\rangle_J \cL^{+-\sigma_3\sigma_4}_{j(-j)34} \nonumber\\
&\qquad\qquad\quad\btimes\cL^{\sigma_4\sigma_5(-\sigma_5)(-\sigma_3)}_{4553}\prod_{m\neq\pm j}\delta_{\mu_m,0}\delta_{\mu_j,1}\delta_{\mu_{-j},1}\delta_{\mu_7,-\sigma_7}\nonumber\\ &\qquad\qquad\quad \btimes E_T\left(-(\tw_j+\tw_{-j},-(\tw_j+\tw_{-j}+\sigma_3\tw_3+\sigma_4\tw_4)\right)\nonumber\\
&\qquad\qquad\;+9\sum_{(8)}\Big\langle \Big[i\lambda_j+\frac{1}{2\tJ_j}\Big] \sqrt{\tJ_7\tJ_{j}}\tJ_3\tJ_5 e^{\sum...}\Big\rangle_J \cL^{+\sigma_2(-\sigma_2)\sigma_4}_{j224} \nonumber\\
&\qquad\qquad\quad\btimes\cL^{\sigma_4\sigma_5(-\sigma_5)\sigma_7}_{4557} \prod_{m\neq j,7} \delta_{\mu_m,0}\delta_{\mu_j,1}\delta_{\mu_7,-\sigma_7}\nonumber\\ &\qquad\qquad\quad \btimes E_T\left(-\tw_j+\sigma_4\tw_4,-\sigma_4\tw_4+\sigma_7\tw_7)\right)\nonumber
\end{align}
\begin{align}
&\qquad\qquad\;+18\sum_{(9)}\Big\langle \Big[i\lambda_j+\frac{1}{2\tJ_j}\Big] \sqrt{\tJ_j\tJ_{-j}}\tJ_3\tJ_5 e^{\sum...}\Big\rangle_J \cL^{+\sigma_2\sigma_3\sigma_4}_{j234} \nonumber\\
&\qquad\qquad\quad\btimes\cL^{\sigma_4(-\sigma_2)-(-\sigma_3)}_{42(-j)3} \prod_{m\neq\pm j} \delta_{\mu_m,0}\delta_{\mu_j,1}\delta_{\mu_{-j},1}\nonumber\\ &\qquad\qquad\quad \btimes E_T\left(-(\tw_j+\tw_{-j},-\sigma_4\tw_4+\sigma_7\tw_7)\right)\bigg\}
\end{align}
where the definitions given in Section \ref{Sec: Feynman} still hold for the summations, with slight variation:
\begin{align}
{\sum}_{(4)} &\doteq \sum_{j=1}^M\sum_{\ul{\sigma}=\left(1,1,\sigma_3,\sigma_3,\sigma_5,-\sigma_5,-\sigma_3\right)}{\sum_{\bk_2...\bk_7}}' \delta_{ \bk_4,-\bk_3}\delta_{ \bk_j,\bk_2}\delta_{ \bk_3,\bk_7}\delta_{ \bk_5,\bk_6}\nonumber\\
{\sum}_{(5)} &\doteq\sum_{j=1}^M\sum_{\ul{\sigma}=\left(1,\sigma_2,-\sigma_2,-1,1,\sigma_6,-\sigma_6\right)}{\sum_{\bk_2...\bk_7}}' \delta_{ \bk_j,-\bk_4}\delta_{ \bk_j,\bk_5}\delta_{ \bk_3,\bk_2}\delta_{ \bk_6,\bk_7}\;\nonumber\\
{\sum}_{(6)} &\doteq \sum_{j=1}^M\sum_{\ul{\sigma}=\left(1,\sigma_2,\sigma_3,\sigma_4,-\sigma_2,1,-\sigma_3\right)}{\sum_{\bk_2...\bk_7}}' \delta_{234}^j\delta_{ \bk_j,\bk_6}\delta_{ \bk_3,\bk_7}\delta_{ \bk_5,\bk_2}\;\nonumber\\
{\sum}_{(7)} &\doteq\sum_{j=1}^M\sum_{\ul{\sigma}=\left(1,-1,\sigma_3,\sigma_4,\sigma_5,-\sigma_5,-\sigma_3\right)}{\sum_{\bk_2...\bk_7}}'\delta_{ -\sigma_4\bk_4,\sigma_3\bk_3}\delta_{ \bk_j,-\bk_2}\delta_{ \bk_3,\bk_7}\delta_{ \bk_5,\bk_6}\; \nonumber\\
{\sum}_{(8)}& \doteq \sum_{j=1}^M\sum_{\ul{\sigma}=\left(1,\sigma_2,-\sigma_2,\sigma_4,\sigma_5,-\sigma_5,\sigma_7\right)}{\sum_{\bk_2...\bk_7}}' \delta_{ \bk_j,\sigma_4\bk_4}\delta_{ \bk_j,\sigma_7\bk_7}\delta_{ \bk_3,\bk_2}\delta_{ \bk_5,\bk_6}\;\nonumber\\
{\sum}_{(9)}& \doteq \sum_{j=1}^M\sum_{\ul{\sigma}=\left(1,\sigma_2,\sigma_3,\sigma_4,-\sigma_2,-1,-\sigma_3\right)}{\sum_{\bk_2...\bk_7}}'\delta_{234}^j\delta_{ \bk_j,-\bk_6}\delta_{ \bk_3,\bk_7}\delta_{ \bk_5,\bk_2}
\end{align}
\item $\mathbf{\cJ_4}\qquad$ The prefactor is of order $O(L^{2d})$ and $\sqrt{J_1...}$ is $O\left(L^{-4d}\right)$. One needs at least a type I diagram with two continuous free wavenumbers to make $\cJ_4$ order $O(1)$. All other diagrams are subleading. In the graph of $\cJ_4$ in Fig.\ref{Fig.: figure24}, $\bk_1$ is pinned to an external blob ($\mu_1=2$). The leading order diagrams are type I, with two free wavenumbers (Figs. \ref{Fig.: figure25} and \ref{Fig.: figure26}).
\begin{itemize}
\item Diagrams 1 and 2 in Fig. \ref{Fig.: figure25} are vanishing, as already shown.
\item Diagrams 3 and 4 (Fig. \ref{Fig.: figure26}) have two free continuous wavenumbers, responsible for two degrees of freedom. Considering the prefactor ($\frac{1}{2}\lambda_1^2+\frac{\lambda_1\mu_1}{2\tJ_1}$, because here $\frac{\mu_1}{2}-1=0$), the $\sqrt{J_1...}$ term and the two free sums, both diagrams give an $O(1)$ contribution to $\cJ_4$.\\\\
\textbf{Diagram 3:}
$\bk_1=\sigma_2\bk_2+\sigma_3\bk_3+\sigma_4\bk_4,\quad \bk_2=\bk_3,\quad \sigma_3=-\sigma_2 \; \Rightarrow \;\bk_1=\sigma_4\bk_4$, $ 
\bk_1=\sigma_7\bk_7 \;\Rightarrow \;\sigma_4\bk_4=\sigma_7\bk_7$, 
$\quad \bk_4=\bk_7, \quad \sigma_4=\sigma_7$.
If $\sigma_4=+1,$ then $\bk_1=\bk_4$. Because of {\it Rule 5}, the left vertex is vanishing and so the diagram is not contributing. If $\sigma_4=-1,$ then $\bk_4=-\bk_1$, $\sigma_7=-1$ and $\bk_7=\bk_4=-\bk_1$. There are 9 graphs equivalent to diagram 3, obtained substituting 4 with 2 and 3 and substituting 7 with 5 and 6, cf.\ (\ref{J4d}).

\textbf{Diagram 4.}
If $\sigma_4=-1,$ then $\bk_4=-\bk_1$, $\sigma_7=1$ and $\bk_7=\bk_1$. Because of {\it Rule 5}, the right vertex is vanishing and the diagram does not contribute.
If $\sigma_4=+1,$ then $\bk_4=\bk_1$. The left vertex vanishes. Therefore, this graph does not contribute. Therefore:
\end{itemize}
\begin{align}
\Big\langle e^{\sum_\bk i \lambda_\bk J_\bk}\cJ_4 \Big\rangle_J \sim  \left(\frac{2\pi}{L}\right)^{2d}&9\sum_{(10)}\Big\langle \Big[-\frac{1}{2}\lambda_j^2+i\frac{\lambda_j}{\tJ_j}\Big]  \tJ_j\tJ_{-j}\tJ_2\tJ_5 e^{\sum_m i \lambda_m\tJ_m} \Big\rangle_J \nonumber\\
&\btimes \cL^{+\sigma_2\sigma_3\sigma_4}_{j234}\cL^{+\sigma_5\sigma_6\sigma_7}_{j567}  \delta_{\mu_j,2}\delta_{\mu_-j,2}\prod_{m\neq\pm j}\delta_{\mu,0}\nonumber\\
&\btimes\Delta_T\left(-\tw_j-\tw_{-j}\right)\Delta_T\left(-\tw_j-\tw_{-j}\right) \label{J4d}
\end{align}
\be {\sum}_{(10)} \doteq \sum_{j=1}^M\sum_{\ul{\sigma}=\left(1,\sigma_2,-\sigma_2,-1,\sigma_5,-\sigma_5,-1\right)}{\sum_{\bk_2...\bk_7}}'\delta_{\bk_1,-\bk_4}\delta_{ \bk_4,\bk_7}\delta_{ \bk_5,\bk_6}\delta_{ \bk_2,\bk_3} \ee

\item $\mathbf{\cJ_5}\qquad$ The order $O(L^{2d})$ from the prefactor and the $O\left(L^{-4d}\right)$ from the $\sqrt{J_1...}$ term lead to a global $O\left(L^{-2d}\right)$. So, we have to seek graphs with two free continuous 
wavenumbers, cf.\ (\ref{J5d}).
\begin{itemize}

\item Diagram 1 in Fig. \ref{Fig.: figure30} (for $B1$) and a similar one for $B_2$, with the arrow referred to 1 reversed, have two free discrete wavenumbers. $\mu_1$ and $\mu_2$ are identically null due to the internal couplings of $\bk_1$ and $\bk_2$. At each vertex, the sum of the two discrete wavenumbers is discrete. Then, because of the vertex condition (``momentum conservation'') only one continuous free wavenumber remains (so as in one dimension, if the sum of the discrete wavenumbers is an integer $n$, the two continuous wavenumbers can be chosen as a real $x$ and the constrained $-n-x$, so that $n+x+(-n-x)=0$). Thus, diagram 1 does not contribute to the leading order of $\cJ_5$.

\item Diagram 3 in Fig. \ref{Fig.: figure31} (referred to $B_1$) has two continuous free wavenumbers (in one dimension, the three continuous wavenumbers are $x$, $y$, $-x-y-n$, if the discrete wavenumber is $n$, so that $n+x+y+(-x-y-n)=0$ at each vertex). This diagram is leading order and its multiplicity is 6.

\item Let us turn our attention to diagram 5 in Fig. \ref{Fig.: figure32}. The condition of the left vertex delta gives $\bk_1=\sigma_3\bk_3$. There are two cases:\\
if $\sigma_3=+1\; \Rightarrow \; \bk_1=\bk_3$, because of {\it Rule 5} this contribution is vanishing;\\
if $\sigma3=-1\; \Rightarrow \; \bk_1=-\bk_3$, $\bk_1=\bk_6$, $\sigma_6=+1$,
the right vertex conditions says: $\bk_2=\sigma_6\bk_6=-\bk_6=-\bk_1$. This is forbidden by the condition on the sum, stating that $\bk_1\neq\bk_2$. Thus, this graph does not represent any physical interaction.

\item A diagram analogous to 5 in Fig. \ref{Fig.: figure32}, but with the ``1'' arrow reversed, is possible for $B_2$. One has that $-\bk_1=\sigma_3\bk_3$.\\
If $\sigma_3=-1$: $\bk_1=\bk_3$, $\sigma_3=-1$, we have a vanishing contribution;\\
If $\sigma_3=+1$: $\bk_3=-\bk_1$, $\bk_6=\bk_1$, $\sigma_6=-1$.
The right vertex condition gives: $\bk_1=\sigma_6\bk_6\; \Rightarrow \; \bk_2=-\bk_6=-\bk_1\; (\neq \bk_1)$. We notice that $\bk_2=\bk_3$, and $\bk_2$ is a sink while $\bk_3$ is a source. The total momentum is conserved. The multiplicity of such a graph is 9.

\item A graph analogous to 5 in Fig. \ref{Fig.: figure32} is also possible for $B_3$, if one reverses the ``2'' arrow.\\
The case $\sigma_3=+1$ requires $\bk_1=\bk_3$, we have a vanishing contribution;\\
If $\sigma_3=-1$, $\bk_3=-\bk_1$, $\bk_6=\bk_1$,\\
$\sigma_6=1$, $\sigma_6\bk_6=-\bk_2\; \Rightarrow\; \bk_2=-\bk_6\; (\neq \bk_1)$.
We have that $\bk_2=\bk_3$, but now $\bk_2$ is the source and $\bk_3$ is the sink. It is the simmetric of the situation previously described for $B_2$. The multiplicity of this contribution is 9.

\item One more contribution comes from a graph analogous to diagram 5 in Fig. \ref{Fig.: figure32} but with 1 and 2 exchanged. In such a case $\mu_2=0$, so it is useless to compute the contribution from $B_3$. Then, one can easily notice that the contribution from $B_2$ is exacly the same as that just calculated for $B_3$ at the previous point, but permuting 1 with 2 and using the right prefactor. The multiplicity of this graph is 9 too.

\item The last non-zero contribution comes from diagram 6 in Fig. \ref{Fig.: figure33}. Here, $\sigma_6=-1$, $\sigma_3=-1$, $\bk_1=-\bk_3$, $\bk_6=\bk_1$ and $\bk_2=\sigma_6\bk_6=-\bk_1$. There is also an identical contribution permuting 1 with 2, so we add a factor $2$ to the multiplicity, which become equal to 18.
\end{itemize}
Finally, the $\cJ_5$ total contribution is the following:
\begin{align}
\Big\langle e^{\sum_\bk i \lambda_\bk J_\bk}&\cJ_5 \Big\rangle_J \sim  \frac{1}{2}\left(\frac{2\pi}{L}\right)^{2d} \nonumber \\
&\btimes\bigg\{6\sum_{(11)}\Big\langle  \Big[-\lambda_j\lambda_{-j}+i\frac{\lambda_j}{\tJ_{-j}}+\frac{1}{4\tJ_j\tJ_{-j}}+\Big]  \sqrt{\tJ_j\tJ_{-j}}\tJ_3\tJ_4\tJ_5\nonumber \\
&\qquad\qquad \btimes e^{\sum_m i \lambda_m \tJ_m}\Big\rangle_J\cL^{+\sigma_3\sigma_4\sigma_5}_{j345}\cL^{-(-\sigma_6)(-\sigma_7)(-\sigma_8)}_{(-j)345} \delta_{\mu_j,1}\delta_{\mu_{-j},1} \nonumber\\
&\qquad\qquad\btimes \prod_{m\neq \pm j}\delta_{\mu,0} \Delta_T\left(\tw^j_{345}\right)\Delta_T\left(\tw^{(-j)345}\right)\nonumber \\
&\quad -9\sum_{(12)}\Big\langle\lambda_j\lambda_{-j} \tJ_j\tJ_{-j}\tJ_4\tJ_7 e^{\sum...}\Big\rangle_J \cL^{-+\sigma_4(-\sigma_4)}_{j(-j)44}\cL^{+-\sigma_7(-\sigma_7)}_{(-j)j77} \nonumber \\
& \qquad\qquad\;\btimes  \delta_{\mu_{-j},1}\prod_{m\neq -j}\delta_{\mu,0}\; \left|\Delta_T\left(-\tw_{-j}-\tw_j\right)\right|^2\nonumber \\
&\quad+9 \sum_{(13)} \Big\langle\frac{\lambda_j}{\tJ_{-j}} \tJ_j\tJ_{-j}\tJ_4\tJ_7e^{\sum...}\Big\rangle_J  \cL^{+-\sigma_4(-\sigma_4)}_{j(-j)44}\cL^{-+\sigma_7(-\sigma_7)}_{(-j)j77} \nonumber \\
& \qquad\qquad\;\btimes \delta_{\mu_{-j},-1} \prod_{m\neq l}\delta_{\mu,0}\; \left|\Delta_T\left(\tw_j+\tw_{-j}\right)\right|^2 \nonumber\\
&\quad-9\sum_{(14)} \Big\langle\lambda_j\lambda_{-j} \tJ_j\tJ_{-j}\tJ_4\tJ_7e^{\sum...}\Big\rangle_J  \cL^{-+\sigma_4(-\sigma_4)}_{jj44}\cL^{+-\sigma_7(-\sigma_7)}_{(-j)(-j)77}\nonumber \\
& \qquad\qquad\;\btimes \delta_{\mu_j,-1} \prod_{m\neq j}\delta_{\mu,0}\; \left|\Delta_T\left(\tw_j+\tw_{-j}\right)\right|^2\nonumber \\
&\quad+18\sum_{(15)} \Big\langle \Big[-\lambda_j\lambda_{-j}+i\frac{2\lambda_j}{\tJ_{-j}}+\frac{1}{\tJ_j\tJ_{-j}}\Big] e^{\sum...}\Big\rangle_J  \tJ_j\tJ_{-j}\tJ_4\tJ_7 \nonumber \\ 
&\qquad\qquad\quad\btimes \cL^{+-\sigma_4(-\sigma_4)}_{j(-j)44}\cL^{-+\sigma_7(-\sigma_7)}_{(-j)j77} \delta_{\mu_j,2}\delta_{\mu_{-j},2}\prod_{m\neq j,l}\delta_{\mu,0} \; \Big(\Delta_T\left(-\tw_j-\tw_{-j}\right)\Big)^2\bigg\} \label{J5d}
\end{align}
\begin{align}
{\sum}_{(11)}& \doteq \sum_{j\neq l}\sum_{\ul{\sigma}=\left(1,1,\sigma_3,\sigma_4,\sigma_5,-\sigma_3,-\sigma_4,-\sigma_5\right)}{\sum_{\bk_3...\bk_8}}'\delta_{\bk_j,-\bk_l}\delta_{ \bk_3,\bk_6}\delta_{ \bk_4,\bk_7}\delta_{ \bk_5,\bk_8} \nonumber\\
{\sum}_{(12)}& \doteq \sum_{j\neq l}\sum_{\ul{\sigma}=\left(1,1,1,\sigma_4,-\sigma_4,-1,\sigma_7,-\sigma_7\right)}{\sum_{\bk_3...\bk_8}}'\delta_{\bk_j,-\bk_l}\delta_{ \bk_3,-\bk_j}\delta_{ \bk_6,\bk_j}\delta_{ \bk_7,\bk_8}\delta_{\bk_4,\bk_5} \nonumber\\
{\sum}_{(13)}& \doteq \sum_{j\neq l}\sum_{\ul{\sigma}=\left(1,-1,-1,\sigma_4,-\sigma_4,1,\sigma_7,-\sigma_7\right)}{\sum_{\bk_3...\bk_8}}'\delta_{\bk_j,-\bk_l}\delta_{ \bk_3,-\bk_j}\delta_{ \bk_6,\bk_j}\delta_{ \bk_7,\bk_8}\delta_{\bk_4,\bk_5} \nonumber\\
{\sum}_{(14)}& \doteq \sum_{j\neq l}\sum_{\ul{\sigma}=\left(-1,1,1,\sigma_4,-\sigma_4,-1,\sigma_7,-\sigma_7\right)}{\sum_{\bk_3...\bk_8}}'\delta_{\bk_j,-\bk_l}\delta_{ \bk_6,-\bk_l}\delta_{\bk_3,\bk_j}\delta_{\bk_7,\bk_8}\delta_{\bk_4,\bk_5}\nonumber\\
{\sum}_{(15)}& \doteq \sum_{j\neq l}\sum_{\ul{\sigma}=\left(1,1,-1,\sigma_4,-\sigma_4,-1,\sigma_7,-\sigma_7\right)}{\sum_{\bk_3...\bk_8}}'\delta_{\bk_j,-\bk_3}\delta_{ \bk_j,-\bk_l}\delta_{\bk_6,\bk_j}\delta_{\bk_7,\bk_8}\delta_{\bk_4,\bk_5}
\end{align}
\end{itemize}



\bibliographystyle{elsarticle-num} 
\bibliography{elsarticle-template-num}




\end{document}